\documentclass[5p, twocolumn, 10 pt]{elsarticle}

\usepackage{hyperref}
\usepackage{amsmath,amssymb}
\usepackage{graphicx}
\usepackage{subcaption}
\usepackage[export]{adjustbox}
\usepackage{tikz}
\usepackage{tikz-3dplot}
\usepackage{bm}
\usepackage{xcolor}
\usepackage{copyrightbox}
\usepackage{array}

\usetikzlibrary{calc,arrows.meta,positioning,backgrounds}

\DeclareMathOperator*{\argmax}{arg\,max}
\DeclareMathOperator*{\argmin}{arg\,min}
\newcommand{\hvec}[1]{\underline{\bm{\mathrm{#1}}}}
\newcommand{\vvec}[1]{\bm{\mathrm{#1}}}
\newcommand{\oB}{\mathrm{B}}
\newcommand{\oC}{\mathrm{C}}
\newcommand{\fB}{\mathcal{B}}
\newcommand{\fC}{\mathcal{C}}

\usepackage{booktabs} 
\newcommand{\ra}[1]{\renewcommand{\arraystretch}{#1}} 
\begin{document}

\tdplotsetmaincoords{-50}{-30}

\begin{frontmatter}

\title{AstroVision: Towards Autonomous Feature Detection and Description\\ for Missions to Small Bodies Using Deep Learning}



\author[gatech]{Travis Driver\corref{mycorrespondingauthor}}
\cortext[mycorrespondingauthor]{Corresponding author}
\ead{travisdriver@gatech.edu}

\author[umich]{Katherine A. Skinner}
\author[gatech]{Mehregan Dor}
\author[gatech]{Panagiotis Tsiotras}

\address[gatech]{Georgia Institute of Technology, Atlanta, Georgia, USA}
\address[umich]{University of Michigan, Ann Arbor, Michigan, USA}


\begin{abstract}
Missions to small celestial bodies rely heavily on optical feature tracking for characterization of and relative navigation around the target body. 
While deep learning has led to great advancements in feature detection and description, training and validating data-driven models for space applications is challenging due to the limited availability of large-scale, annotated datasets. 
This paper introduces AstroVision, a large-scale dataset comprised of 115,970 densely annotated, real images of 16 different small bodies captured during past and ongoing missions. 
We leverage AstroVision to develop a set of standardized benchmarks and conduct an exhaustive evaluation of both handcrafted and data-driven feature detection and description methods.  
Next, we employ AstroVision for end-to-end training of a state-of-the-art, deep feature detection and description network and demonstrate improved performance on multiple benchmarks. 
The full benchmarking pipeline and the dataset will be made publicly available to facilitate the advancement of computer vision algorithms for space applications. 
\end{abstract}


\end{frontmatter}



\section{Introduction}\label{sec:introduction}
There has been an increasing interest in missions to small bodies (e.g., asteroids, comets) due to their great scientific value, with four currently in operation (OSIRIS-REx, Hayabusa2, Lucy, DART) and two scheduled to launch over the next year (Psyche, Janus). 
In addition to planetary protection~\cite{cheng2018} and resource utilization~\cite{mazanek2015,rivkin2019}, small bodies are believed to be remnants from the solar system's formation, and studying their composition could provide insight into the solar system's evolution and the origins of organic life on Earth~\cite{barucci2011}. 

Feature tracking is an integral component of current small body shape reconstruction and relative navigation methodologies. 
However, the current state-of-the-practice relies heavily on humans-in-the-loop. 
Specifically, human operators on the ground manually identify salient surface features from images acquired during an extensive characterization phase, where the definition of saliency usually undergoes multiple iterations~\cite{norman2022autonomous}. 
Extracted features are then combined with \textit{a priori} global shape and spacecraft pose (position and orientation) estimates and used to iteratively construct a collection of digital terrain maps (DTMs), local topography and albedo maps, through a method known as stereophotoclinometry (SPC)~\cite{gaskell2008}. 
DTM construction typically involves extensive human-in-the-loop verification and carefully designed image acquisition plans to achieve optimal results~\cite{barnouin2020,palmer2022practical}.
These topographic features, along with global shape models, are critical for precision navigation and orbit determination for ground-based maneuvering and planning during data acquisition phases~\cite{antreasian2022scitech}.
Moreover, upon satisfying strict accuracy and resolution requirements, a catalog of DTMs can be uplinked to the spacecraft and correlated with onboard images to produce an onboard navigation solution for execution of safety-critical maneuvers~\cite{bhaskaran2011}, e.g., during the OSIRIS-REx Touch-And-Go (TAG) sample collection event~\cite{norman2022autonomous}.
While this manual approach has achieved much success, its reliance on extensive human involvement for extended durations limits mission capabilities and increases operational costs~\cite{quadrelli2015,nesnas2021,getzandanner2022scitech}. 

While automated feature tracking methods have been investigated to reduce reliance on current human-in-the-loop practices for missions to small bodies~\cite{dennison2021,morrell2020}, these works have focused exclusively on traditional \textit{handcrafted} features (e.g., SIFT~\cite{Lowe_2004IJCV}). 
More recently, feature detection and description methods that leverage deep \textit{convolutional neural networks} (CNNs) have been shown to significantly outperform handcrafted methods when applied to terrestrial imagery, especially in scenarios involving considerable change in illumination, scale, and perspective~\cite{Detone_2018CVPRW,dusmanu2019d2,revaud2019r2d2,luo2020cvpr}. 
However, transferring recent advances in deep learning to small body science applications is challenging due to the unavailability of relevant, annotated data~\cite{song2022}. 
To the best our knowledge, there exists no large-scale, annotated dataset comprised entirely of \textit{real} small body images. 
Indeed, previous work has relied entirely on simulated data~\cite{pugliatti2021,zhou2021exp,zhou2021sim}, small sets (i.e., $<$150 images) of manually annotated real imagery~\cite{fuchs2015}, or datasets restricted to a single body~\cite{lee2020}.
Moreover, operation in space presents a unique set of environmental (e.g., dynamic hard lighting, self-similar surface features) and operational (e.g., significant scale and perspective change during approach) challenges that are likely not adequately captured in available datasets based on terrestrial imagery.

This paper presents \textit{AstroVision}, a large-scale dataset comprised of 115,970 densely annotated, real images of 16 different small bodies from both legacy and ongoing deep space missions to bridge the \textit{terrestrial-to-extraterrestrial} domain gap and facilitate the study of deep learning for autonomous navigation in the vicinity of a small body.

The contributions of this paper are as follows:
(i) AstroVision is a \textit{first-of-a-kind} dataset for vision-based tasks in the vicinity of a small body with special emphasis on feature tracking applications;
(ii) we perform an exhaustive evaluation of both \textit{handcrafted} and \textit{data-driven} keypoint detection and feature description pipelines under challenging conditions on real imagery;
(iii) we employ AstroVision for end-to-end training of a state-of-the-art, deep feature detection and description network and demonstrate improved performance with respect to our benchmarks.
We make our dataset, benchmarking pipeline, and trained models publicly available at \href{https://github.com/astrovision}{\texttt{https://github.com/astrovision}}.


\section{Background}\label{sec:background}
In the following subsections, we detail the feature tracking process (Section \ref{sec:feature-tracking}) and feature-based pose estimation methodologies (Section \ref{sec:feature-pose}). 
For completeness,
we also provide a brief overview of structure-from-motion in Section \ref{sec:sfm}.  



\begin{figure}[tb!]
\centering
\begin{subfigure}[t]{\linewidth}
  \centering
  \includegraphics[width=\linewidth]{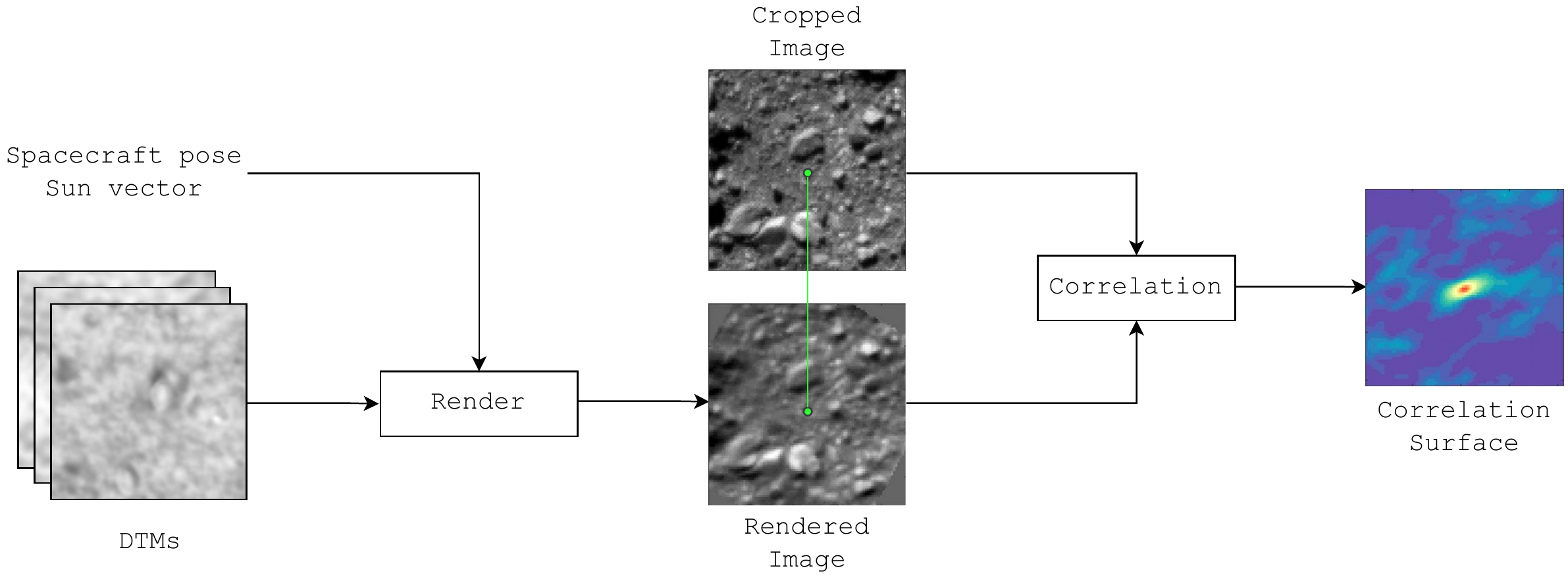}
  \caption{\textbf{DTM-based feature tracking.} DTMs are rendered by leveraging \textit{a priori} spacecraft pose and sun vector information, along with a photometric model, which is subsequently correlated with the input image to register a match. Adapted from \cite{olds2022psj}.}
  \label{fig:dtm-based-tracking}
\end{subfigure}\\
\bigskip
\begin{subfigure}[t]{\linewidth}
  \centering
  \includegraphics[width=\linewidth]{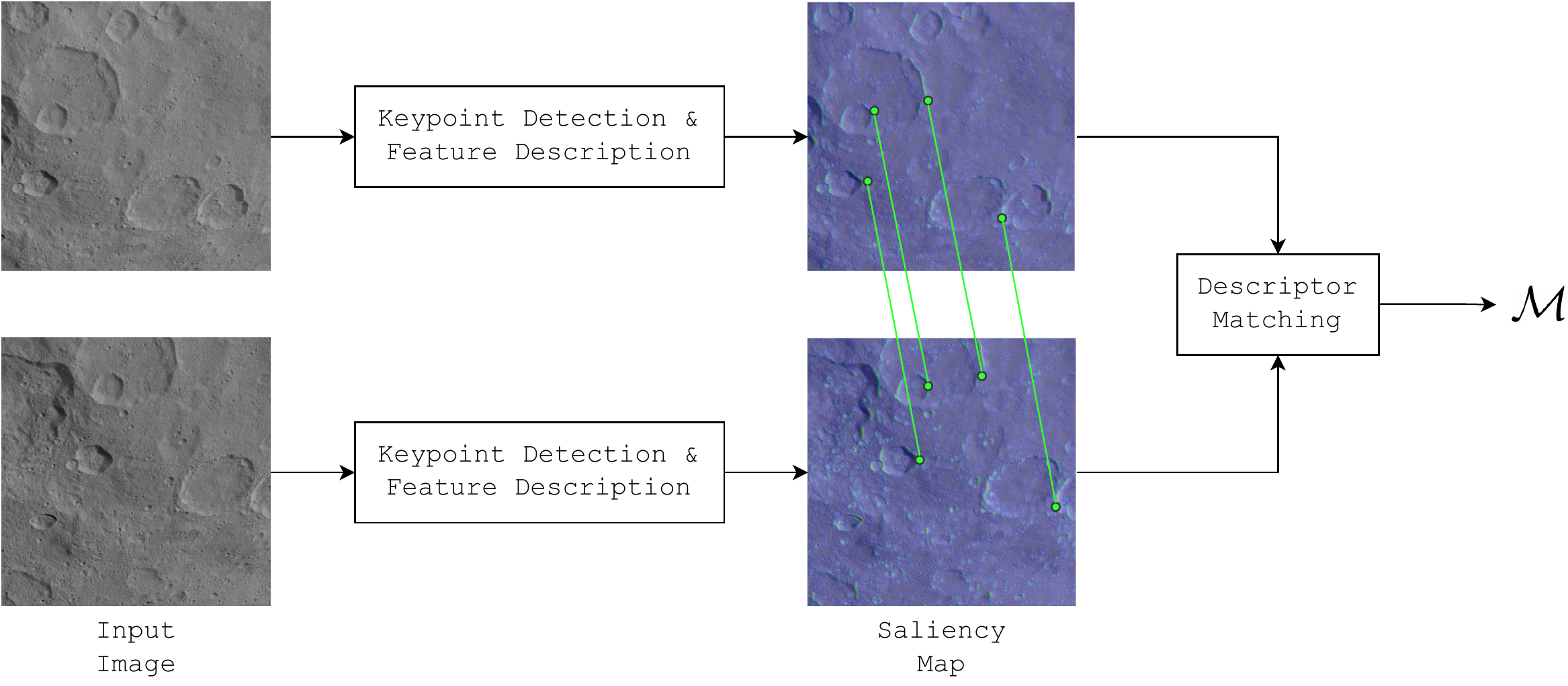}
  \caption{\textbf{Keypoint-based feature tracking.} Keypoints, extracted from each images' saliency map, and their associated descriptors abstract away the image, and tracking is performed by matching local descriptors between images.}
  \label{fig:keypoint-based-tracking}
\end{subfigure}%
\caption{\textbf{Feature tracking paradigms.}}
\label{fig:feature-tracking}
\end{figure}

\subsection{Feature Tracking} \label{sec:feature-tracking}

Robust tracking of salient image features is a critical component of current small body relative navigation methods, as the apparent displacement of tracked features between images can be leveraged to estimate the relative pose of the spacecraft as it moves around the body. 
In the context of optical feature tracking, saliency typically refers to the ability to detect and and precisely localize the feature under multiple viewing conditions (i.e., \textit{repeatability}) and to the distinctiveness of the feature to ensure accurate matching between images (i.e., \textit{reliability})~\cite{dusmanu2019d2,revaud2019r2d2}. 
The current state-of-the-practice for small body feature tracking leverages high-fidelity DTMs of salient surface regions as local feature representations, which require extensive human involvement and mission operations planning for accurate construction~\cite{barnouin2020,palmer2022practical}. 
Criteria for selecting salient features typically undergoes multiple iterations through testing and development of the DTMs~\cite{norman2022autonomous}.
Next, each DTM is combined with \textit{a priori} estimates of the spacecraft's pose and Sun pointing vector, along with a photometric model, to yield a photorealistic rendering of the DTM with respect to the input image. 
Finally, tracking is performed by comparing the rendering against the input image near the expected feature location using normalized cross-correlation, where a match is declared if a significant correlation peak is detected~\cite{lorenz2017,norman2022autonomous}. 
This process is illustrated in Figure \ref{fig:dtm-based-tracking}. 
The relative pose of the spacecraft when the image was taken can be computed using the registered matches and the \textit{a priori} DTM position estimates. 
Therefore, this DTM-based method relies on the fidelity of the \textit{a priori} data products and can only be utilized after the target body has been adequately observed and reconstructed at the required resolutions~\cite{bhaskaran2011}.

In this work we instead investigate approaches to feature tracking that rely on \textit{autonomous} keypoint detection and feature description. 
Consider two images $I: \Omega \rightarrow \mathbb{R}$ and $I': \Omega' \rightarrow \mathbb{R}$ with pixel domains $\Omega\subset \mathbb{R}^2$ and $\Omega'\subset \mathbb{R}^2$, respectively. 
\textit{Keypoints} $\vvec{p}_k \in \Omega$ ($\vvec{p}'_k \in \Omega'$) localize salient regions in the image, which are typically extracted from a saliency map $S: \Omega \rightarrow \mathbb{R}$.
Saliency can be predefined (e.g., corners) and localized using image filtering methods or learned from data (see Section \ref{sec:features}).

Feature description is the task of forming a latent representation of the local image data at detected keypoints, where the latent representation commonly takes the form of a $d$-dimensional vector $\vvec{d}_k \in \mathbb{R}^d$ referred to as the \textit{descriptor} associated with the keypoint $\vvec{p}_k$. 
Consider, for instance, corresponding keypoints $\{\vvec{p}_k\}_{k\in K}$ and $\{\vvec{p}'_k\}_{k\in K'}$ with correspondences defined by $\mathcal{M} := \left\{(k, \tau(k)) \mid \tau: K \leftrightarrow K'\right\}$. 
The overarching goal of feature description is to compute descriptors such that
\begin{equation}\label{eq:desc_obj}
    d(\vvec{d}_l, \vvec{d}'_{l'}) < \min\left(\min_{k \neq l} d(\vvec{d}_k, \vvec{d}'_{l'}), \min_{k' \neq l'} d(\vvec{d}_l, \vvec{d}'_{k'})\right)
\end{equation}
for all $(l, l') \in \mathcal{M}$, where $d(\cdot,\cdot)$ is some distance metric. 
In words, feature description seeks to assign a descriptor to each keypoint such that descriptors of corresponding keypoints are closer together than those of other non-corresponding keypoints. 
Common metrics $d(\cdot,\cdot)$ include the Euclidean distance, or the Hamming distance for binary descriptors~\cite{Rublee_2011ICCV}. 
We give an overview of different keypoint detection and feature description methodologies based on both handcrafted filtering approaches and deep learning in Section \ref{sec:features}.

Finally, feature tracking is conducted through detection of keypoints and matching of their corresponding descriptors between images. 
The objective defined in \eqref{eq:desc_obj} elicits a straightforward descriptor matching criterion referred to as \textit{mutual nearest-neighbors} (MNN):
\begin{multline}
    \mathcal{M} := \left\{(l, l') \mid d(\vvec{d}_l, \vvec{d}'_{l'}) < \min_{k \neq l} d(\vvec{d}_k, \vvec{d}'_{l'}) \right\} \\
    \bigcap \left\{(l, l') \mid d(\vvec{d}_l, \vvec{d}'_{l'}) < \min_{k' \neq l'} d(\vvec{d}_l, \vvec{d}'_{k'}) \right\}.
\end{multline}
In this work we leverage MNN with the Euclidean distance metric for feature matching between images. 
This keypoint-based tracking process is illustrated in Figure \ref{fig:keypoint-based-tracking}. 
Exploiting recently developed matching approaches based on deep learning ~\cite{sarlin2020cvpr} will be the subject of future work. 


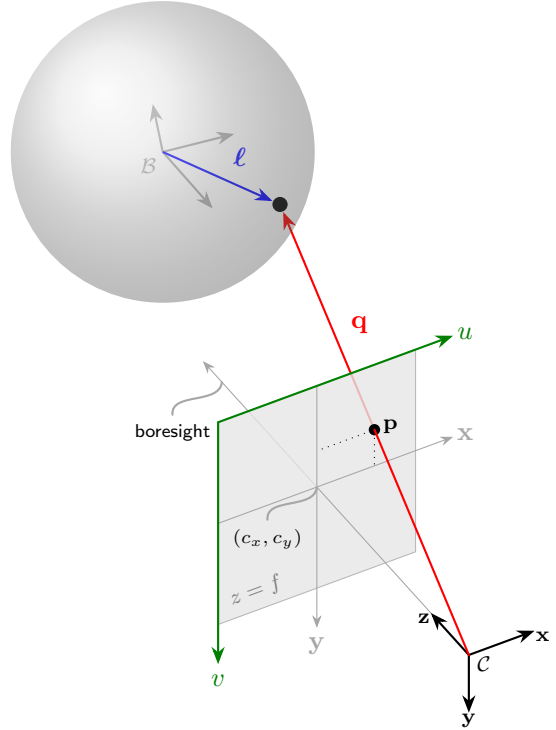
\begin{figure}[tb!]
    \centering
    \begin{tikzpicture}
  [
    tdplot_main_coords,
    >=Stealth,
    my dashed/.style={dashed, thick, ->, shorten >=-15pt, shorten <=-15pt, every node/.append style={font=\footnotesize}},
    my box/.style={thin, gray!70},
    my blue/.style={blue, style={fill, circle, inner sep=0pt, minimum size=#1*3.5pt, anchor=center, outer sep=0pt}},
    my label/.append style={midway, font=\scriptsize},
    my vectors/.style={black, {Stealth[scale=.75]}-{Stealth[scale=.75]}},
    my red/.style={thick, red, line cap=round},
    my grey/.style={gray!70},
    description/.style={draw=gray!70, thick, line cap=round, every node/.style={align=center, font=\scriptsize\sffamily, anchor=north}},
  ]
  \draw [->, my grey] (0,4,0) -- (0,7,0);
  \coordinate (o) at (0,0,0);
  \path [draw=gray!70, text=gray, fill=gray!20, opacity=0.8, text opacity=1, font=\footnotesize] (-1.5,4,1.75) coordinate (a) -- ++(0,0,-3.5) coordinate (b) -- ++(3,0,0) coordinate (c) -- ++(0,0,3.5) coordinate (d) -- cycle node [pos=.95, above, sloped, anchor=south west] {$z=f$} ;
  \draw [my grey] (0,0,0) -- (0,4,0);
  \draw [thick, ->, every node/.style={font=\footnotesize, inner sep=0pt}] (o) node [anchor=north west] {$\ \mathcal{C}$} (o) edge node [pos=1, anchor=north east] {$\vvec{z}$} ++(0,1,0) edge node [pos=1, anchor=north] {$\vvec{y}$} ++(0,0,1) -- ++(1,0,0) node [anchor=north west] {$\vvec{x}$};
  \coordinate (p2) at (1,4,-.5);
  \draw [my grey, thin, ->, shorten >=-15pt] ($(b)!1/2!(c)$) -- ($(d)!1/2!(a)$) node [below=15pt, anchor=north] {$\vvec{y}$};
  \draw [my grey, thin, ->, shorten >=-15pt] ($(b)!1/2!(a)$) -- ($(d)!1/2!(c)$) node [above right=17pt, anchor=north west] {$\vvec{x}$};
  \draw [->, thick, shorten >=-15pt, shorten <=-15pt, green!50!black, <->] (a) node [below=15pt, anchor=north] {$v$} -- (b) -- (c) node [above right=17pt, anchor=north west] {$u$};
  \path [black, every node/.style={font=\footnotesize, inner sep=0pt}] (p2) node [above right, anchor=south west] {$\vvec{p}$};
  \path (p2) ++(-.125,0,0) coordinate (q2) ++(0,0,-.125) coordinate (r2);
  \node[] (d1) at (r2) {};
  \node[black, circle, fill, minimum width=0.15cm, inner sep=0pt] (s2) at (r2) {};
  \node[black, circle, fill, minimum width=0.20cm, inner sep=0pt] (s3) at ($2.00*($(s2)$)$) {};
  \scoped[on background layer]{\draw [my blue=1.75] ($($1.75*($(s2)-(0,4,0)$)$)+(0,7,0)$) -- ++($1.75*($(r2)-(s2)$)$) node (d2) [label={[label distance=5pt, left]above:{}}] {};}
  \draw [black, dotted, thin] ($(s2)+(0,0,.6)$) -- (s2) node [below, my label, sloped] {};
  \draw [black, dotted, thin] ($(s2)-(.8,0,0)$) -- (s2) node [below, my label, sloped] {};
  \draw [my red] (o) -- (d1.center);
  \scoped[on background layer]{\draw [->, my red] (d1.center) -- (s3);}
  \node[red] (lc) at ($0.85*($(d2.center)$)$) [above right=7pt, anchor=north west] {$\vvec{q}$};
  \path [description] (0,4,0) [out=-95, in=95, font=\footnotesize] to (-.75,4,.25) node {$(c_x, c_y)$} (0,6.5,0) [out=-95, in=95, font=\footnotesize] to (-.75,6.5,.25) node {boresight};

  \coordinate (o2) at (1.7,11,0);
  \draw [my grey, thick, ->, every node/.style={font=\footnotesize, inner sep=2pt}] (o2) node [anchor=north east] {$\ \mathcal{B}$} (o2) edge node [pos=1, anchor=north east] {} ++(0.38,.92,0.0) edge node [pos=1, anchor=north] {} ++(.35,-.71,.61) -- ++(1,-.15,0) node [anchor=south east] {};
  \draw [blue, ->, thick] (o2) -- (s3);
  \node[blue] (lb) at ($(o2)+0.28*($(s3)-(o2)$)$) [above right=15pt, anchor=north west] {$\vvec{\ell}$};
  \shade[ball color = gray!40, opacity = 0.4] (o2) circle (2cm);
  
  
\end{tikzpicture}
    \caption{\textbf{Camera model geometry.}}
    \label{fig:camera-model}
\end{figure}

\subsection{Feature-based Pose Estimation}  \label{sec:feature-pose}

Consider a spacecraft equipped with a monocular camera navigating around a target small body. 
The relative pose between cameras can be estimated by tracking the apparent motion of salient surface \textit{landmarks} between images. 
Formally, let $\fB$ denote some body-fixed frame of the small body with origin $\oB$, and let $\fC_i$ denote the camera frame at time index $i$ with origin $\oC_i$.
Moreover, let $\vvec{\ell}^\fB_k = \left[\ell_{x,k}^\fB\ \ell_{y,k}^\fB\ \ell_{z,k}^\fB\right]^\top \in \mathbb{R}^3$ denote the vector from $\oB$ to the $k$th surface landmark expressed in $\mathcal{B}$,
let $\vvec{q}^{\fC_i}_k = \left[q_{x,k}^{\fC_i}\ q_{y,k}^{\fC_i}\ q_{z,k}^{\fC_i}\right]^\top \in \mathbb{R}^3$ denote the vector from $\oC_i$ to the $k$th landmark expressed in $\fC_i$, 
and let $\vvec{p}^{(i)}_k = \left[u^{(i)}_k\ v^{(i)}_k\right]^\top \in \mathbb{R}^2$ denote the 2D image coordinates of the $k$th landmark  observed by camera $\mathcal{C}_i$, i.e., the keypoint.

A landmark can be \textit{forward-projected} onto the image plane via
%
\begin{align} \label{eq:fproj}
    \hvec{p}^{(i)}_k = \Pi\left(\vvec{\ell}^\fB_{k}, T_{\fC_i\fB}; K\right) &= \frac{1}{d_{k}^{\fC_i}}
    \left[K\,|\,\bm{0}^{3\times1}\right] T_{\fC_i\fB} \hvec{\ell}^{\mathcal{B}}_{k} \nonumber \\
    &= \frac{1}{d_{k}^{\fC^i}} K \vvec{q}^{\fC_i}_k
\end{align}
where $d_{k}^{\fC_i} = q_{z,k}^{\fC_i}$ is the landmark depth in $\fC_i$, $\hvec{\ell}^{\mathcal{B}}_{k} = \left[\left(\vvec{\ell}^\fB_k\right)^\top\ 1\right]^\top \in \mathbb{P}^3$ and $\hvec{p}^{(i)}_k = \left[\left(\vvec{p}^{(i)}_k\right)^\top\ 1\right]^\top \in \mathbb{P}^2$ denote the homogeneous coordinates of $\vvec{\ell}^{\fB}_k$ and $\vvec{p}^{(i)}_k$, respectively, $T_{\fC_i\fB} \in \mathrm{SE}(3)$ denotes the relative pose of $\mathcal{B}$ with respect to $\mathcal{C}_i$:
\begin{equation}
    T_{\fC_i\fB} = 
    \begin{bmatrix}
    R_{\fC_i\fB} & \vvec{r}_{\oB\oC_i}^{\fC_i} \\
    \bm{0}^{1\times3} & 1
    \end{bmatrix},
\end{equation}
and $K$ is the camera calibration matrix:
\begin{equation}
K = 
    \begin{bmatrix}
    f_x & 0 & c_x \\
    0 & f_y & c_y \\
    0 & 0 & 1
    \end{bmatrix}
\end{equation}
where $f_x$ and $f_y$ are the \textit{focal lengths} in the $x$- and $y$-directions of the camera frame, and $(c_x, c_y)$ is the \textit{principal point} of the camera. 
The geometry of the pinhole camera model is illustrated in Figure \ref{fig:camera-model}. 
Conversely, a 2D keypoint may be \textit{backward-projected} into 3D coordinates via
%
\begin{align} \label{eq:bproj}
    \hvec{\ell}_{k}^{\fB} = \Pi^{-1}\left(\vvec{p}^{(i)}_k, d^{\fC_i}_k, T_{\fC_i\fB}; K\right) &= T_{\fC_i\fB}^{-1} \begin{bmatrix} d^{\fC_i}_k K^{-1} \hvec{p}^{(i)}_{k} \\ 1 \end{bmatrix} \nonumber \\
    &= T_{\fB\fC_i} \hvec{q}_k^{\fC_i}.
\end{align}
Then, given corresponding keypoints $\vvec{p}_k^{(i)}$ and $\vvec{p}_k^{(j)}$ observed by cameras $\fC_i$ and $\fC_j$, respectively, the \textit{essential matrix} $E := [\vvec{r}_{\oC_i\oC_j}^{\fC_j}]_\times R_{\fC_j\fC_i}$ satisfies
\begin{equation} \label{eq:ess_mat}
    \left(\hvec{p}_k^{(j)}\right)^\top K^{-\top}E K^{-1}\hvec{p}_k^{(i)} = 0,
\end{equation}
where we have assumed a shared camera matrix $K$ for simplicity, and $[\cdot]_\times$ denotes the skew-symmetric matrix cross-product matrix, defined for any $\vvec{r}\in \mathbb{R}^3$, such that  
\begin{equation}
 [\vvec{r}]_\times = 
    \begin{bmatrix}
    0 & -r_z & r_y \\
    r_z & 0 & -r_x \\
    -r_y & r_x & 0
    \end{bmatrix}.
\end{equation}
The well-known five-point algorithm~\cite{nister2004} can be used to solve for $E$ given five or more correspondences. 
Finally, $R_{{\mathcal{C}_j}{\mathcal{C}_i}}$ and $\vvec{r}_{\oC_i\oC_j}^{\fC_j}$ (up to some unknown scale) can be estimated using singular value decomposition (SVD) of the associated essential matrix and by imposing the \textit{Cheirality constraint}, i.e., triangulating the landmark associated with keypoints $\vvec{p}_k^{(i)}$, $\vvec{p}_k^{(j)}$ and enforcing that the associated landmark lies in front of the cameras~\cite{forsyth2011modern}. 


\subsection{Structure-from-Motion} \label{sec:sfm}

In the structure-from-motion (SfM) or simultaneous localization and mapping (SLAM) setting, we are interested in \textit{simultaneously} estimating a collection of camera poses $\mathcal{T} := \left\{T_{\fC_i\fB}\in \mathrm{SE}(3) \mid i = 1,\ldots,m\right\}$ and a network of landmarks (the \textit{map}) $\mathcal{L} := \{\vvec{\ell}^\fB_k \in \mathbb{R}^3 \mid k = 1,\ldots,n\}$. 
Note that the SfM solution is innately expressed in some arbitrary body-fixed frame since most SfM techniques assume operation in a static scene, typically referred to as the``world" frame~\cite{cadena2016tr}. 
SfM seeks the maximum \textit{a-posteriori} (MAP) estimate of the poses $\mathcal{T}$ and landmarks $\mathcal{L}$, given the (independent) keypoint \textit{measurements} $\mathcal{P} := \{\Hat{\vvec{p}}^{(i)}_k \in \mathbb{R}^2 \mid i=1,\ldots,m, k=1,\ldots,n\}$:
\begin{align}
    \mathcal{T}^*, \mathcal{L}^* &= \argmax_{\mathcal{T}, \mathcal{L}} p\left(\mathcal{T}, \mathcal{L}\mid \mathcal{P}\right) \\
    &\propto \argmax_{\mathcal{T}, \mathcal{L}} p\left(\mathcal{T}, \mathcal{L}\right) p\left(\mathcal{P} \mid \mathcal{T}, \mathcal{L}\right) \\
    &= p\left(\mathcal{T}, \mathcal{L}\right) \prod_{i} \prod_{k} p\left(\Hat{\vvec{p}}^{(i)}_k \mid T_{\fC_i\fB}, \vvec{\ell}^\fB_k\right). \label{eq:map}
\end{align}
By assuming measurements $\Hat{\vvec{p}}^{(i)}_k$ are corrupted by zero-mean Gaussian noise, i.e., $\Hat{\vvec{p}}^{(i)}_k = \vvec{p}^{(i)}_k + \vvec{\eta}^{(i)}_k$ where $\vvec{\eta}^{(i)}_k \sim \mathcal{N}(\vvec{0},\Sigma^{(i)}_k)$, we get 
\begin{equation}
    p\left(\Hat{\vvec{p}}^{(i)}_k \mid T_{\fC_i\fB}, \vvec{\ell}^\fB_k\right) \propto \exp\left\{\|\Hat{\hvec{p}}_k^{(i)} - \Pi\left(\vvec{\ell}^\mathcal{B}_{k}, T_{\mathcal{C}_i\mathcal{B}}; K\right)\|_{\Sigma^{(i)}_k}^2\right\},
\end{equation}
where $\|\vvec{e}\|^2_{\Sigma} := \vvec{e}^\top\Sigma^{-1}\vvec{e}$. 
The MAP estimate can be formulated as the solution to a nonlinear least-squares problem by taking the negative logarithm of \eqref{eq:map}:
\begin{equation} \label{eq:ba}
    \mathcal{T}^*, \mathcal{L}^* = \argmin_{\mathcal{T}, \mathcal{L}} \sum_{i,k} \|\Hat{\hvec{p}}_k^{(i)} - \Pi\left(\vvec{\ell}^\fB_k, T_{\fC_i\fB}; K\right)\|_{\Sigma^{(i)}_k}^2,
\end{equation}
where we have omitted the priors $p\left(\mathcal{T}, \mathcal{L}\right)$ for conciseness and generality, which can be ignored if no prior information is assumed (i.e., $p\left(\mathcal{T}, \mathcal{L}\right) = const.$) or can encode relative pose constraints via known dynamical models~\cite{tweddle2015jfr}.
This process is commonly referred to as \textit{Bundle Adjustment} (BA). 
Note that the optimization process of SPC decouples estimation of the poses and landmarks, i.e., \textit{a priori} landmark position and camera pose estimates are passed back-and-forth between the pose determination and DTM construction steps, respectively, until convergence~\cite{gaskell2008}. 

In this work, we focus on two-view pose estimation by estimating the essential matrix using the five-point algorithm. 
Future work will focus on incorporating our feature detection and description methods into a full SfM pipeline.


\section{Related Work}\label{sec:related_work}

In this section, we give an overview of both handcrafted and data-driven feature detection and description methods (Section \ref{sec:features}), and then discuss existing datasets and benchmarks for vision tasks in the vicinity of a small body (Section \ref{sec:space-vision}) and data-driven relative navigation techniques (Section \ref{sec:space-deep-nav}).

\subsection{Feature Detection and Description} \label{sec:features}

Many computer vision algorithms rely on local image features. 
The seminal work of David Lowe's Scale Invariant Feature Transform (SIFT)~\cite{Lowe_2004IJCV} laid the foundation for the field, where he outlined a rigorous framework for identifying and describing image features.
SIFT follows a \textit{detect-then-describe} paradigm, whereby a series of predetermined (or \textit{handcrafted}) filters are applied to the image for keypoint localization, followed by pooling and normalization of image gradients to form the descriptor. 
SIFT aims to extract features that are invariant to changes in scale, illumination, and rotation. 
Keypoints are extracted from local extrema of the saliency map derived by convolving the  difference of Gaussians (DoG) kernel with the input image, as the DoG function provides a close approximation to the scale-normalized Laplacian of Gaussian function which has been shown to be scale invariant~\cite{lindeberg1994scale}. 
This detection scheme generally results in keypoints centered around large gradients in the image (e.g., edges, corners).
Descriptors are then computed by pooling gradients in a local window of each keypoint into histograms according to their orientation, where a canonical orientation is assigned to each keypoint according to the dominant gradient orientation.
The oriented histograms are then concatenated and normalized to form the descriptor vector. 
Speeded-up Robust Features (SURF) built upon the success of SIFT to enable more efficient feature detection and description by leveraging integral images to eliminate the need for computing the DoG~\cite{Bay_2008CVIU}.
Oriented FAST and Rotated BRIEF (ORB) has become a popular alternative to SIFT, especially for SLAM applications~\cite{Rublee_2011ICCV}. 
ORB is based on Features from Accelerated Segment Test (FAST) detectors~\cite{rosten2005iccv} and Binary Robust Independent Elementary Features (BRIEF) descriptors~\cite{calonder2010eccv} and outputs binary descriptor vectors, enabling more efficient matching. 

More recently, feature detection and description methods that leverage deep \textit{convolutional neural networks} (CNNs) have achieved state-of-the-art performance and have been shown to outperform handcrafted methods, especially in scenarios involving significant illumination, scale, and perspective change~\cite{Ono_2018NIPS,Detone_2018CVPRW,revaud2019r2d2,luo2020cvpr}.
The first data-driven methods focused on individual components of the full image processing pipeline, including keypoint detection~\cite{verdie2015cvpr}, orientation estimation~\cite{Yi_2016CVPR}, and feature description~\cite{SimoSerra_2015ICCV}.
Yi et al.~\cite{Yi_2016ECCV} developed the first complete learning-based pipeline, Learned Invariant Feature Transform (LIFT). 
LIFT uses a patch-based Siamese training architecture and implements each component of the traditional feature detector and descriptor scheme sequentially using CNNs. 
The approach relies on an incremental training procedure to pretrain each subnetwork component individually, with a final training phase that optimizes over the entire network end-to-end. 
LFNet~\cite{Ono_2018NIPS} proposed a sequential two-stage approach: the first stage learns keypoint detection and the second stage learns feature description. 
SuperPoint~\cite{Detone_2018CVPRW} developed a network composed of separate interest point and descriptor decoders that operate on a spatially reduced representation of the input image from a shared encoder network. 
Simulated data of simple geometric shapes is used to pre-train the interest point detector which is then combined with a random homographic warping procedure to train the network end-to-end in a \textit{self-supervised} fashion. 

Towards joint detection and description, the seminal work of D2-Net~\cite{dusmanu2019d2} proposed a \textit{detect-and-describe} approach that trains a single deep CNN to detect and describe salient image features. 
\textit{Reliability} (or \textit{distinctiveness}) of descriptors is enforced through a triplet margin ranking loss term which is weighted according to soft detection scores to jointly enforce \textit{repeatability} of detections. 
R2D2~\cite{revaud2019r2d2} leverages the detect-and-describe paradigm to perform simultaneous feature detection and description, but repeatability and reliability are enforced in separate terms in the loss function. 
Repeatability is enforced through maximization of the cosine similarity of the detection scores of corresponding image patches, while reliability of the descriptors is learned through maximizing a differentiable approximation of the average precision~\cite{he2018cvpr} between corresponding patch descriptors. 
ASLFeat~\cite{luo2020cvpr} builds upon the success of D2-Net and proposes a multi-level detection scheme to generate detection scores that enable more accurate keypoint localization, and leverages deformable convolutional networks (DCNs)~\cite{zhu2019deformable} to model local geometric variations in the image and learn more transformation invariant features. 
ASLFeat is trained using the BlendedMVS~\cite{yao2020blendedmvs} and GL3D~\cite{shen2018accv} datasets, which contain 125,623 high-resolution images of 543 different scenes annotated with depth information using scene reconstructions from a dense SfM pipeline.
Although the training data is exceptionally comprehensive, we seek to capitalize on the recent success of deep feature detection and description methods by training these models on domain-relevant data to increase feature tracking performance for missions to small bodies.


\subsection{Datasets and Benchmarks for Vision Tasks in the Vicinity of a Small Body} \label{sec:space-vision}

Morrell et. al~\cite{morrell2020} and Dennison et. al~\cite{dennison2021} conduct an extensive evaluation of handcrafted feature extraction methods on synthetic images of comet 67P and asteroid 433 Eros, respectively, where SIFT demonstrates superior overall performance with respect to the algorithms studied. 
While the results are promising, the experiments were conducted in a controlled, simulated environment of a single target body, and their benchmarks were not made publicly available. 
Conversely, we benchmark both handcrafted and data-driven feature detection and description methods on real imagery of multiple small bodies with different surface characteristics and under varying illumination, scale, and perspective. 

With respect to small body image datasets, we are only aware of the work by Zhou et. al~\cite{zhou2021exp,zhou2021sim}, which includes images of both mock-up and computer generated asteroid models. 
The authors fabricate in-house models to represent arbitrary small bodies as opposed to leveraging available models of asteroids observed from past or current small body missions. 
The authors do not apply their learned models on real mission imagery. 
In our work, we train and test our approach on real imagery.


\subsection{Data-driven Relative Navigation} \label{sec:space-deep-nav}

Fuchs et. al~\cite{fuchs2015} train a random forest classifier on patches extracted from 119 images of the comets Hartley 2 and Tempel 1. 
However, significant performance degradation is observed when applied to unseen bodies, demonstrating the necessity to train models on data from a diverse set of small body instances.
Pugliatti et. al~\cite{pugliatti2021} employ a custom U-Net for segmentation of small body images into a constrained set of classes (i.e., terminator, boulders, craters, surface, background) using synthetic images of 7 different small bodies (e.g., 101955 Bennu, 21 Lutetia).
However, the performance suffers when applied to real images.

Data-driven crater detection has received much attention, especially for lunar applications.  
Wang et. al~\cite{wang2018} leverages a lightweight CNN architecture pre-trained on Martian crater samples to extract feature maps, which are then fed into a fully convolutional architecture to perform crater detection. 
Detected craters are then matched against an \textit{a priori} database for localization.
Silburt et. al~\cite{silburt2019} implement a custom U-Net architecture to detect and identify craters from digital elevation maps (DEMs). 
Lee et al.~\cite{lee2020} employ a CNN-based object detector to discriminate between a catalog of handpicked lunar surface landmarks while also predicting landmark detection probabilities as a function of the Sun's relative azimuth and elevation. 
The reliance on a catalog of known landmarks for navigation and the specification of craters as the most salient features limit the range of applications of this technology. 
Instead of explicitly specifying the features-of-interest beforehand, we allow the network to learn the most salient features for a wide variety of surface characteristics. 


\section{The AstroVision Dataset}\label{sec:dataset}
\begin{figure*}[hbtp]
\centering
\input{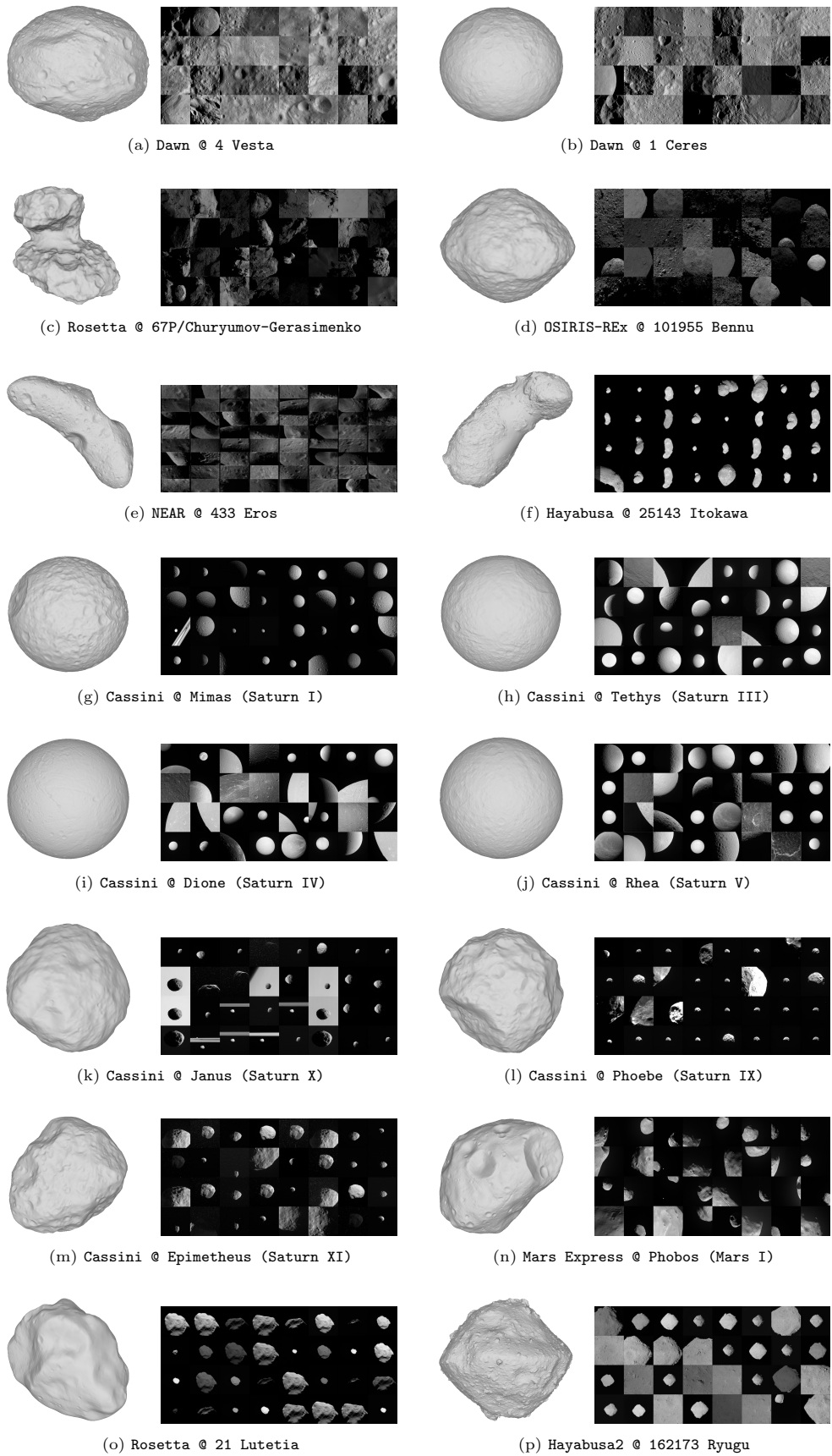}
\caption{\textbf{AstroVision image datasets}. Shape model references are provided in Table \ref{tab:mission-info}.}
\label{fig:astrovision-datasets}
\end{figure*}

In this section, we present our novel small body image dataset, referred to as AstroVision, for training and evaluation of keypoint detection and feature description methods. 
AstroVision features over 110,000 real images of 16 small bodies from 8 missions, as shown in Figure \ref{fig:astrovision-datasets}.
We describe the full data generation pipeline of AstroVision in the following subsections. 
Next, we develop a novel benchmarking suite (Section \ref{sec:evaluation}) and train a deep feature detection and description network (Section \ref{sec:results}) using our novel dataset.

\begin{table}[tb!]
\footnotesize
\centering
\ra{1.5}
\caption{\textbf{Dataset information.}}
\begin{adjustbox}{width=\linewidth}
\begin{tabular}{@{}lllrr@{}}
\toprule
Mission & Target & Type & \# Images & Shape Model Ref. \\
\midrule
Dawn~\cite{russell2011dawn} & 1 Ceres & Asteroid (G-type) & 38540 & Park et al.~\cite{park2019high} \\
                            & 4 Vesta & Asteroid (V-type) & 17504 & Gaskell et al.~\cite{gaskell2012spc} \\
\midrule
Cassini~\cite{dougherty2009saturn} & Dione (Saturn IV) & Icy Moon & 1381 & Gaskell~\cite{gaskell2020dione} \\
& Epimetheus (Saturn XI) & Icy Moon & 133 & Daly et al.~\cite{daly2018spc} \\
& Janus (Saturn X) & Icy Moon & 184 & Daly et al.~\cite{daly2018spc} \\
& Mimas (Saturn I) & Icy Moon & 307 & Gaskell~\cite{gaskell2020mimas} \\
& Phoebe (Saturn IX) & Icy Moon & 96 & Daly et al.~\cite{daly2018spc} \\
& Rhea (Saturn V) & Icy Moon & 665 & Daly et al.~\cite{gaskell2020tethys} \\
& Tethys (Saturn III) & Icy Moon & 751 & Daly et al.~\cite{gaskell2020tethys} \\
\midrule
Hayabusa~\cite{fujiwara2006rubble} & 25143 Itokawa & Asteroid (S-type) & 603 & Park et al.~\cite{park2019high} \\
\midrule
Hayabusa2~\cite{tsuda2020hayabusa2} & 162173 Ryugu & Asteroid (C-type) & 788 & Gaskell et al.~\cite{gaskell2021itokawa} \\
\midrule
Mars Express~\cite{bibring2006mars} & Phobos (Mars I) & Moon & 890 & Gaskell~\cite{gaskell2020phobos} \\
\midrule
NEAR~\cite{cheng1997near} & 433 Eros & Asteroid (S-type) & 11156 & Gaskell~\cite{gaskell2021eros} \\
\midrule
OSIRIS-REx~\cite{lauretta2017osiris} & 101955 Bennu & Asteroid (B-type) & 16618 & Barnouin et al.~\cite{barnouin2019bennu} \\
\midrule
Rosetta~\cite{taylor2017rosetta,schulz2012rosetta} & 67P/C-G & Comet & 26314 & Gaskell et al.~\cite{gaskell2014chury} \\
& 21 Lutetia & Asteroid (M-type) & 40 & Jorda et al.~\cite{jorda2013lutetia} \\
\toprule
\textbf{TOTALS}: 8 missions & 16 bodies & & \multicolumn{2}{l}{115,970 images} \\
\bottomrule
\end{tabular}
\end{adjustbox}
\label{tab:mission-info}
\end{table}

\subsection{Image and Ancillary Data Extraction}

AstroVision leverages publicly available images and ancillary data (i.e., camera pose, camera calibration, shape models) from both legacy and active small body science missions provided through NASA's Planetary Data System (PDS)~\cite{pds} and maintained by NASA's Navigation and Ancillary Information Facility. 
High-fidelity shape models (i.e., watertight, 3D triangular surface meshes) are developed as part of the relative navigation pipeline of small body missions, as they are critical for characterization of the body and relative navigation in subsequent phases.
Specifically, shape models for these missions are typically developed using SPC~\cite{gaskell2008}. 
SPC leverages feature correspondences between images captured during an extended characterization phase procured by human operators on the ground. 
A network of landmarks is estimated using stereophotogrammetry and subsequently densified using photometric stereo techniques via \textit{a priori} camera pose and sun pointing estimates and a reflectance model. 
The process yields high quality shape models that are precisely registered to the images and provide the foundation for our small body image dataset. 
For more details about the shape reconstruction and state estimation process, we refer the reader to \cite{gaskell2008}, \cite{bhaskaran2011}, and \cite{barnouin2020}.
Moreover, information and references for the various missions, images, and shape models used in this work are provided in Table \ref{tab:mission-info}.

Images provided by PDS are commonly stored using the Flexible Image Transport System (FITS), the standard data format used in astronomy, with pixel intensity values in units of either radiance ($\mathrm{W}\, \mathrm{s}^{-1}\, \mathrm{m}^{-2}$) or reflectance (unitless). 
We linearly scale pixel intensities to $[0, 1]$ before converting to a grayscale Portable Network Graphics (PNG) image. 
Photometrically calibrated (e.g., flat field and dark current correction) images were utilized when available. 
Moreover, we provide undistorted images to ensure alignment with the depth maps by leveraging geometric distortion estimates derived during a meticulous calibration procedure conducted both on the ground and during flight by mission scientists. 
See \ref{sec:photo_calib} for specific calibration details for each mission.


\subsection{Data Generation}

\begin{figure*}[htb!]
\centering
\input{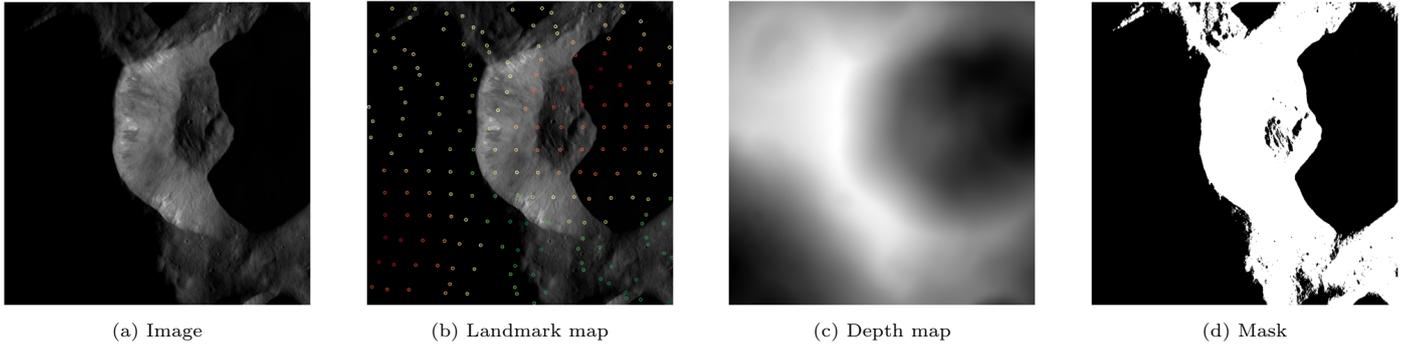}
\caption{\textbf{Example of AstroVision data products.}}
\label{fig:data-products}
\end{figure*}

The suite of AstroVision data products includes a landmark map, a depth map, and a mask for each image as shown in Figure \ref{fig:data-products}.
The \textit{landmark map} provides a consistent, discrete set of tie-points for sparse correspondence computation and is derived by forward-projecting vertices from a medium-resolution (i.e., $\sim$800k facets) shape model onto the image plane. 
We classify visible landmarks by tracing rays\footnote{Ray tracing uses the Trimesh library: \href{https://trimsh.org/}{https://trimsh.org/}} from the landmarks toward the camera origin and recording landmarks whose line-of-sight ray does not intersect the 3D model.
The \textit{depth map} provides a dense representation of the imaged surface and is computed by backward-projecting rays at each pixel in the image and recording the depth of the intersection between the ray and a high-resolution (i.e., $\sim$3.2 million facets) shape model. 
Finally, the mask provides an estimate of the non-occluded portions of the imaged surface. 

In order to generate the visibility masks, both global and dynamic intensity thresholding was used. 
For the more recent missions (i.e., Dawn, Hayabusa2, OSIRIS-REx, Rosetta), global thresholding was used. 
For some of the legacy missions (i.e., Cassini, Hayabusa, NEAR, Mars Express), variable vignetting was observed, primarily influenced by exposure time. 
Therefore, Otsu's method~\cite{otsu1979} was employed to compute a dynamic threshold for these instances.
While illuminated pixels could have been computed by tracing the Sun's incident light ray, estimating the mask independently of the ground truth scene geometry proved to be a useful tool for algorithmic outlier rejection, in addition to an extensive manual cleaning process. 
Specifically, we compute the ratio of the intersection area between the intensity mask and depth map and the total area of the mask as an alignment measure between the shape model and image, where a nominal value of $0.97$ was empirically chosen.
Moreover, we found that utilizing these intensity masks during training led to significant performance increases, which will be discussed further in Section \ref{sec:masking}.


\section{Small Body Feature Benchmarks}\label{sec:evaluation}
In this section, we conduct a comprehensive evaluation of existing feature detection and description methods using the proposed AstroVision dataset.
First, we detail our suite of performance metrics and verification procedures.  
Then, we present and discuss the benchmarking results.

\begin{table*}[htp!]
\footnotesize
\centering
\ra{1.5}
\caption{\textbf{AstroVision feature benchmarks}. Feature performance with respect to precision (P), recall (R), accuracy (A), and pose AUC in percentages. \textbf{First} and \underline{second} best results are bolded and underlined, respectively. See Section \ref{sec:metrics} for metric definitions.}
\begin{adjustbox}{width=0.95\linewidth}
\begin{tabular}{@{}lrlrrrr|rrr@{}}
\toprule
 Dataset                        &             &               &            &        &       &        & \multicolumn{3}{c}{AUC} \\
\cmidrule(lr){8-10}
 (Mean GSD, Median GSD)                        & \# Images   & Feature       & \# Matches &    P &   R &   A & @$5^{\circ}$ & @$10^{\circ}$ & @$20^{\circ}$ \\
\midrule 
 Cassini @ Epimetheus (Saturn XI) & 133         & ORB           &     \textbf{895} & 17.1 & 17.9 & 57.5 &     \textbf{2.9} &      \underline{9.2} &     \underline{14.9} \\
 (326.7 m/pixel, 255.4  m/pixel)  &             & SIFT          &     204 & \textbf{32.5} & \textbf{36.6} & 54.7 &     \underline{2.7} &      \textbf{9.5} &     \textbf{15.0} \\
                                  &             & SuperPoint    &     396 & 13.6 & 26.1 & 59.2 &     2.6 &      7.5 &     12.8 \\
                                  &             & R2D2          &     \underline{423} & 25.3 & 26.1 & \textbf{77.1} &     \textbf{2.9} &      9.1 &     14.7 \\
                                  &             & ASLFeat       &     386 & \underline{27.4} & \underline{29.0} & \underline{74.7} &     \underline{2.7} &      8.2 &     13.7 \\
\midrule 
 Cassini @ Mimas (Saturn I)       & 307         & ORB           &       \textbf{843} &   4.7 &   4.3 &   52.4 &       0.0 &        0.0 &        0.1 \\
 (1,176.2 m/pixel, 943.8 m/pixel) &             & SIFT          &       340 &  \underline{14.3} &  \underline{15.1} &   41.1 &       \textbf{0.2} &        \textbf{0.2} &        \textbf{0.4} \\
                                  &             & SuperPoint    &       121 &   8.6 &  10.4 &   50.5 &       0.0 &        0.0 &        0.0 \\
                                  &             & R2D2          &       209 &  13.8 &   8.8 &   \textbf{75.5} &       \underline{0.1} &        \underline{0.1} &        0.1 \\
                                  &             & ASLFeat       &       \underline{372} &  \textbf{21.8} &  \textbf{15.7} &   \underline{65.3} &       \textbf{0.2} &        \textbf{0.2} &        \underline{0.3} \\
\midrule 
 Dawn @ 1 Ceres                   & 3624        & ORB           &      1437 &  29.6 &  44.5 &   64.9 &       3.7 &        8.9 &       15.9 \\
 (122.0 m/pixel, 35.4 m/pixel)    &             & SIFT          &      \textbf{1656} &  42.3 &  \underline{72.2} &   69.4 &      \textbf{28.8} &       \textbf{44.3} &       \textbf{56.6} \\
                                  &             & SuperPoint    &       442 &  42.9 &  \textbf{75.7} &   70.1 &      \underline{13.1} &       \underline{28.3} &       \underline{43.5} \\
                                  &             & R2D2          &       954 &  \textbf{50.0} &  52.8 &   \textbf{85.8} &       8.9 &       20.0 &       32.4 \\
                                  &             & ASLFeat       &      \underline{1535} &  \underline{48.4} &  67.8 &   \underline{80.2} &      12.9 &       27.1 &       42.4 \\
\midrule 
 Dawn @ 4 Vesta                   & 2006        & ORB           &      \underline{1465} &  17.8 &  28.0 &   59.8 &       2.6 &        6.2 &       11.1 \\
 (63.3 m/pixel, 21.2 m/pixel)     &             & SIFT          &      1350 &  37.1 &  52.3 &   64.0 &      \textbf{17.9} &       \underline{28.7} &       \underline{38.8} \\
                                  &             & SuperPoint    &       506 &  38.7 &  \underline{55.0} &   65.8 &      11.3 &       21.3 &       32.7 \\
                                  &             & R2D2          &       926 &  \underline{55.9} &  46.7 &   \textbf{86.9} &      11.4 &       22.3 &       34.1 \\
                                  &             & ASLFeat       &      \textbf{1526} &  \textbf{59.1} &  \textbf{66.2} &   \underline{84.3} &      \underline{17.6} &       \textbf{32.0} &       \textbf{46.0} \\
\midrule 
 Hayabusa @ 25143 Itokawa         & 603         & ORB           &       \textbf{767} &   2.7 &   2.6 &   43.9 &       0.9 &        1.6 &        2.9 \\
 (95.5 cm/pixel, 78.7 cm/pixel)   &             & SIFT          &       217 &   4.8 &   5.0 &   35.8 &       1.9 &        3.3 &        4.8 \\
                                  &             & SuperPoint    &        79 &   7.3 &  \textbf{12.7} &   42.3 &       1.7 &        3.1 &        5.4 \\
                                  &             & R2D2          &       \underline{339} &  \underline{10.7} &   9.4 &   \textbf{67.0} &       \textbf{2.6} &        \textbf{4.6} &        \textbf{8.0} \\
                                  &             & ASLFeat       &       338 &  \textbf{13.5} &  \underline{11.3} &   \underline{47.5} &       \underline{2.2} &        \underline{4.2} &        \underline{7.6} \\
\midrule 
 OSIRIS-REx @ 101955 Bennu        & 1789        & ORB           &      \textbf{1581} &   4.9 &   5.2 &   54.0 &       0.3 &        0.8 &        1.6 \\
 (21.9 cm/pixel, 9.9 cm/pixel)    &             & SIFT          &      1317 &  13.7 &  15.3 &   55.2 &       \underline{5.6} &        \underline{8.8} &       11.8 \\
                                  &             & SuperPoint    &       747 &  18.1 &  \underline{20.3} &   55.4 &       3.8 &        7.3 &       11.1 \\
                                  &             & R2D2          &       502 &  \underline{29.3} &  18.3 &   \textbf{84.7} &       4.2 &        8.6 &       \underline{13.8} \\
                                  &             & ASLFeat       &      \underline{1378} &  \textbf{33.1} &  \textbf{30.9} &   \underline{68.7} &       \textbf{8.0} &       \textbf{14.4} &       \textbf{20.9} \\
\midrule 
 Rosetta @ 67P                    & 3039        & ORB           &      \textbf{1426} &  10.4 &   9.5 &   52.4 &       0.2 &        0.7 &        1.6 \\
 (5.5 m/pixel, 2.4 m/pixel)       &             & SIFT          &      \underline{1168} &  15.7 &  16.6 &   44.7 &       \underline{2.4} &        \underline{4.8} &        \underline{7.7} \\
                                  &             & SuperPoint    &       485 &  17.6 &  \underline{20.7} &   49.9 &       1.6 &        3.6 &        6.4 \\
                                  &             & R2D2          &       634 &  \underline{20.2} &  16.5 &   \textbf{79.3} &       1.9 &        3.9 &        7.1 \\
                                  &             & ASLFeat       &      1147 &  \textbf{25.0} &  \textbf{24.0} &   \underline{62.8} &       \textbf{3.4} &        \textbf{6.4} &       \textbf{10.6} \\
\midrule 
 Rosetta @ 21 Lutetia             & 40          & ORB           &       486 &  12.1 &  12.9 &   45.5 &       1.3 &        1.9 &        4.3 \\
 (230.5 m/pixel, 228.1 m/pixel)   &             & SIFT          &       283 &  23.7 &  \underline{31.7} &   46.6 &       \underline{5.9} &        \underline{9.8} &       15.9 \\
                                  &             & SuperPoint    &       381 &  26.7 &  30.7 &   55.5 &       4.2 &        8.0 &       \underline{16.2} \\
                                  &             & R2D2          &       \underline{588} &  \underline{33.2} &  25.6 &   \textbf{74.7} &       3.1 &        6.0 &       13.3 \\
                                  &             & ASLFeat       &       \textbf{970} &  \textbf{42.9} &  \textbf{35.0} &   \underline{71.9} &       \textbf{6.0} &       \textbf{12.1} &       \textbf{23.8} \\
\bottomrule
\end{tabular}
\end{adjustbox}\\
\label{tab:matching-metrics-all}
\end{table*}


\subsection{Performance Metrics}\label{sec:metrics}

We evaluate the matching performance on a per image pair basis using the standard metrics precision, recall, and accuracy. 
First, \textit{precision} defines the inlier ratio of the putative matches (as determined by our verification process decsribed in the following section):
\begin{equation}
    \text{precision} = \frac{\#\ \text{correct matches}}{\#\ \text{putative matches}}.
\end{equation}
Second, \textit{recall} describes the number of identified ground truth matches:
\begin{equation}
    \text{recall} = \frac{\#\ \text{correct matches}}{\#\ \text{ground truth matches}}.
\end{equation}
Third, \textit{accuracy} measures the matching performance with respect to the total number of computed features: 
\begin{equation}
    \text{accuracy} = \frac{\#\ \text{correct matches}\ \&\ \text{nonmatches}}{\#\ \text{features}}.
\end{equation}
We classify correct nonmatches as keypoints which were not included in the set of putative or ground truth matches, where we take the minimum of the number of such keypoints in each image in the pair~\cite{dennison2021}.

Finally, we compute the maximum of the angular error between the estimated and ground truth pose orientation and (unit) translation in degrees. 
Specifically, the angle of rotation between the estimated $\Tilde{\vvec{q}}_{\fC_j\fC_i}$ and ground truth $\vvec{q}_{\fC_j\fC_i}$ relative orientation quaternions
\begin{equation} \label{eq:err_q}
    \epsilon_q := \cos^{-1}(\langle \Tilde{\vvec{q}}_{\fC_j\fC_i}, \vvec{q}_{\fC_j\fC_i} \rangle^2 - 1)
\end{equation}
is used as a metric~\cite{huynh2009metrics} for the orientation error, and 
\begin{equation}
    \epsilon_t := \cos^{-1}\left(\Tilde{\vvec{r}}_{\oC_i\oC_j}^{\fC_j} \cdot \vvec{r}_{\oC_i\oC_j}^{\fC_j} \middle/ \|\Tilde{\vvec{r}}_{\oC_i\oC_j}^{\fC_j}\| \|\vvec{r}_{\oC_i\oC_j}^{\fC_j}\|\right)
\end{equation}
provides a measure of the translation error. 
The final pose error metric is taken to be $\epsilon := \max(\epsilon_q, \epsilon_t)$. 
The normalized cumulative error curve for $\epsilon$ is computed for each test sequence and the area under the curve (AUC) is reported for thresholds of 5$^{\circ}$, 10$^{\circ}$ and 20$^{\circ}$. 
We compute AUC using the explicit integration procedure of \cite{sarlin2020cvpr} rather than coarse histograms.


\subsection{Implementation}\label{sec:eval_implementation}

We evaluated the performance of ORB~\cite{Rublee_2011ICCV} and SIFT~\cite{Lowe_2004IJCV} as two representatives of \textit{handcrafted} features. 
Three state-of-the-art \textit{data-driven} features were selected that leverage different learning approaches (previously detailed in Section \ref{sec:background}): SuperPoint~\cite{Detone_2018CVPRW}, R2D2~\cite{revaud2019r2d2}, and ASLFeat~\cite{luo2020cvpr}.
We use the OpenCV implementations of ORB and SIFT and the open source implementations and pretrained models of the learned features made available by the respective authors. 
Each feature is limited to detect 5,000 keypoints and descriptors. 

Given a set of keypoints and descriptors, putative matches are computed using MNN. 
Matches are verified by first backward-projecting (via Equation \eqref{eq:bproj}) each keypoint in the first image into 3D world coordinates using the ground truth calibration and depth map. 
The 3D points are then forward-projected (via Equation \eqref{eq:fproj}) into the second image, and matches are verified by checking that the projected image coordinates are within some distance $\gamma$ to the keypoint of its matched feature, where we empirically chose a value of $\gamma = 5$ pixels. 
Ground truth matches are estimated in a similar way for computing recall, where a ground truth match is registered if there exists a keypoint within $\gamma = 5$ pixels of the projected image coordinate.  

Finally, poses are computed from the putative matches by first estimating the essential matrix using the five-point method~\cite{nister2004}, implemented in OpenCV’s \texttt{findEssentialMat} function, and RANSAC with an inlier threshold of 1 pixel, followed by SVD of the essential matrix to determine the relative pose, implemented in OpenCV's \texttt{recoverPose} function. 
Evaluation is conducted for 2$N$ randomly generated image pairs with at least 20\% overlap with respect to the landmark map, where $N$ is the number of images in the respective test dataset rounded up to the nearest multiple of 100, and metrics are averaged over all the image pairs.


\begin{figure*}[hp!]
\centering
\begin{subfigure}[t]{0.20\linewidth}
  \centering
  \begin{subfigure}[t]{\linewidth}
    \includegraphics[width=.935\linewidth,right]{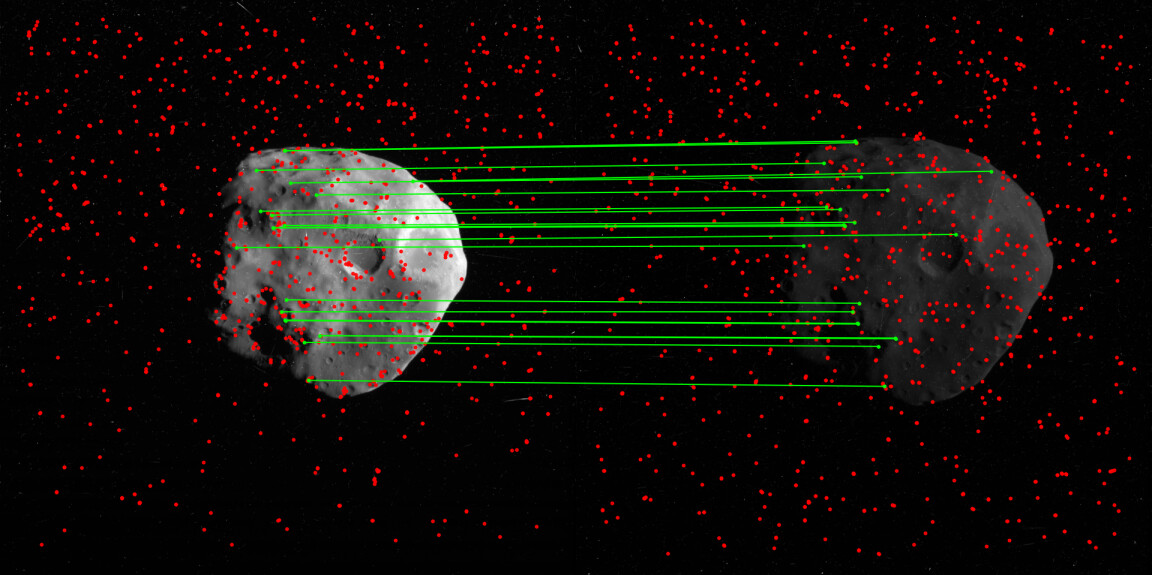}
  \end{subfigure}\\
  \vspace{1pt}
  \begin{subfigure}[t]{\linewidth}
    \begin{tabular}{c c}
         \hspace{-10.85pt}
         \rotatebox[origin=c]{90}{\tiny{(a) \texttt{Epimetheus}}}
         \hspace{-18pt}
         &
         \includegraphics[valign=m,width=.935\linewidth]{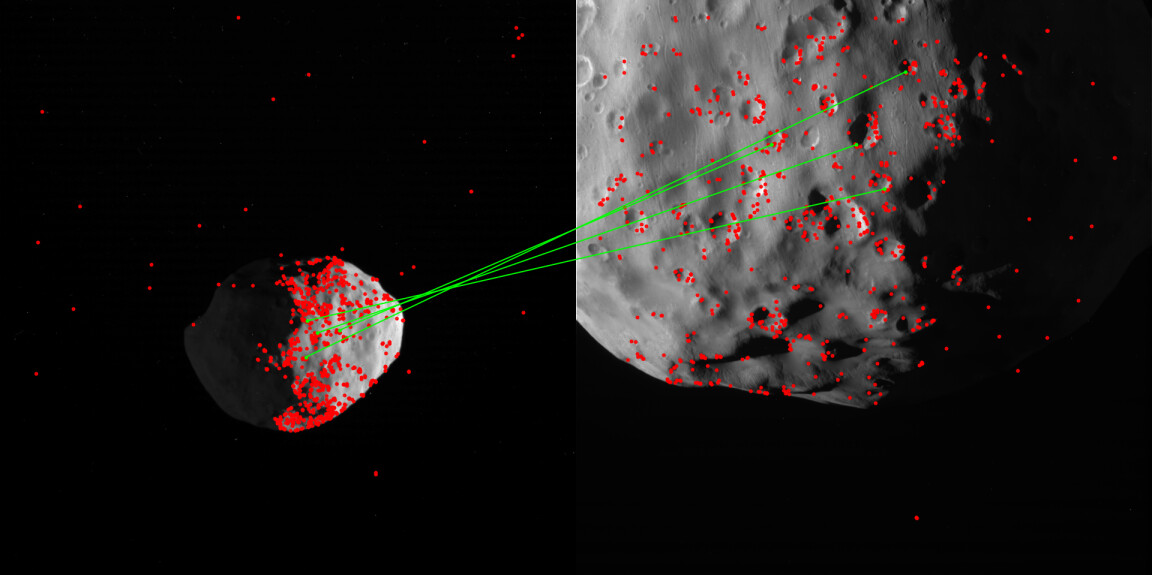} 
    \end{tabular}
    \refstepcounter{subfigure}\label{fig:eval-epim}
  \end{subfigure}\\
  \vspace{3pt}
  \begin{subfigure}[t]{\linewidth}
    \begin{tabular}{c c}
         \hspace{-10.85pt}
         \rotatebox[origin=c]{90}{\tiny{(b) \texttt{Mimas}}}
         \hspace{-18pt}
         &
         \includegraphics[valign=m,width=.935\linewidth]{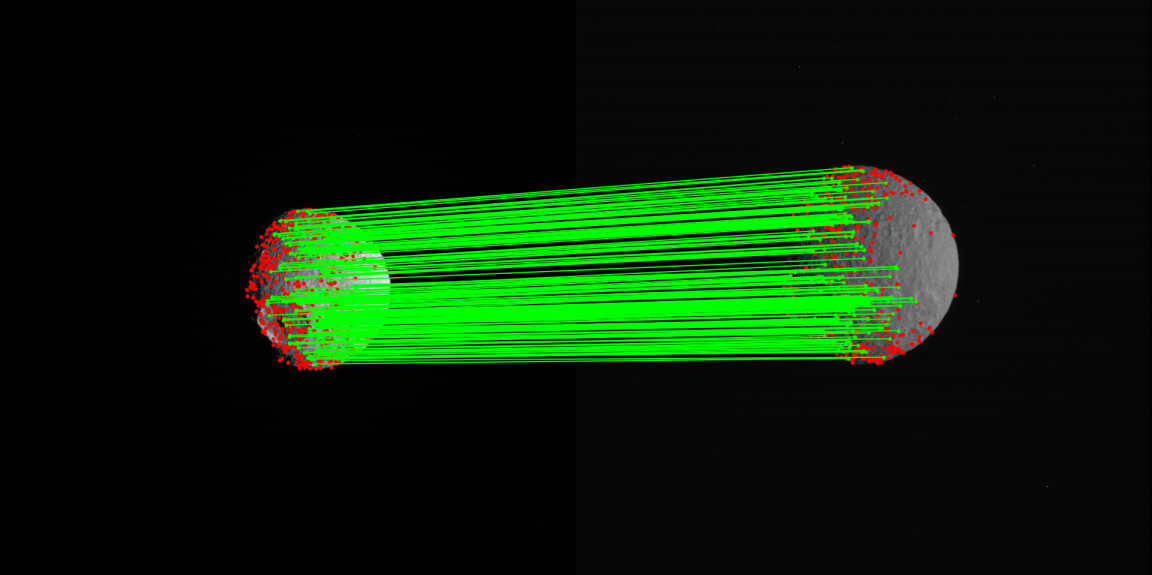}
    \end{tabular}
    \refstepcounter{subfigure}\label{fig:eval-mimas}
  \end{subfigure}\\
  \vspace{3pt}
  \begin{subfigure}[t]{\linewidth}
    \includegraphics[width=.935\linewidth,right]{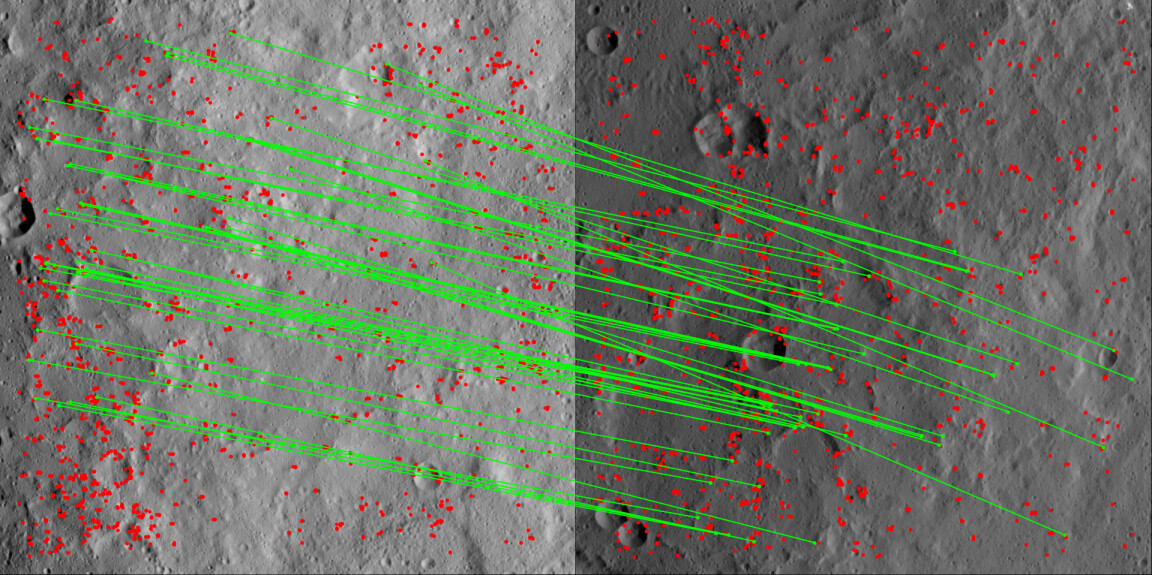}
  \end{subfigure}\\
  \vspace{1pt}
  \begin{subfigure}[t]{\linewidth}
    \begin{tabular}{c c}
         \hspace{-10.85pt}
         \rotatebox[origin=c]{90}{\tiny{(c) \texttt{1 Ceres}}}
         \hspace{-18pt}
         &
         \includegraphics[valign=m,width=.935\linewidth]{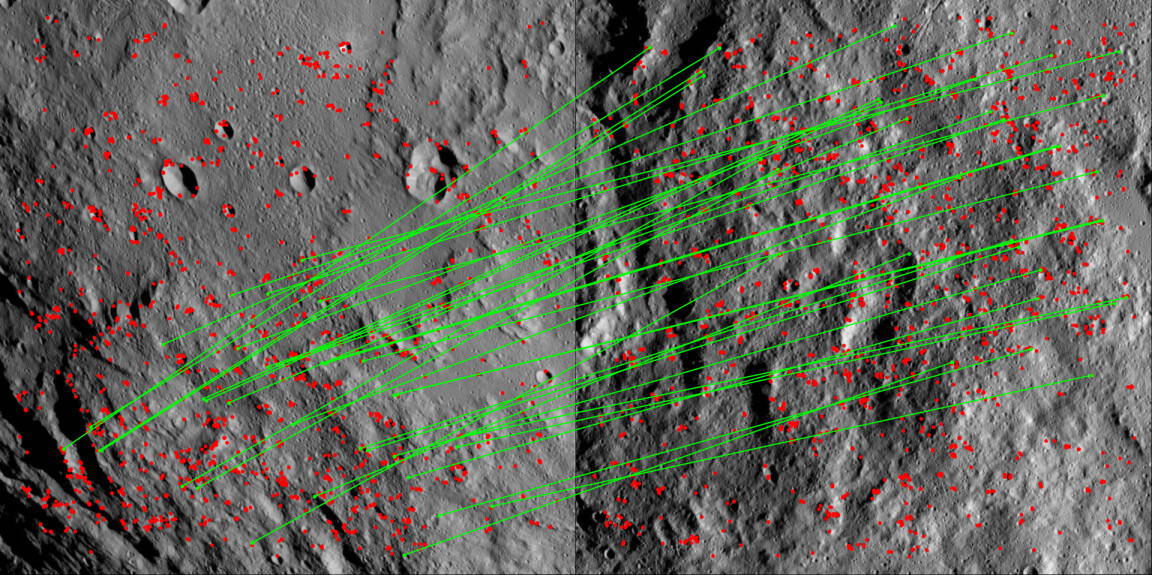}
    \end{tabular}
    \refstepcounter{subfigure}\label{fig:eval-ceres}
  \end{subfigure}\\
  \vspace{3pt}
  \begin{subfigure}[t]{\linewidth}
    \begin{tabular}{c c}
         \hspace{-10.85pt}
         \rotatebox[origin=c]{90}{\tiny{(d) \texttt{4 Vesta}}}
         \hspace{-18pt}
         &
         \includegraphics[valign=m,width=.935\linewidth]{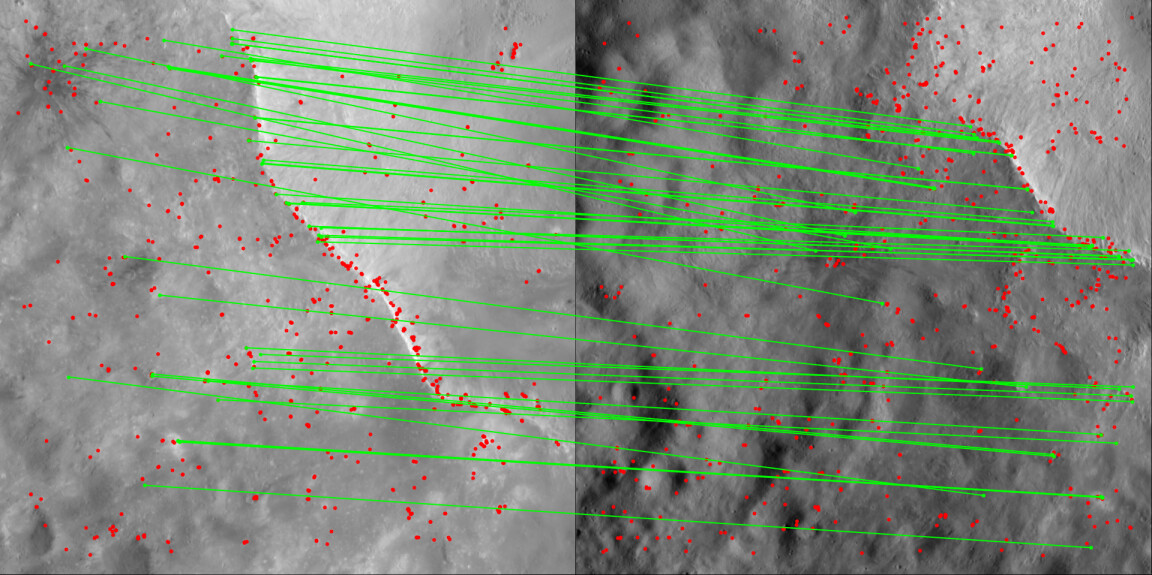}
    \end{tabular}
    \refstepcounter{subfigure}\label{fig:eval-vesta}
  \end{subfigure}\\
  \vspace{3pt}
  \begin{subfigure}[t]{\linewidth}
    \begin{tabular}{c c}
         \hspace{-10.85pt}
         \rotatebox[origin=c]{90}{\tiny{(e) \texttt{25143 Itokawa}}}
         \hspace{-18pt}
         &
         \includegraphics[valign=m,width=.935\linewidth]{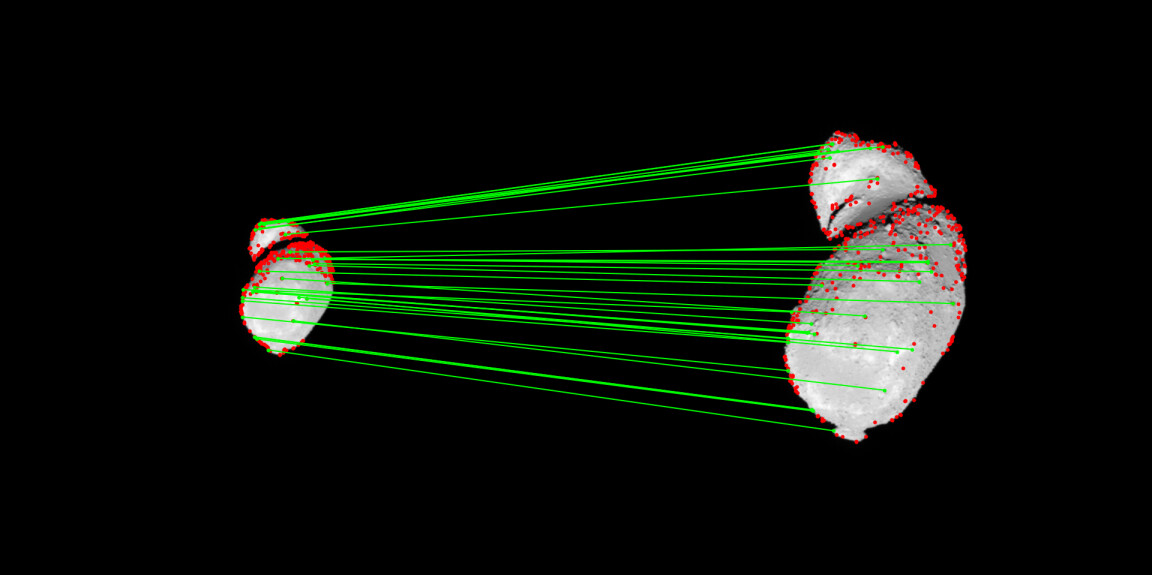}
    \end{tabular}
    \refstepcounter{subfigure}\label{fig:eval-itokawa}
  \end{subfigure}\\
  \vspace{3pt}
  \begin{subfigure}[t]{\linewidth}
    \includegraphics[width=.935\linewidth,right]{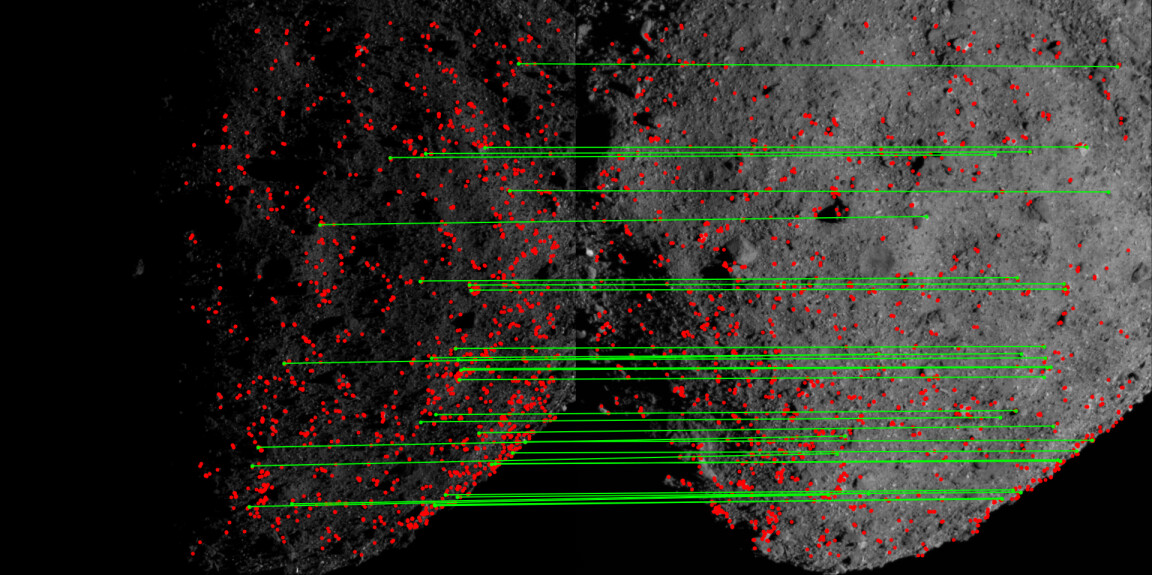}
  \end{subfigure}\\
  \vspace{1pt}
  \begin{subfigure}[t]{\linewidth}
    \begin{tabular}{c c}
         \hspace{-11pt}
         \rotatebox[origin=c]{90}{\tiny{(f) \texttt{101955 Bennu}}}
         \hspace{-18pt}
         &
         \includegraphics[valign=m,width=.935\linewidth]{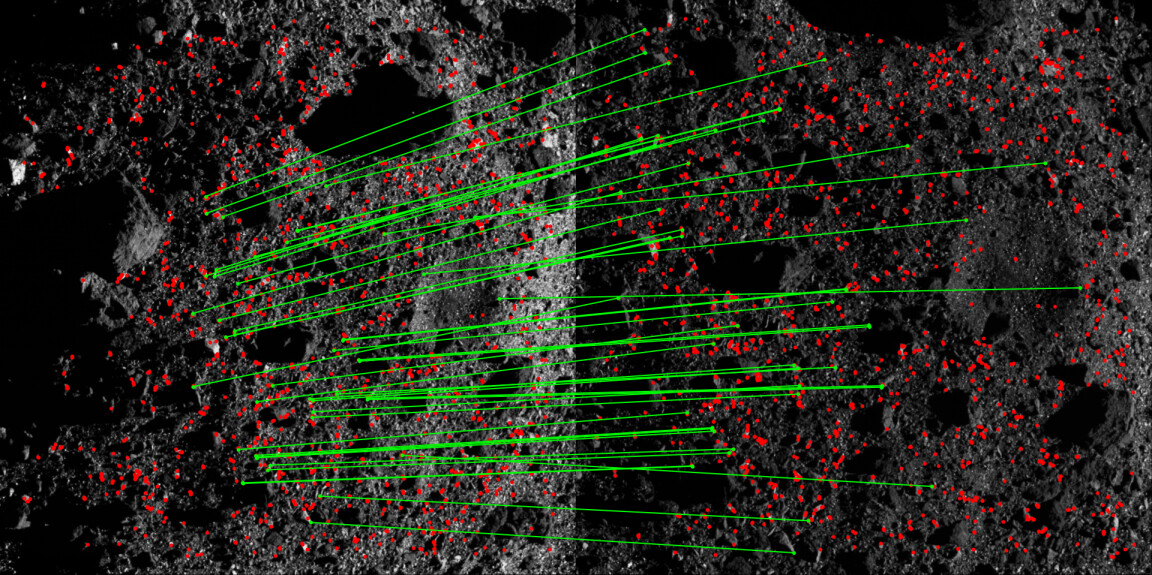}
    \end{tabular}
    \refstepcounter{subfigure}\label{fig:eval-bennu}
  \end{subfigure}\\
  \vspace{3pt}
  \begin{subfigure}[t]{\linewidth}
    \includegraphics[width=.935\linewidth,right]{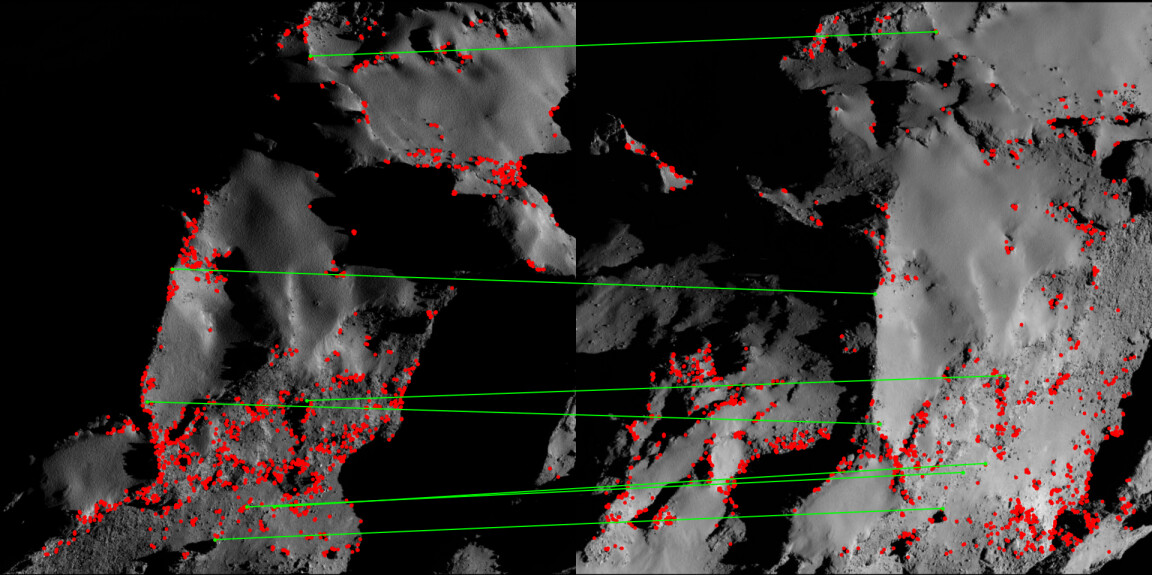}
  \end{subfigure}\\
  \vspace{1pt}
  \begin{subfigure}[t]{\linewidth}
    \begin{tabular}{c c}
         \hspace{-10.85pt}
         \rotatebox[origin=c]{90}{\tiny{(g) \texttt{67P}}}
         \hspace{-18pt}
         &
         \includegraphics[valign=m,width=.935\linewidth]{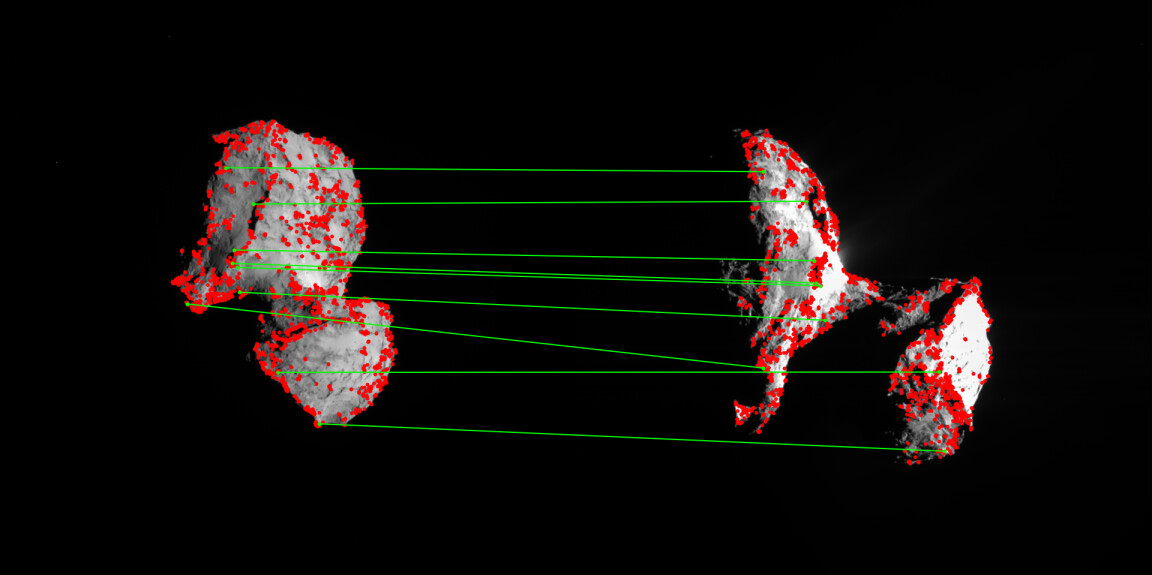}
    \end{tabular}
    \refstepcounter{subfigure}\label{fig:eval-chury}
  \end{subfigure}\\
  \vspace{3pt}
  \begin{subfigure}[t]{\linewidth}
    \begin{tabular}{c c}
         \hspace{-10.85pt}
         \rotatebox[origin=c]{90}{\tiny{(h) \texttt{21 Lutetia}}}
         \hspace{-18pt}
         &
         \includegraphics[valign=m,width=.935\linewidth]{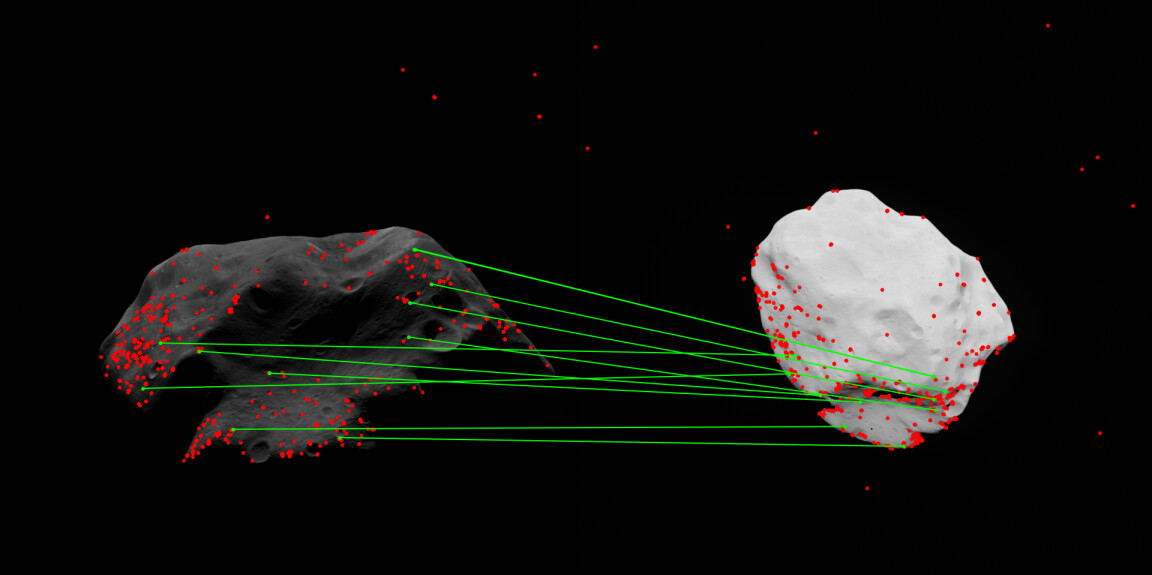}
    \end{tabular}
    \refstepcounter{subfigure}\label{fig:eval-lutetia}
  \end{subfigure}
  \caption*{\large{ORB}}
\end{subfigure}
\hspace{1pt}
\begin{subfigure}[t]{0.20\linewidth}
  \centering
  \begin{subfigure}[t]{\linewidth}
    \includegraphics[width=.935\linewidth]{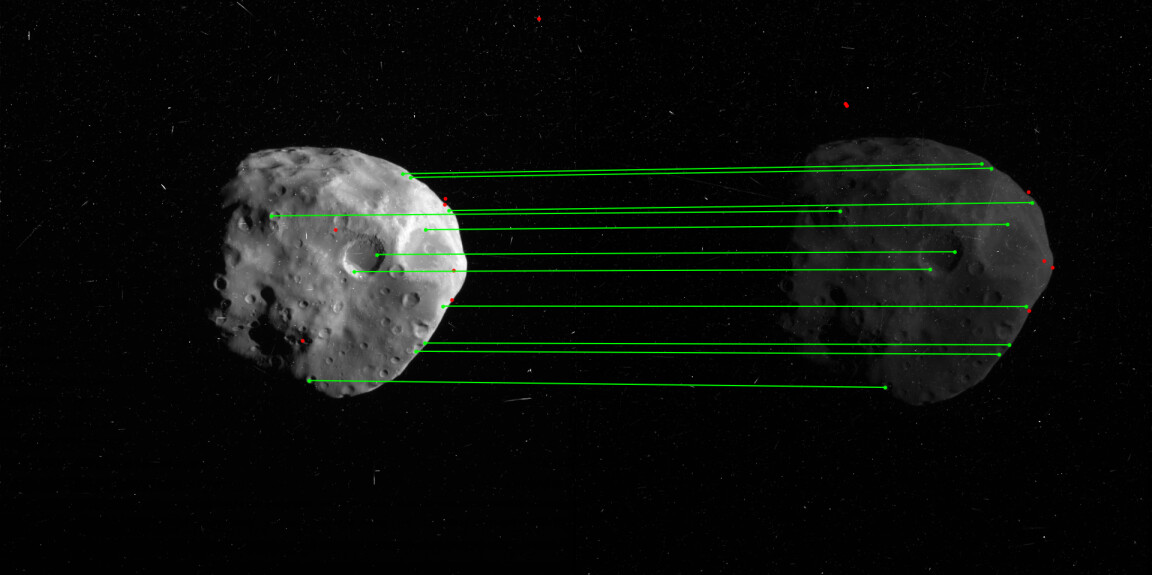}
  \end{subfigure}\\
  \vspace{1pt}
  \begin{subfigure}[t]{\linewidth}
    \includegraphics[width=.935\linewidth]{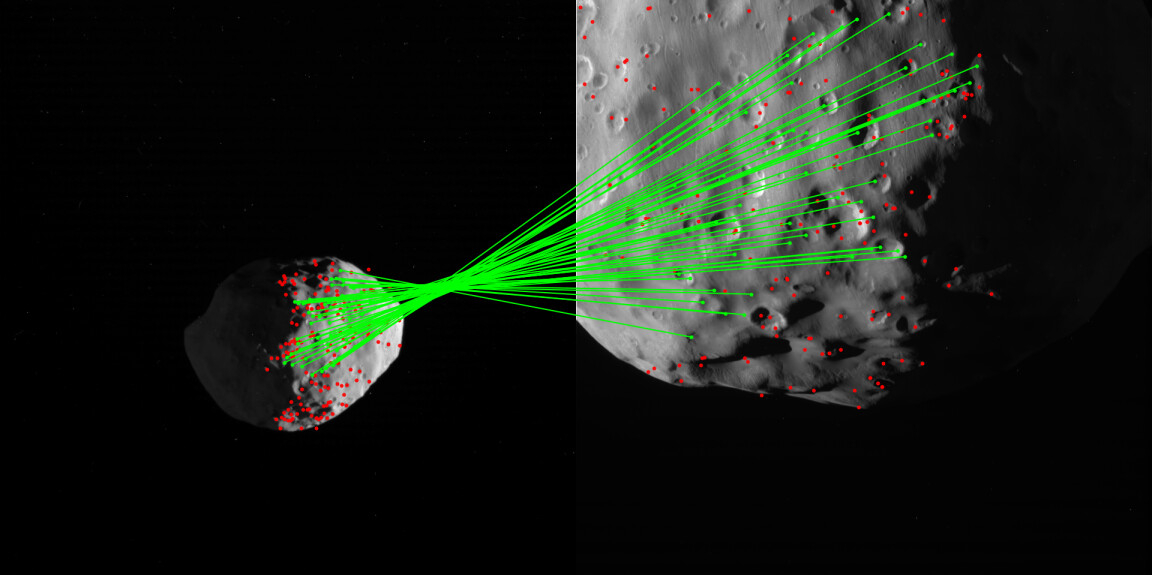}
  \end{subfigure}\\
  \vspace{4pt}
  \begin{subfigure}[t]{\linewidth}
    \includegraphics[width=.935\linewidth]{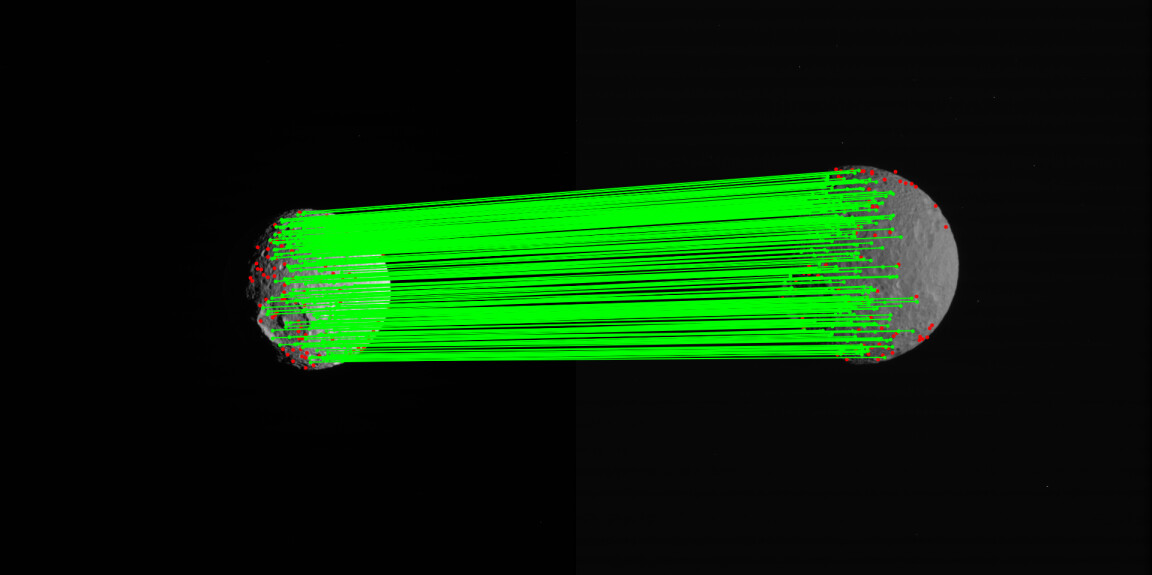}
  \end{subfigure}\\
  \vspace{4pt}
  \begin{subfigure}[t]{\linewidth}
    \includegraphics[width=.935\linewidth]{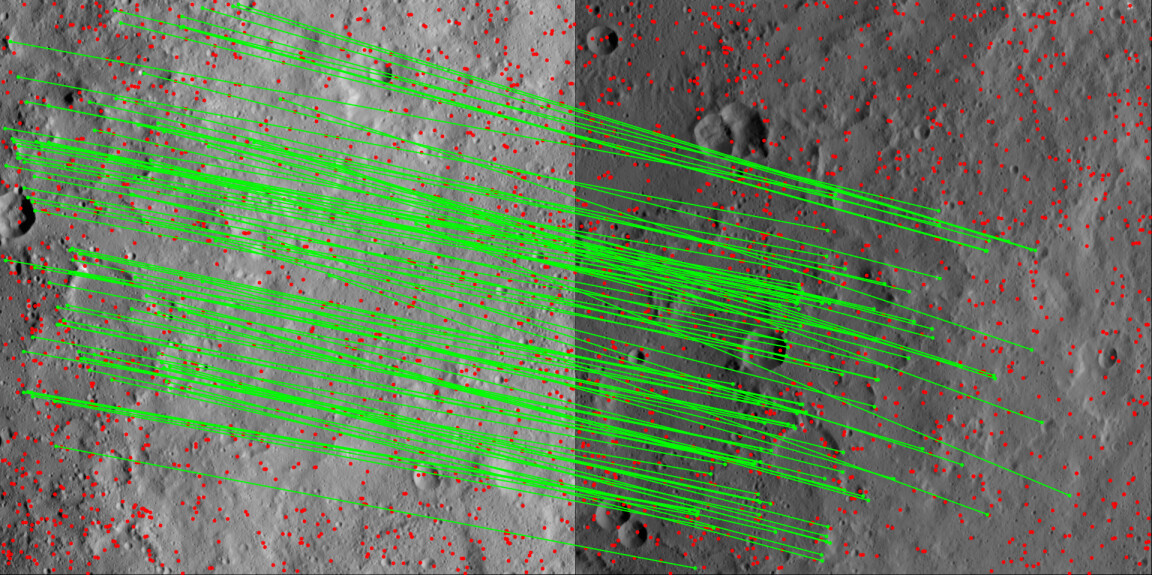}
  \end{subfigure}\\
  \vspace{1pt}
  \begin{subfigure}[t]{\linewidth}
    \includegraphics[width=.935\linewidth]{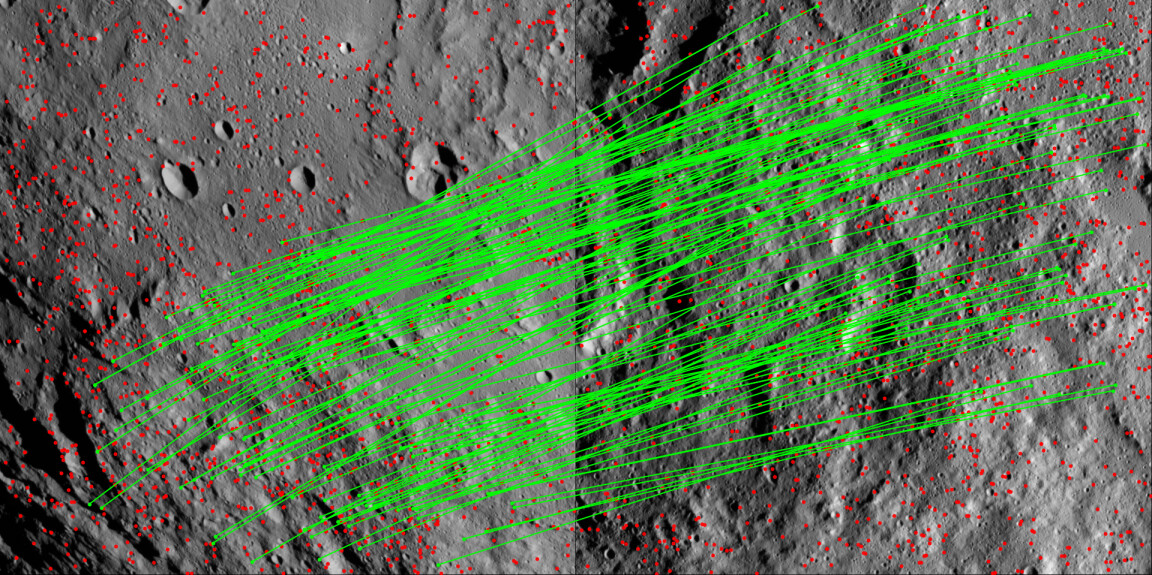}
  \end{subfigure}\\
  \vspace{4pt}
  \begin{subfigure}[t]{\linewidth}
    \includegraphics[width=.935\linewidth]{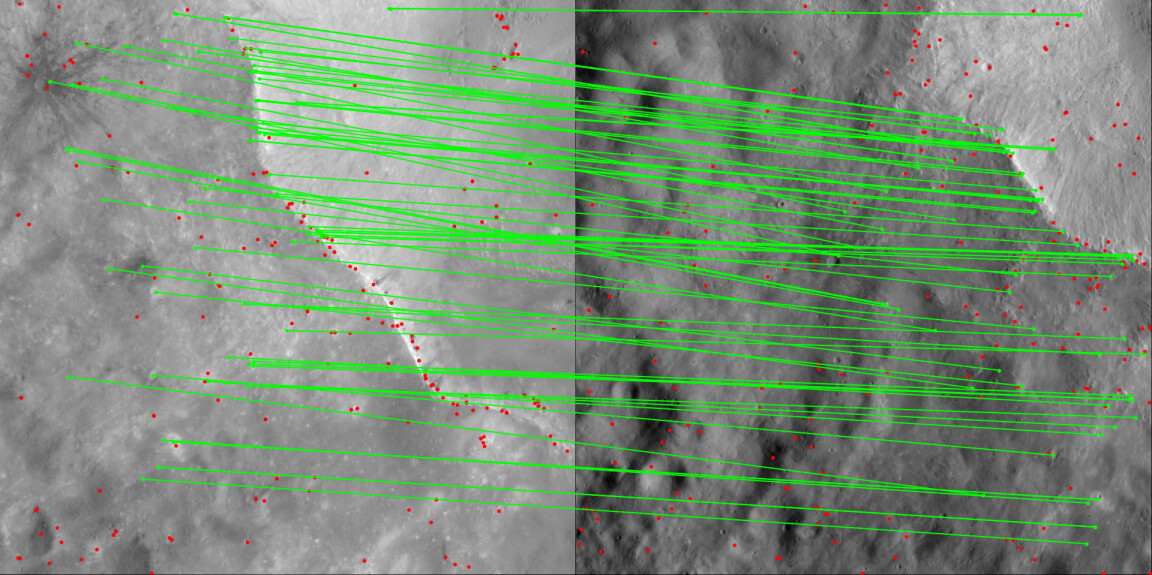}
  \end{subfigure}\\
  \vspace{4pt}
  \begin{subfigure}[t]{\linewidth}
    \includegraphics[width=.935\linewidth]{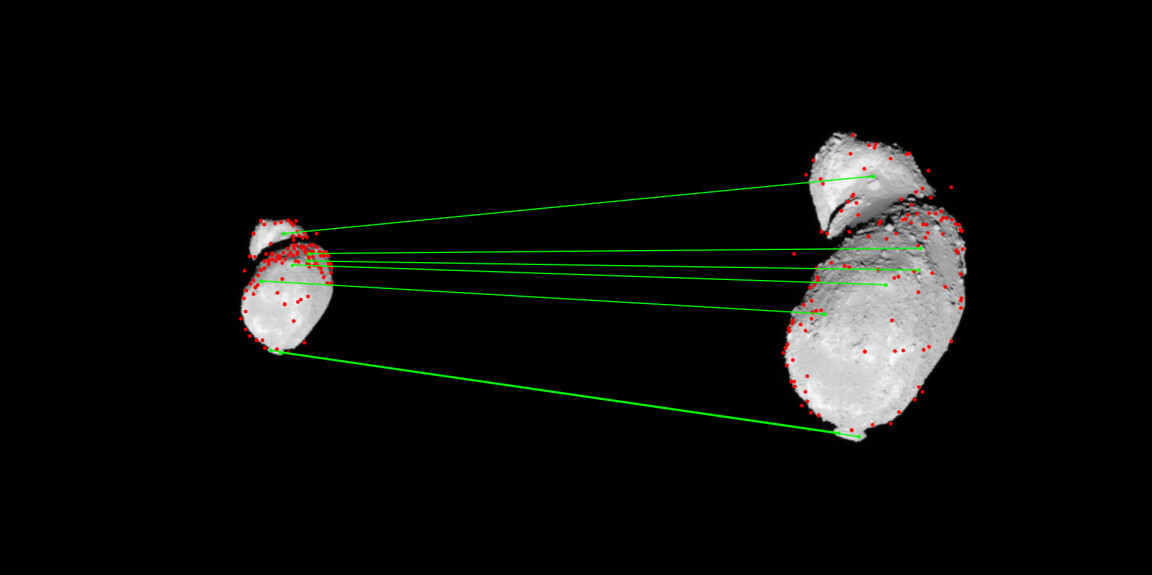}
  \end{subfigure}\\
  \vspace{4pt}
  \begin{subfigure}[t]{\linewidth}
    \includegraphics[width=.935\linewidth]{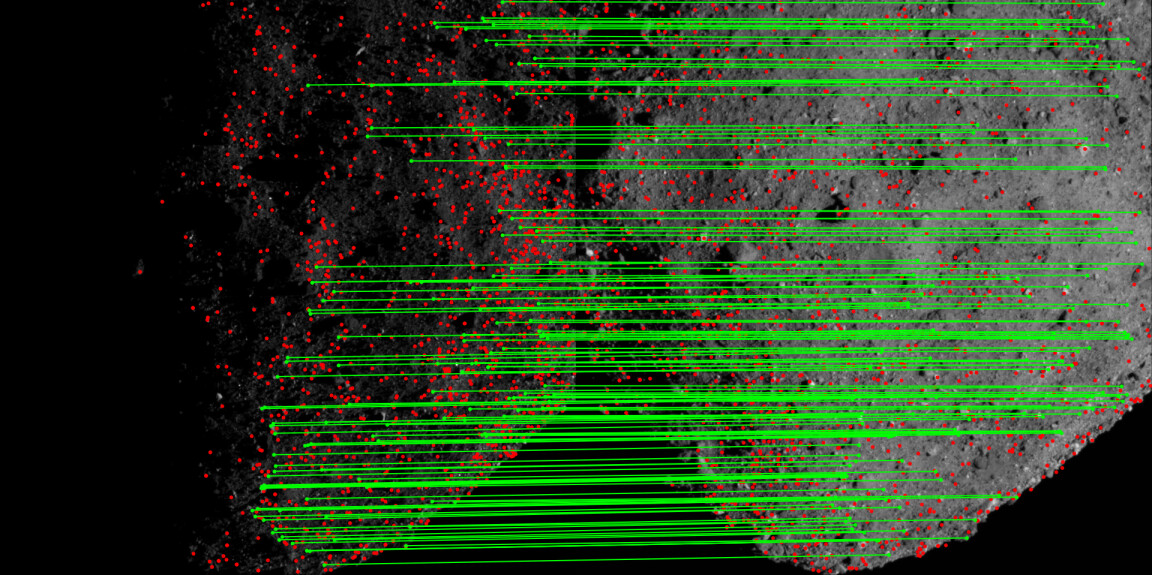}
  \end{subfigure}\\
  \vspace{1pt}
  \begin{subfigure}[t]{\linewidth}
    \includegraphics[width=.935\linewidth]{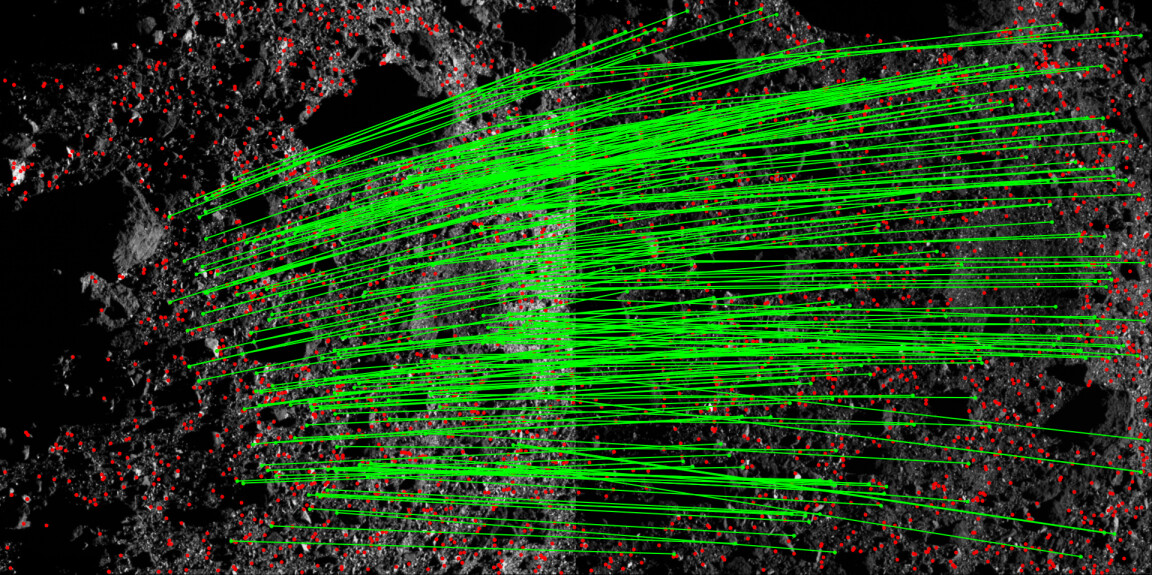}
  \end{subfigure}\\
  \vspace{4pt}
  \begin{subfigure}[t]{\linewidth}
    \includegraphics[width=.935\linewidth]{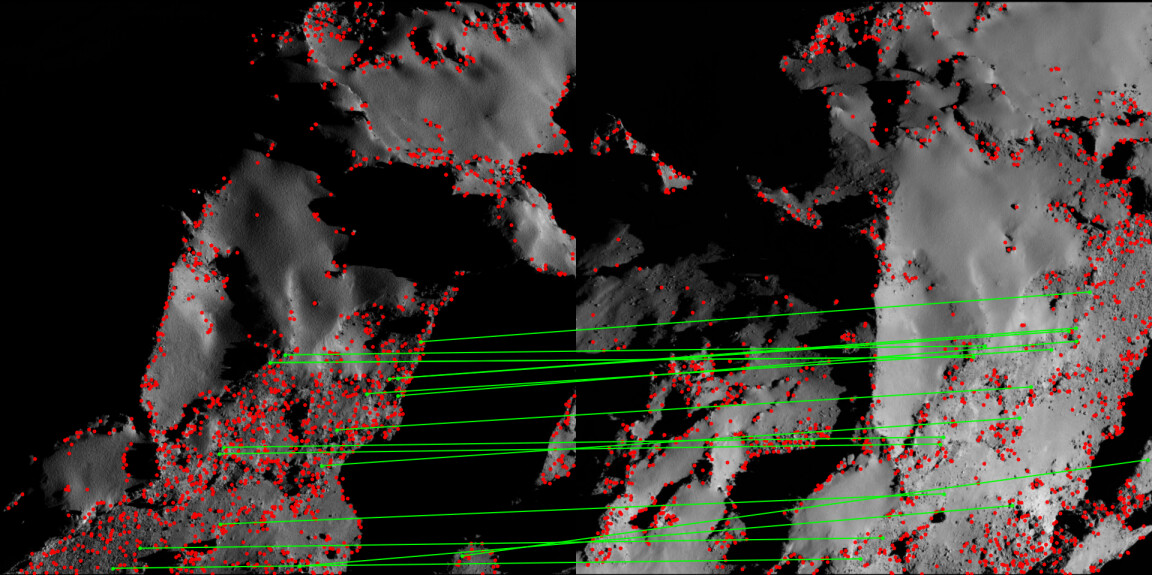}
  \end{subfigure}\\
  \vspace{1pt}
  \begin{subfigure}[t]{\linewidth}
    \includegraphics[width=.935\linewidth]{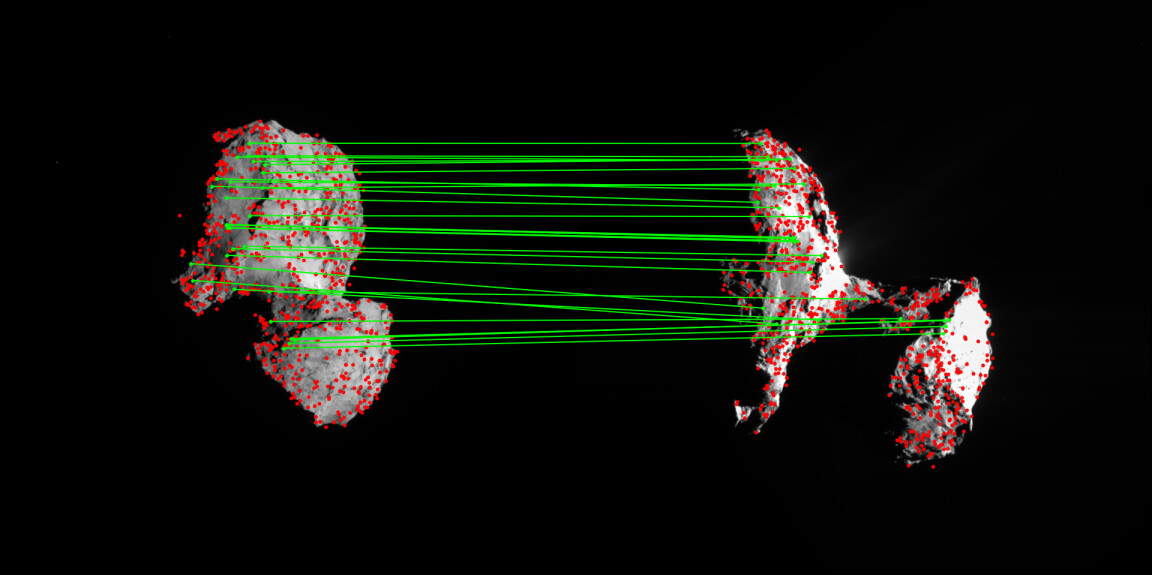}
  \end{subfigure}\\
    \vspace{4pt}
  \begin{subfigure}[t]{\linewidth}
    \includegraphics[width=.935\linewidth]{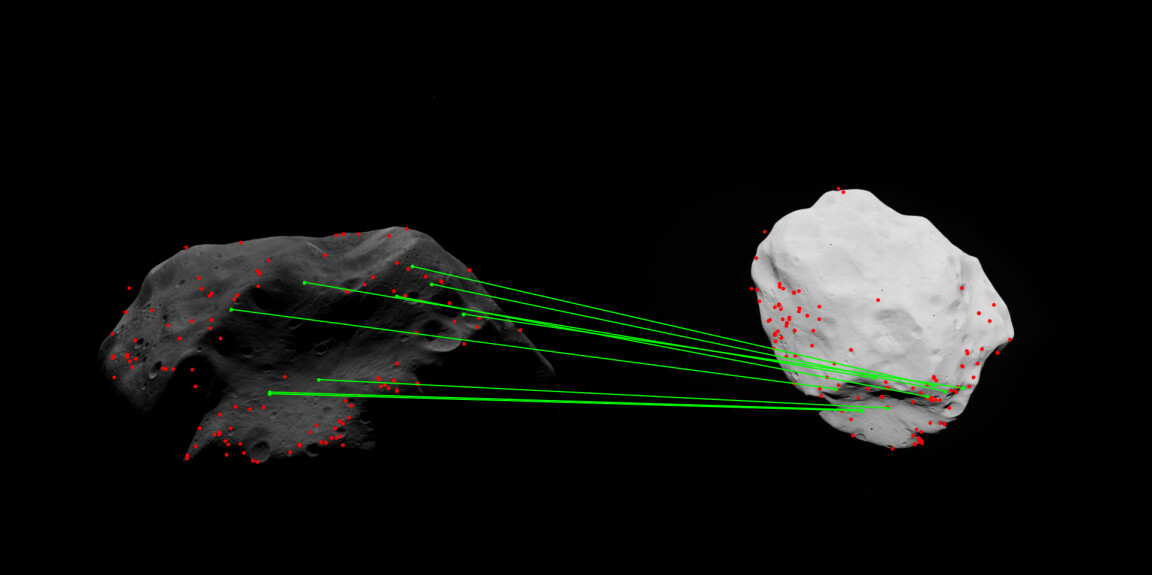}
  \end{subfigure}%
  \caption*{\large{SIFT}}
\end{subfigure}
\hspace{-5.5pt}
\begin{subfigure}[t]{0.20\linewidth}
  \centering
  \begin{subfigure}[t]{\linewidth}
    \includegraphics[width=.935\linewidth]{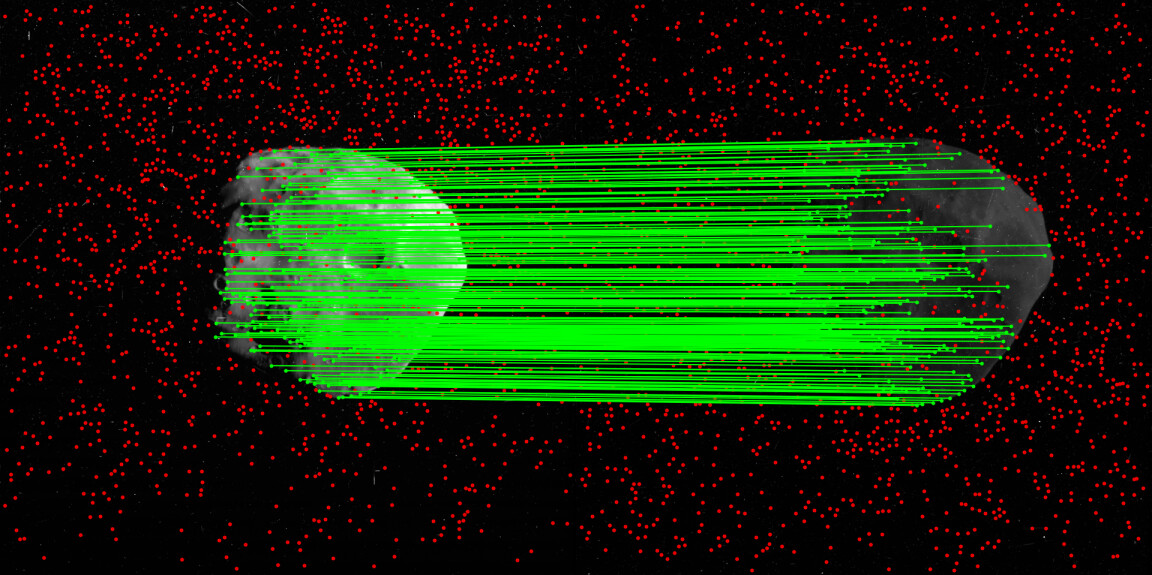}
  \end{subfigure}\\
  \vspace{1pt}
  \begin{subfigure}[t]{\linewidth}
    \includegraphics[width=.935\linewidth]{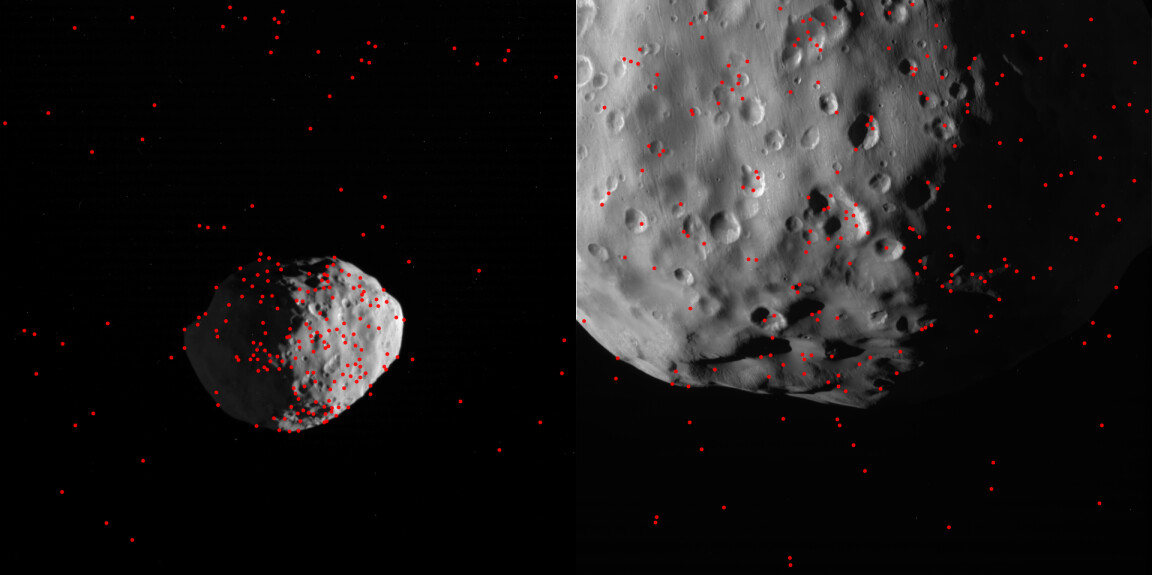}
  \end{subfigure}\\
  \vspace{4pt}
  \begin{subfigure}[t]{\linewidth}
    \includegraphics[width=.935\linewidth]{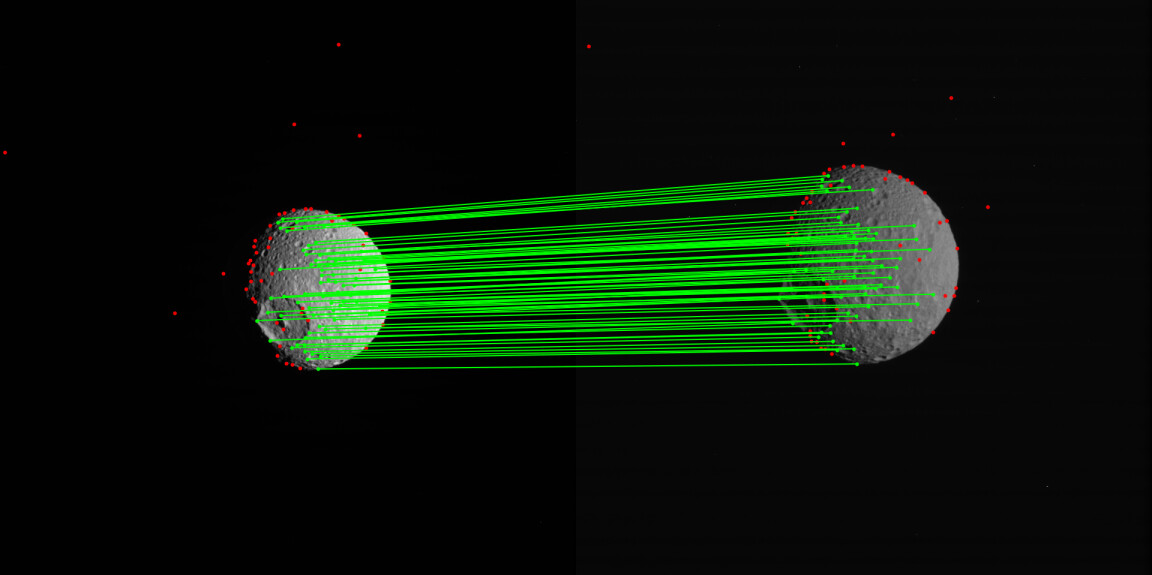}
  \end{subfigure}\\
  \vspace{4pt}
  \begin{subfigure}[t]{\linewidth}
    \includegraphics[width=.935\linewidth]{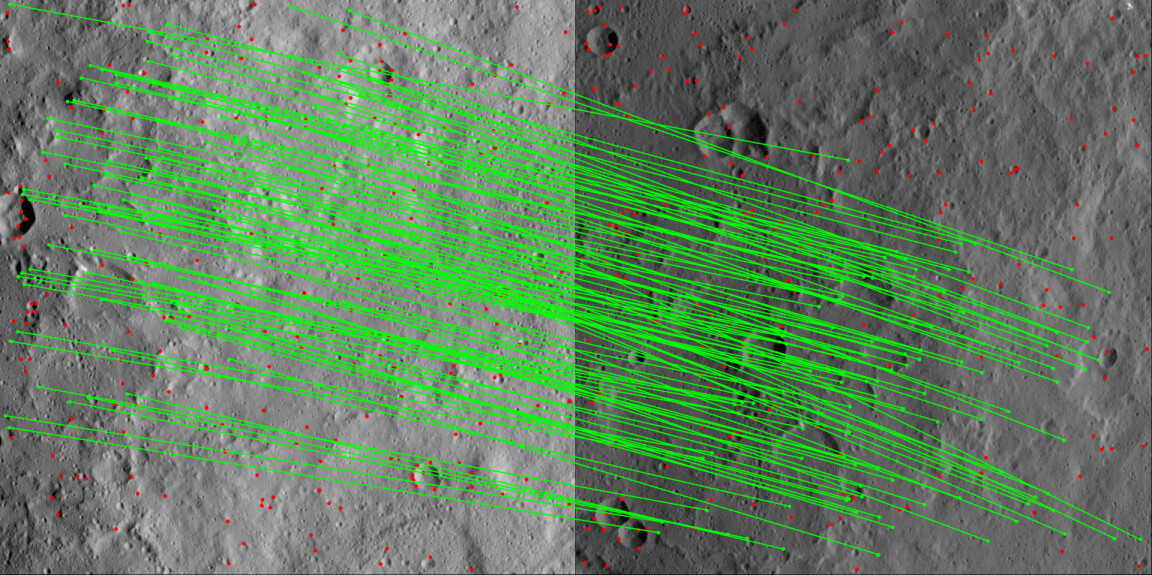}
  \end{subfigure}\\
  \vspace{1pt}
  \begin{subfigure}[t]{\linewidth}
    \includegraphics[width=.935\linewidth]{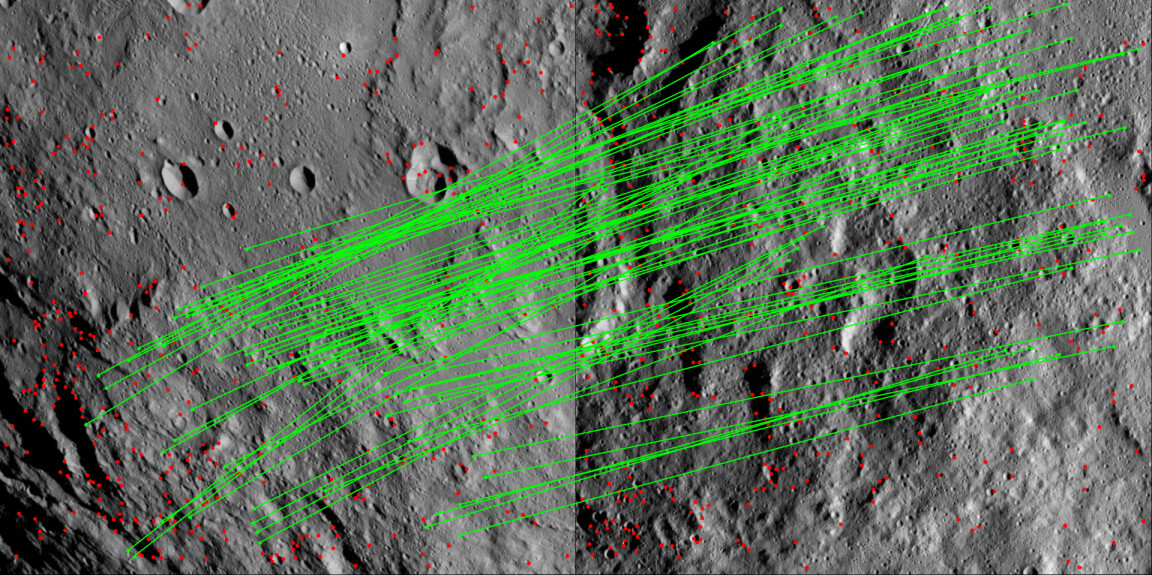}
  \end{subfigure}\\
  \vspace{4pt}
  \begin{subfigure}[t]{\linewidth}
    \includegraphics[width=.935\linewidth]{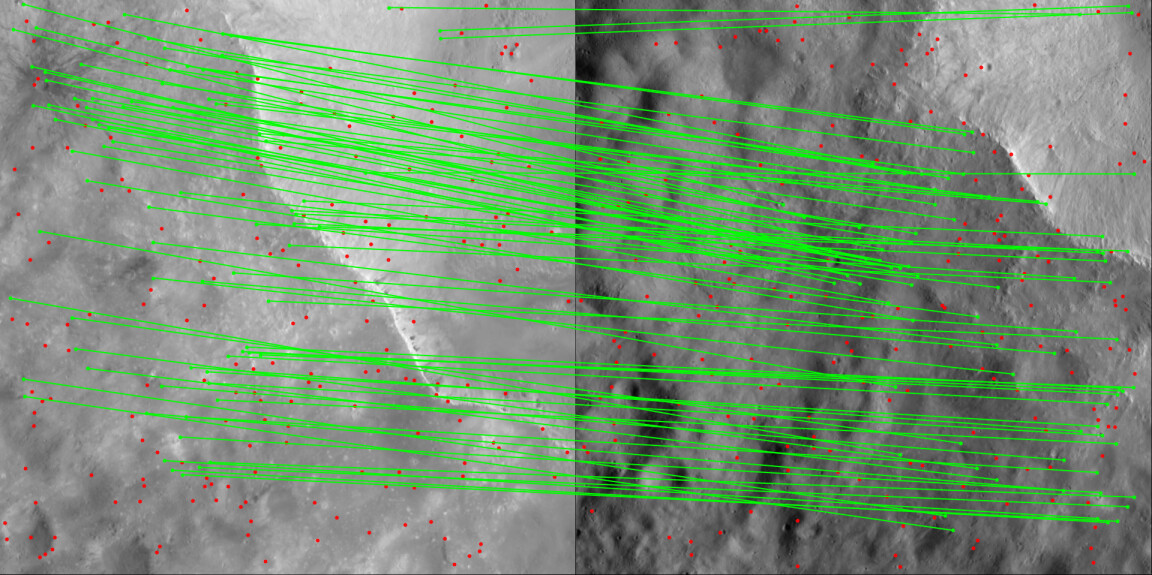}
  \end{subfigure}\\
  \vspace{4pt}
  \begin{subfigure}[t]{\linewidth}
    \includegraphics[width=.935\linewidth]{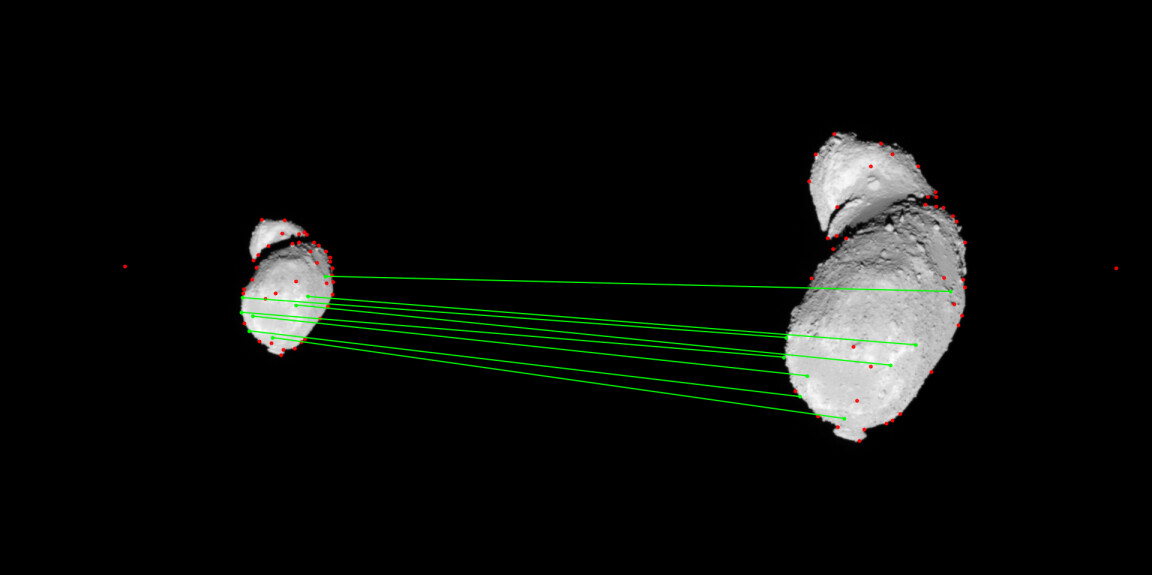}
  \end{subfigure}\\
  \vspace{4pt}
  \begin{subfigure}[t]{\linewidth}
    \includegraphics[width=.935\linewidth]{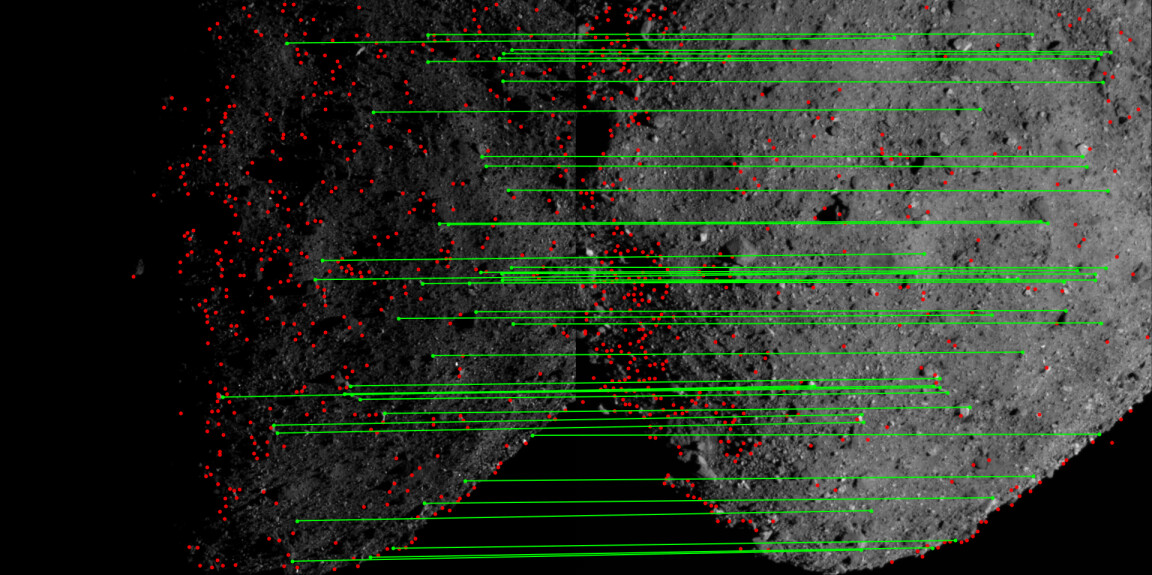}
  \end{subfigure}\\
  \vspace{1pt}
  \begin{subfigure}[t]{\linewidth}
    \includegraphics[width=.935\linewidth]{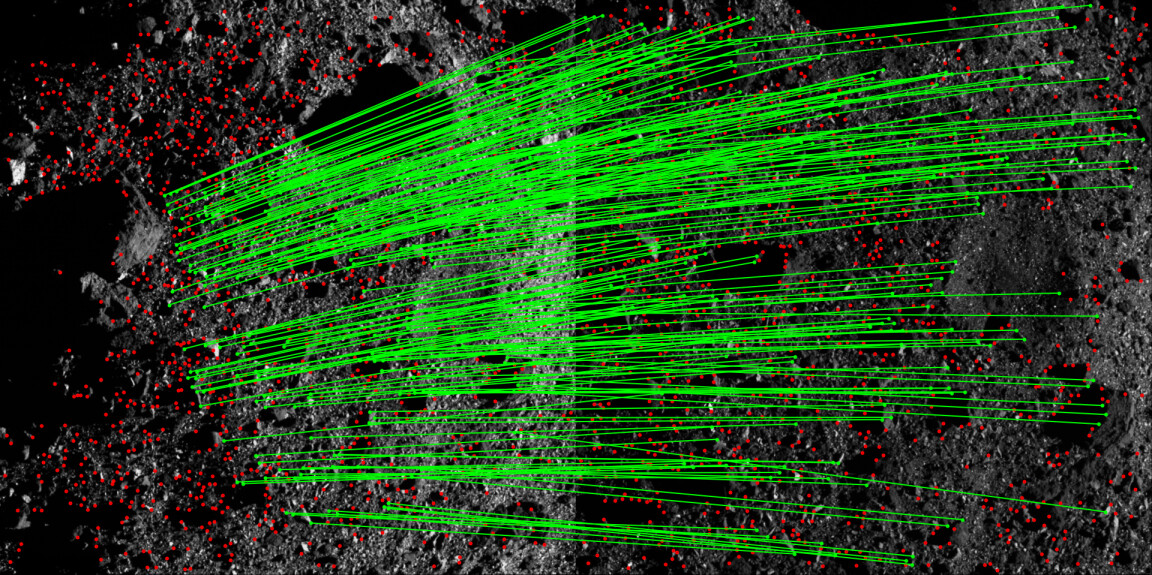}
  \end{subfigure}\\
  \vspace{4pt}
  \begin{subfigure}[t]{\linewidth}
    \includegraphics[width=.935\linewidth]{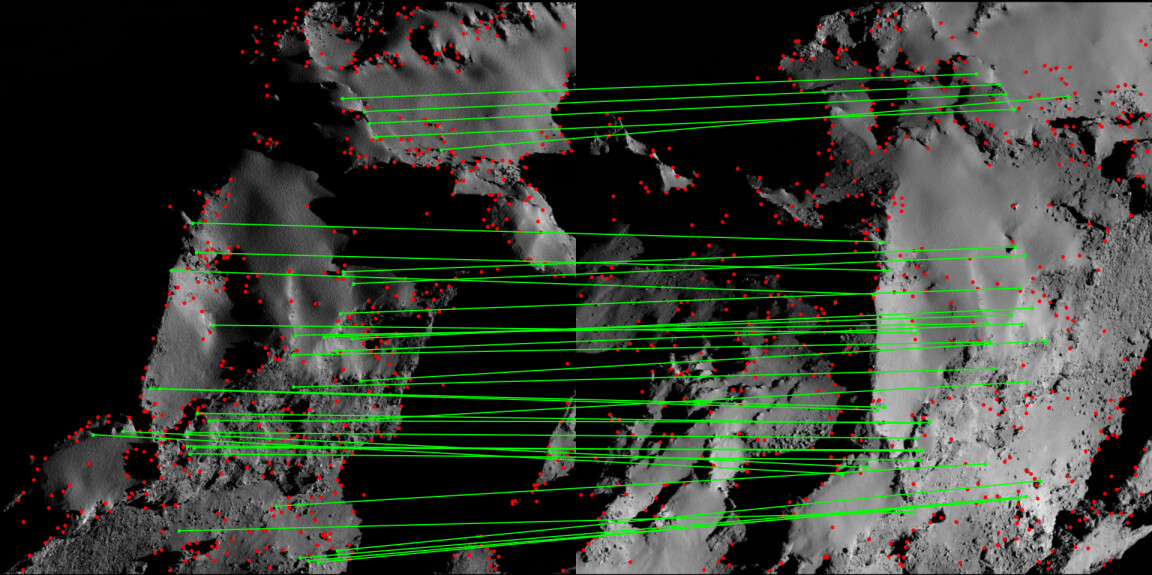}
  \end{subfigure}\\
  \vspace{1pt}
  \begin{subfigure}[t]{\linewidth}
    \includegraphics[width=.935\linewidth]{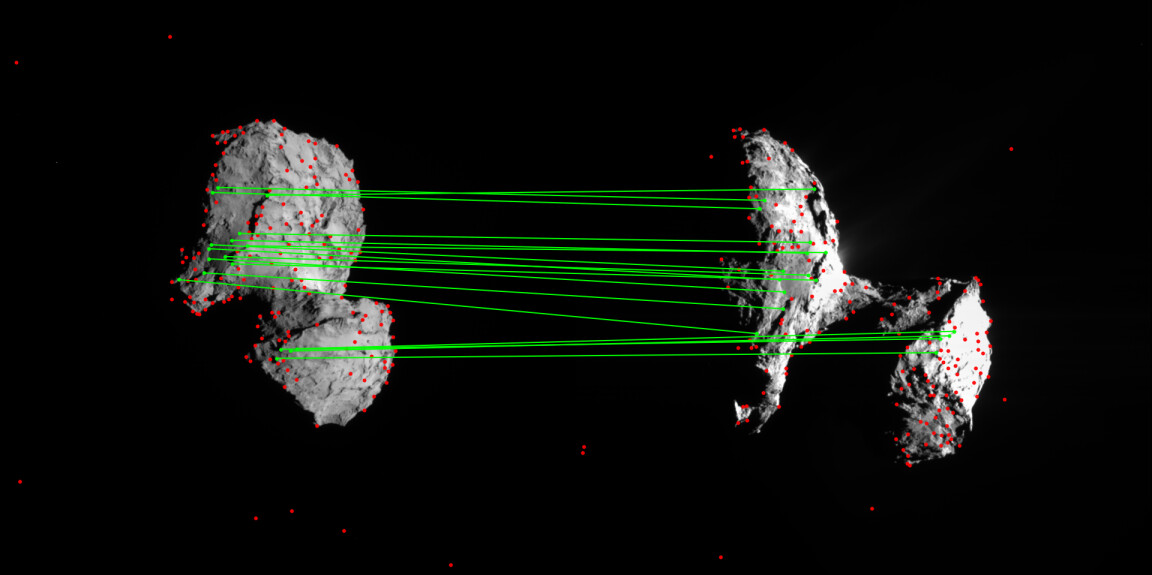}
  \end{subfigure}\\
    \vspace{4pt}
  \begin{subfigure}[t]{\linewidth}
    \includegraphics[width=.935\linewidth]{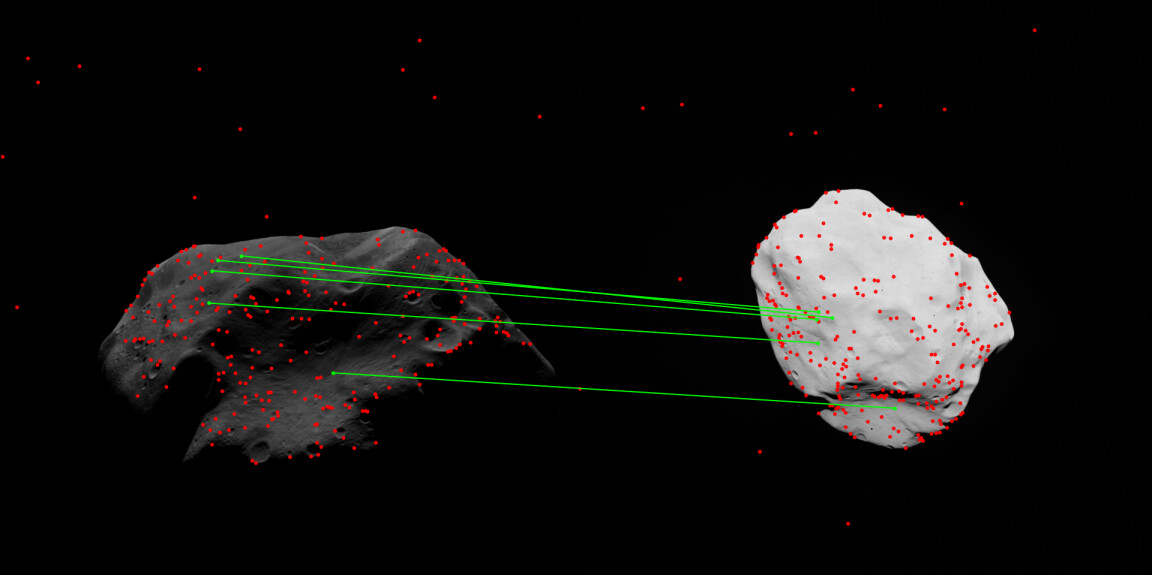}
  \end{subfigure}%
  \caption*{\large{SuperPoint}}
\end{subfigure}
\hspace{-5.5pt}
\begin{subfigure}[t]{0.20\linewidth}
  \centering
  \begin{subfigure}[t]{\linewidth}
    \includegraphics[width=.935\linewidth]{figures/match_plots/cas_epim_34_25_sift.jpg}
  \end{subfigure}\\
  \vspace{1pt}
  \begin{subfigure}[t]{\linewidth}
    \includegraphics[width=.935\linewidth]{figures/match_plots/cas_epim_132_119_sift.jpg}
  \end{subfigure}\\
  \vspace{4pt}
  \begin{subfigure}[t]{\linewidth}
    \includegraphics[width=.935\linewidth]{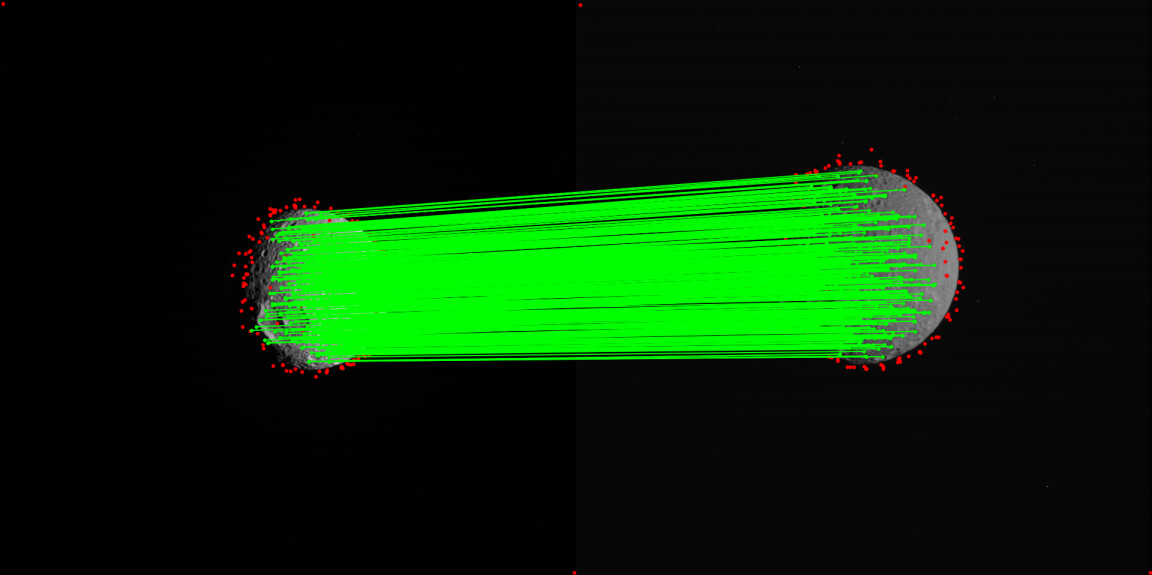}
  \end{subfigure}\\
  \vspace{4pt}
  \begin{subfigure}[t]{\linewidth}
    \includegraphics[width=.935\linewidth]{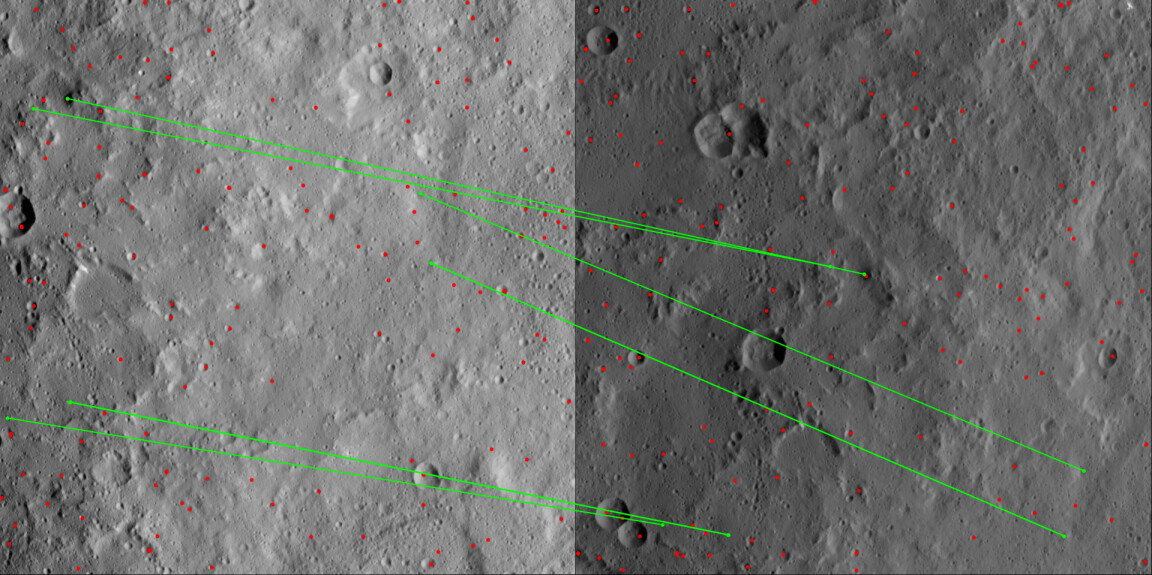}
  \end{subfigure}\\
  \vspace{1pt}
  \begin{subfigure}[t]{\linewidth}
    \includegraphics[width=.935\linewidth]{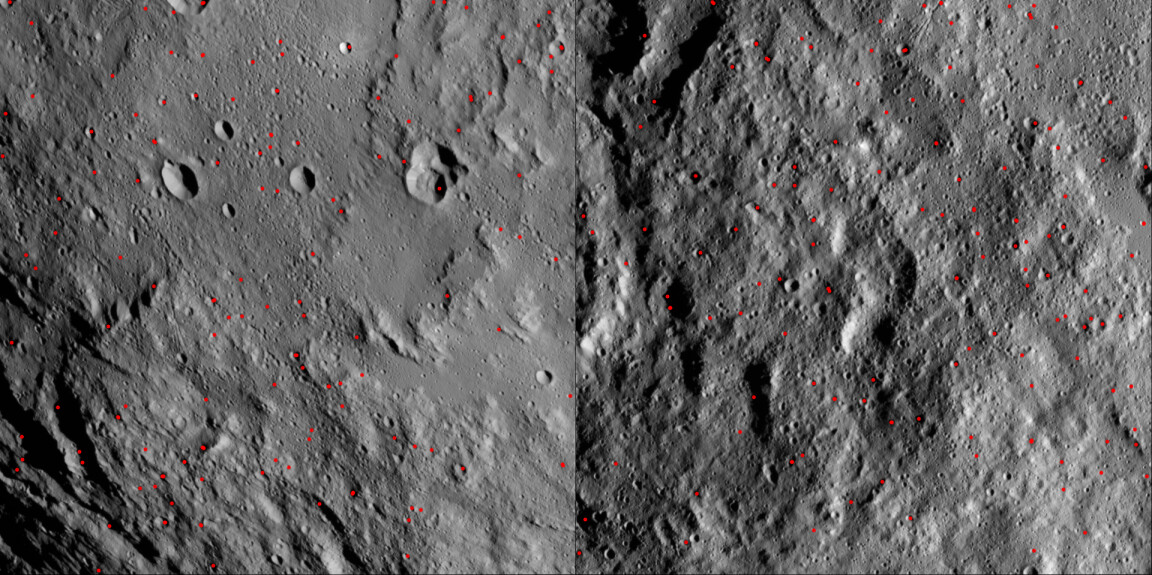}
  \end{subfigure}\\
  \vspace{4pt}
  \begin{subfigure}[t]{\linewidth}
    \includegraphics[width=.935\linewidth]{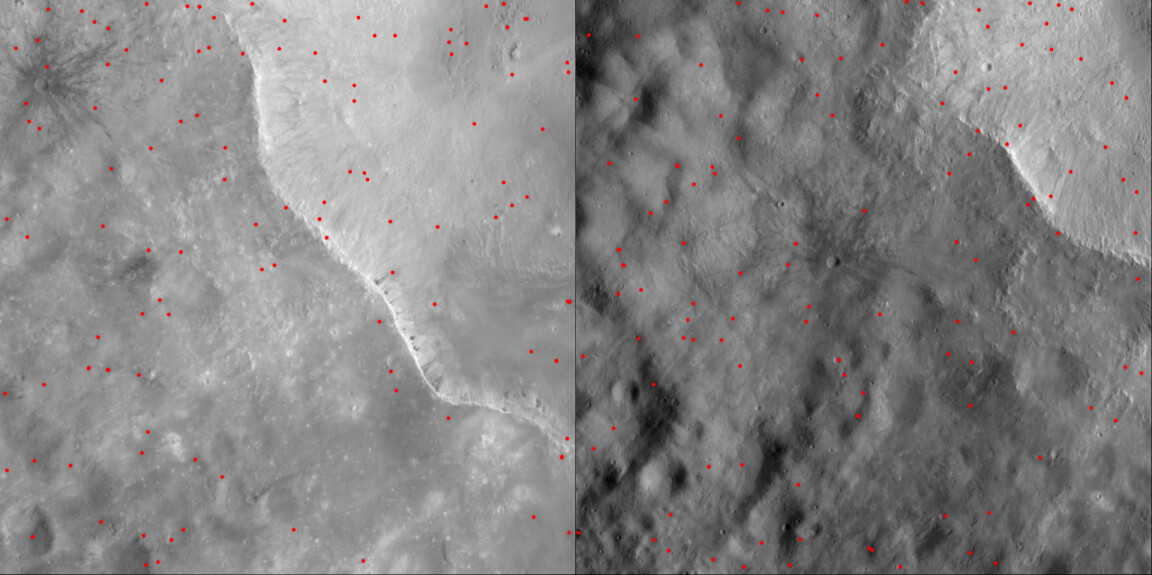}
  \end{subfigure}\\
  \vspace{4pt}
  \begin{subfigure}[t]{\linewidth}
    \includegraphics[width=.935\linewidth]{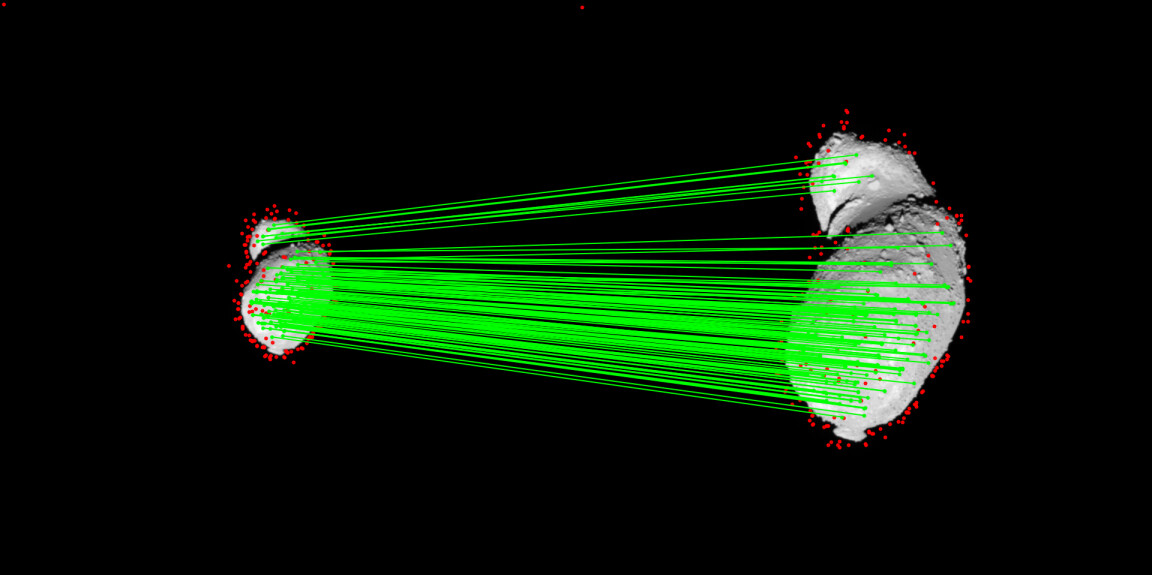}
  \end{subfigure}\\
  \vspace{4pt}
  \begin{subfigure}[t]{\linewidth}
    \includegraphics[width=.935\linewidth]{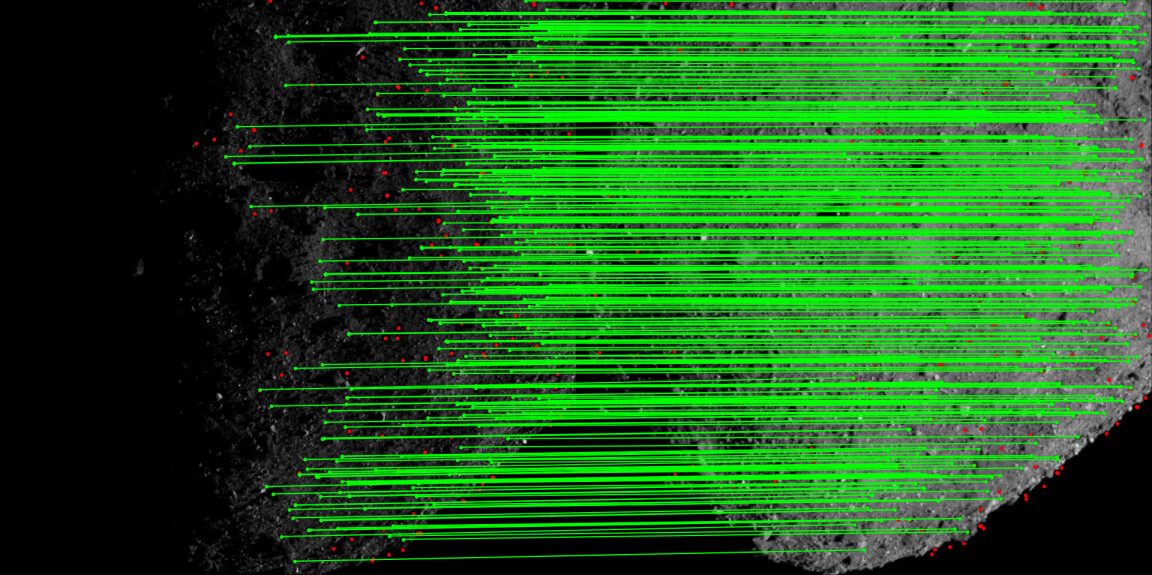}
  \end{subfigure}\\
  \vspace{1pt}
  \begin{subfigure}[t]{\linewidth}
    \includegraphics[width=.935\linewidth]{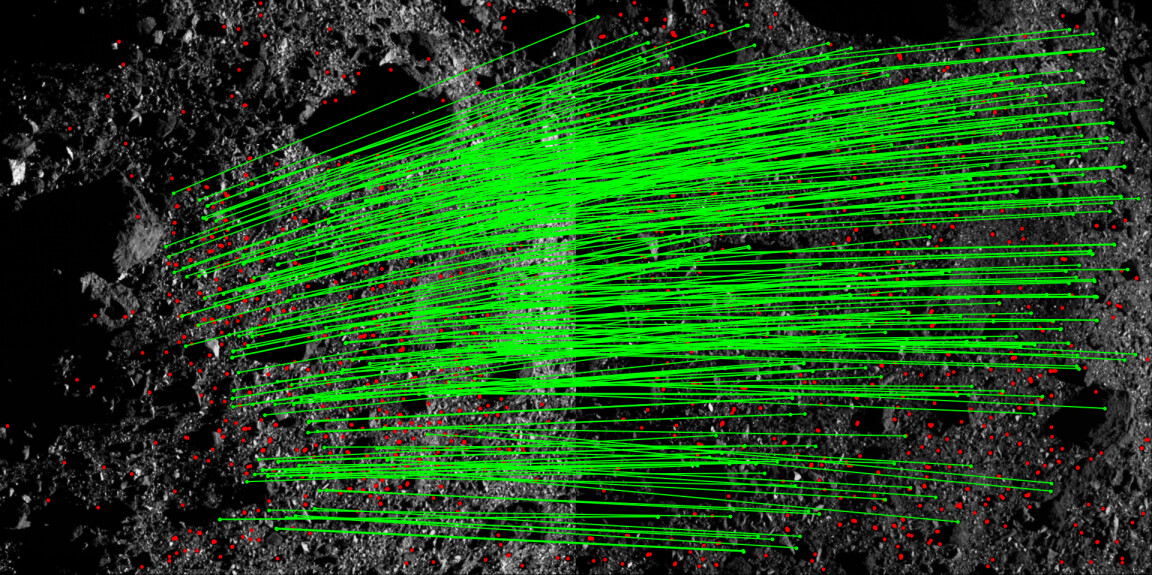}
  \end{subfigure}\\
  \vspace{4pt}
  \begin{subfigure}[t]{\linewidth}
    \includegraphics[width=.935\linewidth]{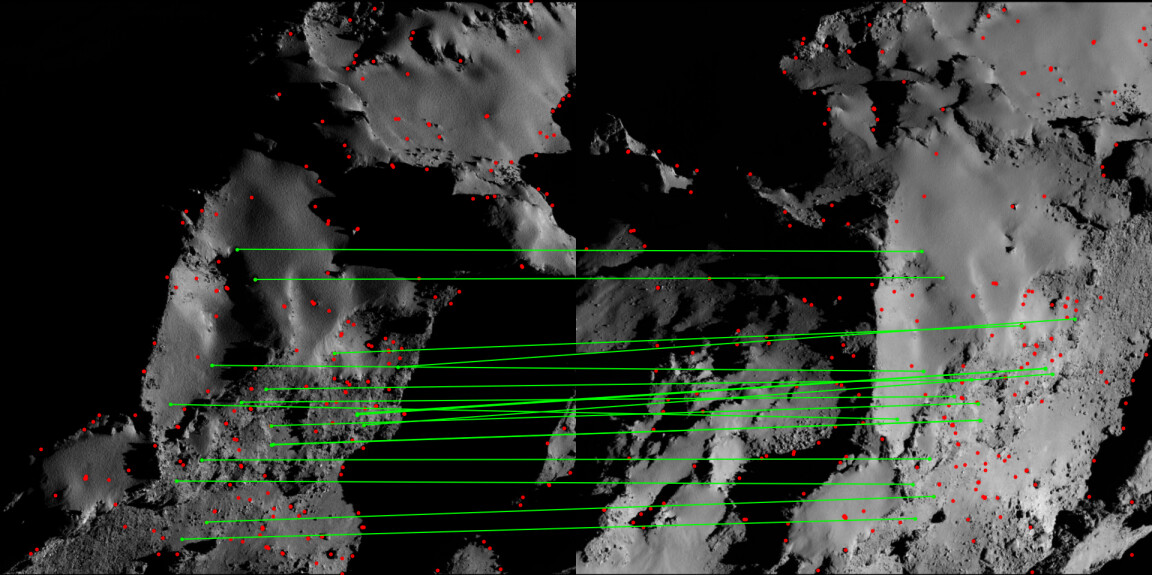}
  \end{subfigure}\\
  \vspace{1pt}
  \begin{subfigure}[t]{\linewidth}
    \includegraphics[width=.935\linewidth]{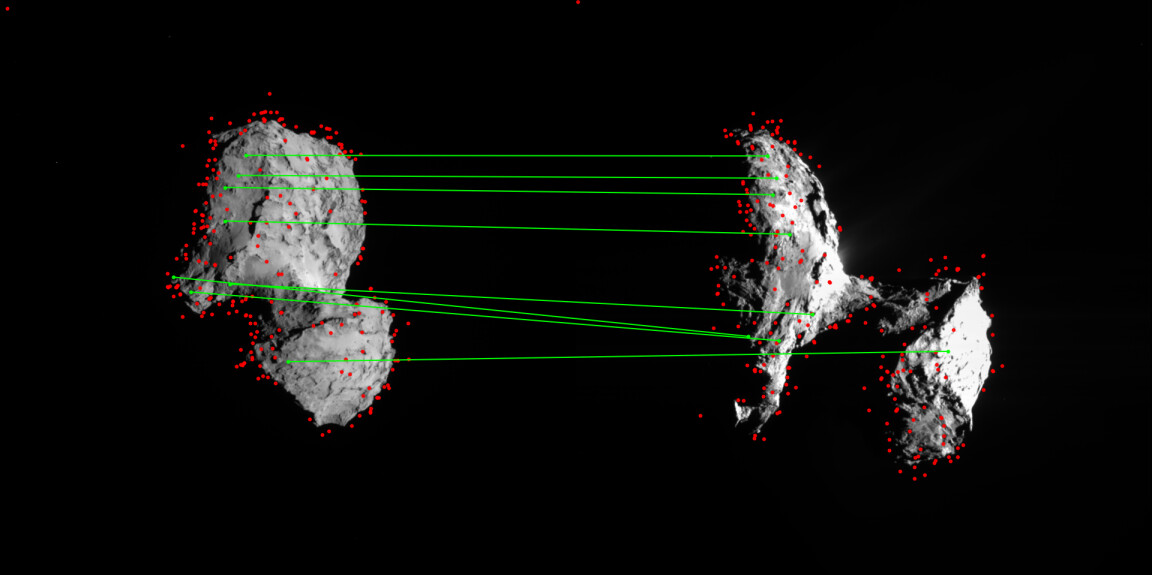}
  \end{subfigure}\\
    \vspace{4pt}
  \begin{subfigure}[t]{\linewidth}
    \includegraphics[width=.935\linewidth]{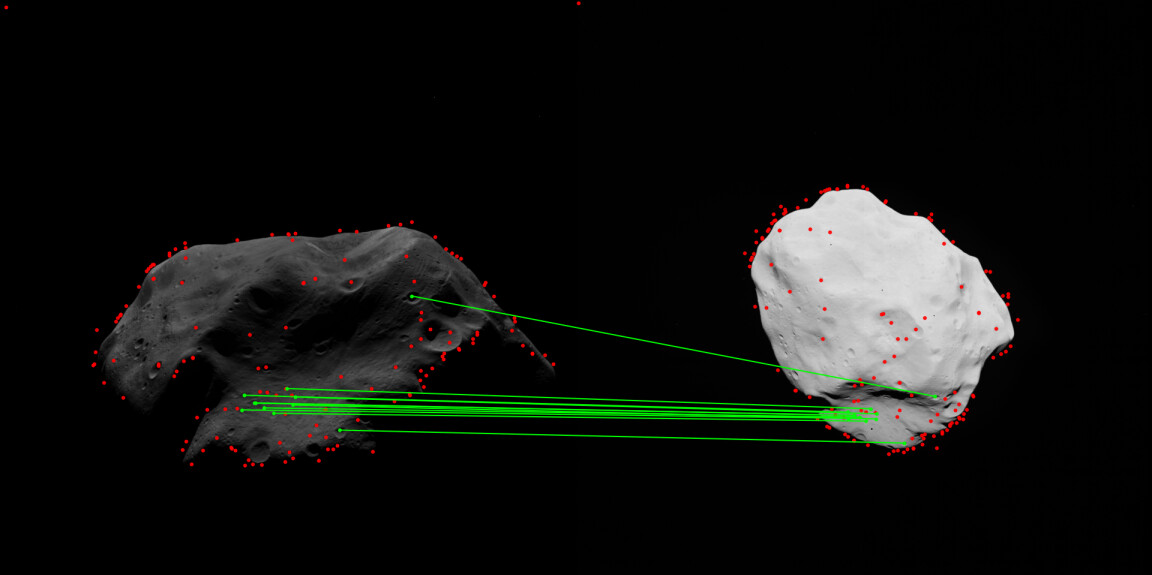}
  \end{subfigure}%
  \caption*{\large{R2D2}}
\end{subfigure}
\hspace{-5.5pt}
\begin{subfigure}[t]{0.20\linewidth}
  \centering
  \begin{subfigure}[t]{\linewidth}
    \includegraphics[width=.935\linewidth]{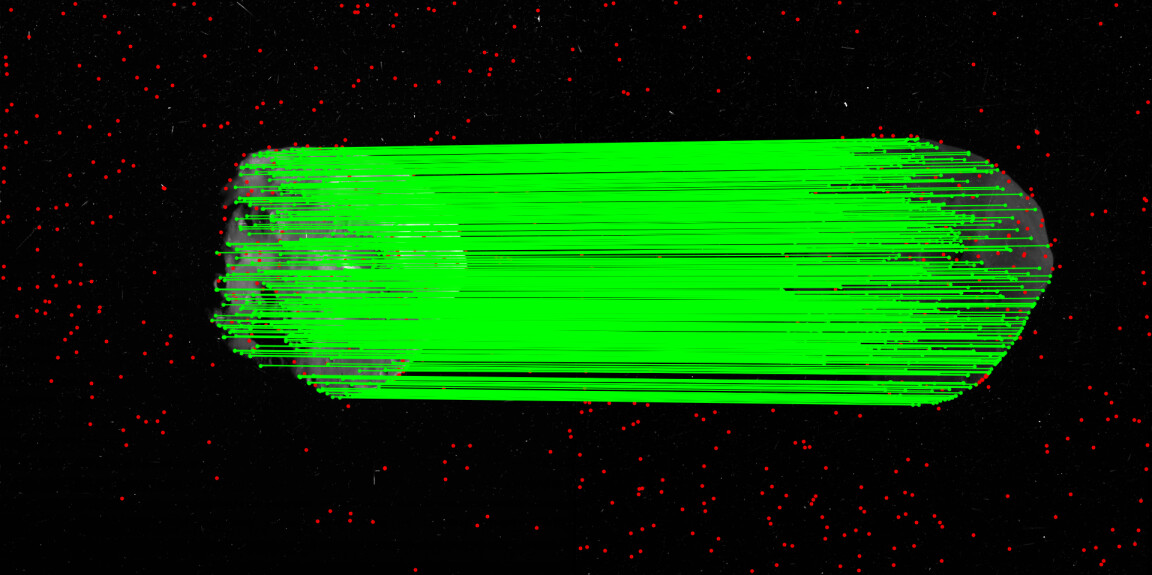}
  \end{subfigure}\\
  \vspace{1pt}
  \begin{subfigure}[t]{\linewidth}
    \includegraphics[width=.935\linewidth]{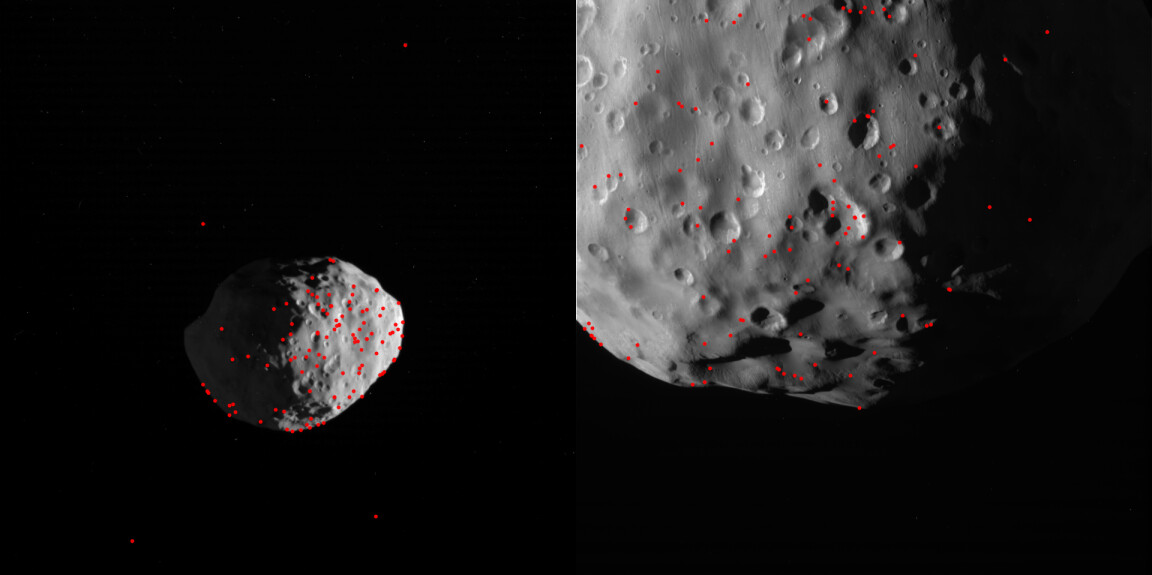}
  \end{subfigure}\\
  \vspace{4pt}
  \begin{subfigure}[t]{\linewidth}
    \includegraphics[width=.935\linewidth]{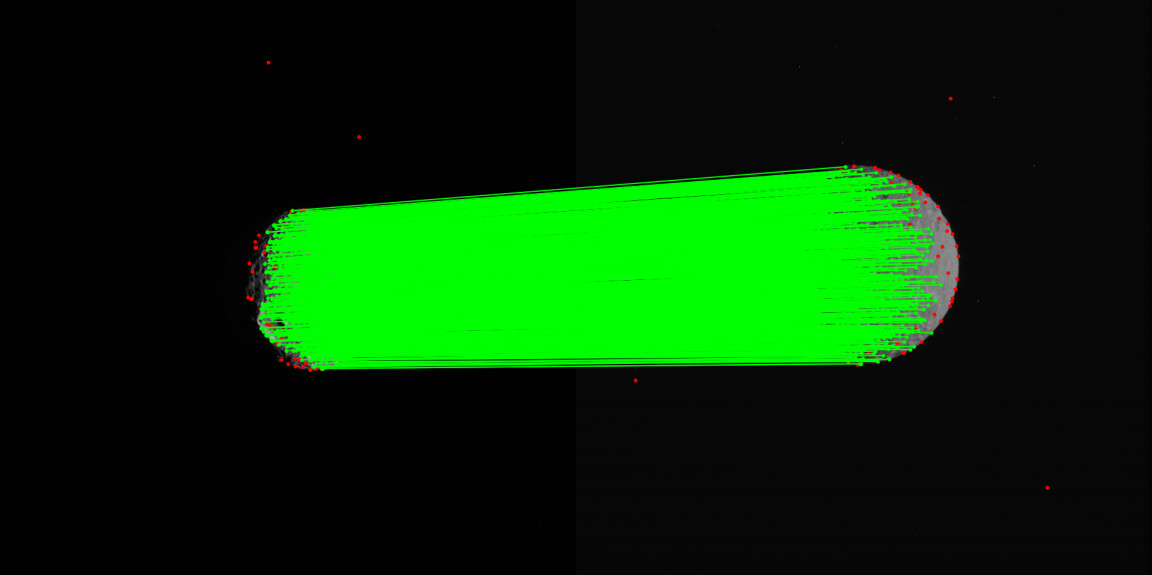}
  \end{subfigure}\\
  \vspace{4pt}
  \begin{subfigure}[t]{\linewidth}
    \includegraphics[width=.935\linewidth]{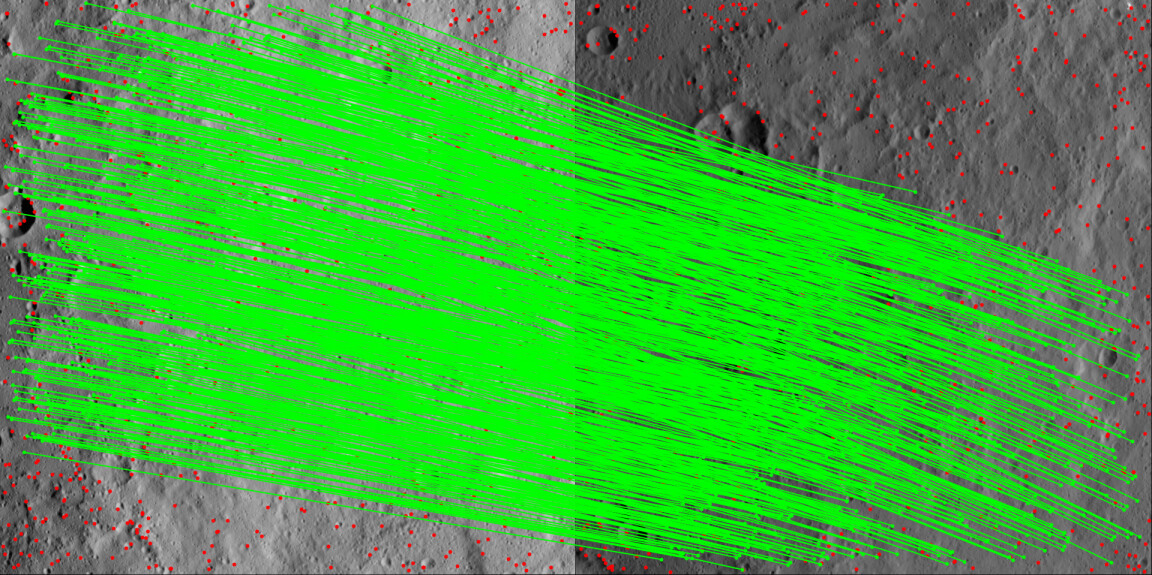}
  \end{subfigure}\\
  \vspace{1pt}
  \begin{subfigure}[t]{\linewidth}
    \includegraphics[width=.935\linewidth]{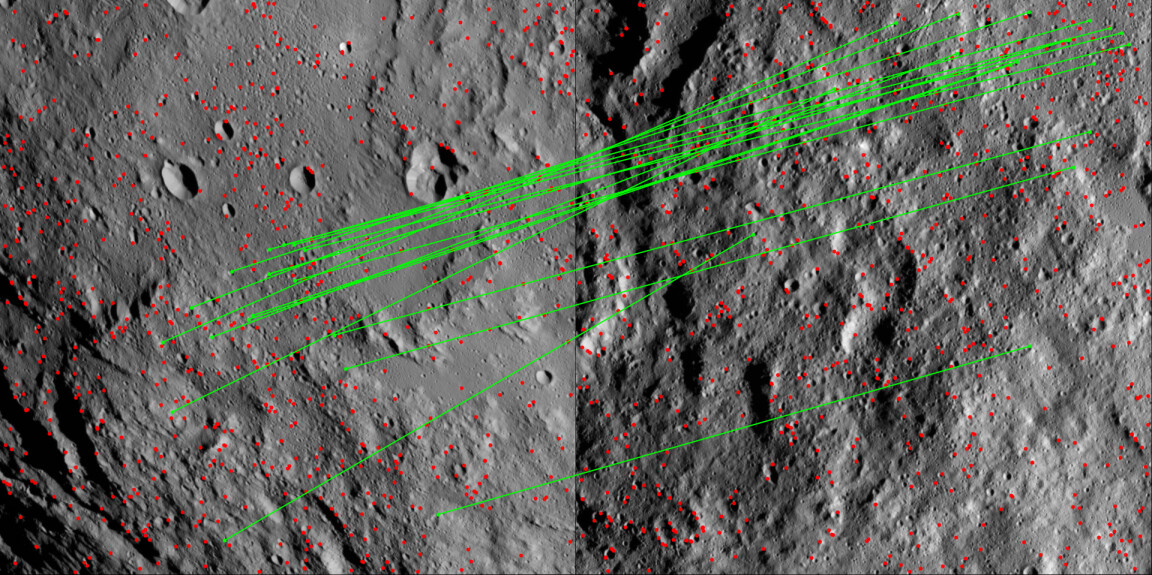}
  \end{subfigure}\\
  \vspace{4pt}
  \begin{subfigure}[t]{\linewidth}
    \includegraphics[width=.935\linewidth]{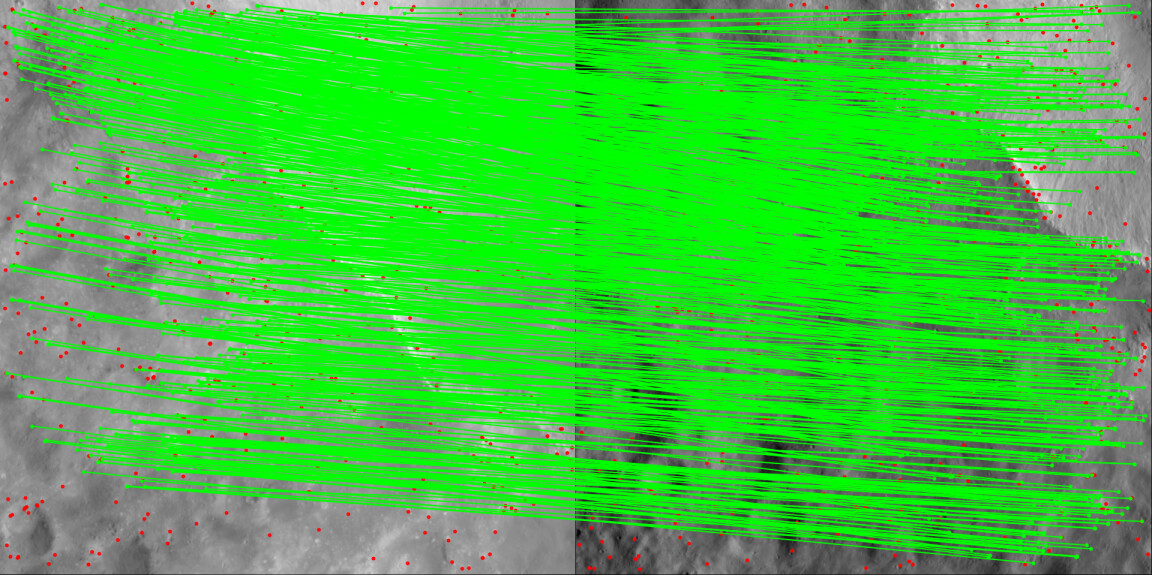}
  \end{subfigure}\\
  \vspace{4pt}
  \begin{subfigure}[t]{\linewidth}
    \includegraphics[width=.935\linewidth]{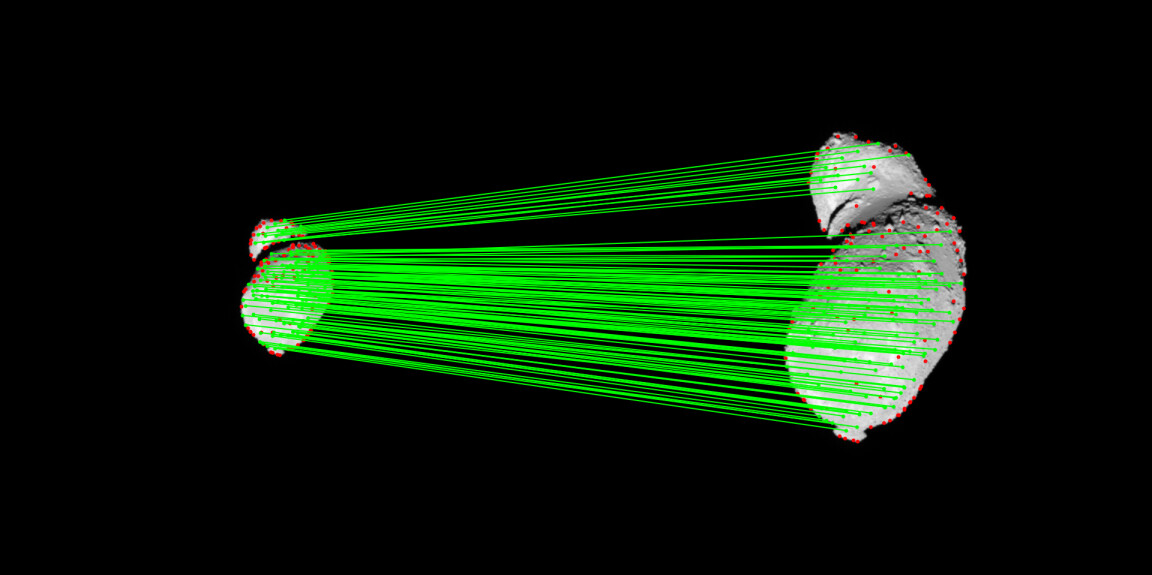}
  \end{subfigure}\\
  \vspace{4pt}
  \begin{subfigure}[t]{\linewidth}
    \includegraphics[width=.935\linewidth]{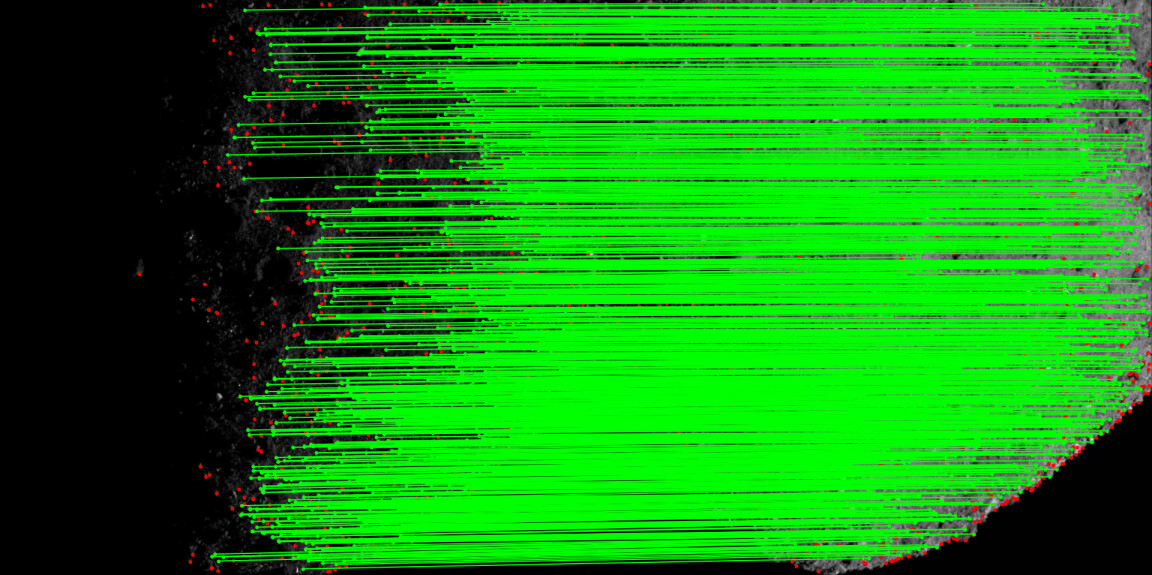}
  \end{subfigure}\\
  \vspace{1pt}
  \begin{subfigure}[t]{\linewidth}
    \includegraphics[width=.935\linewidth]{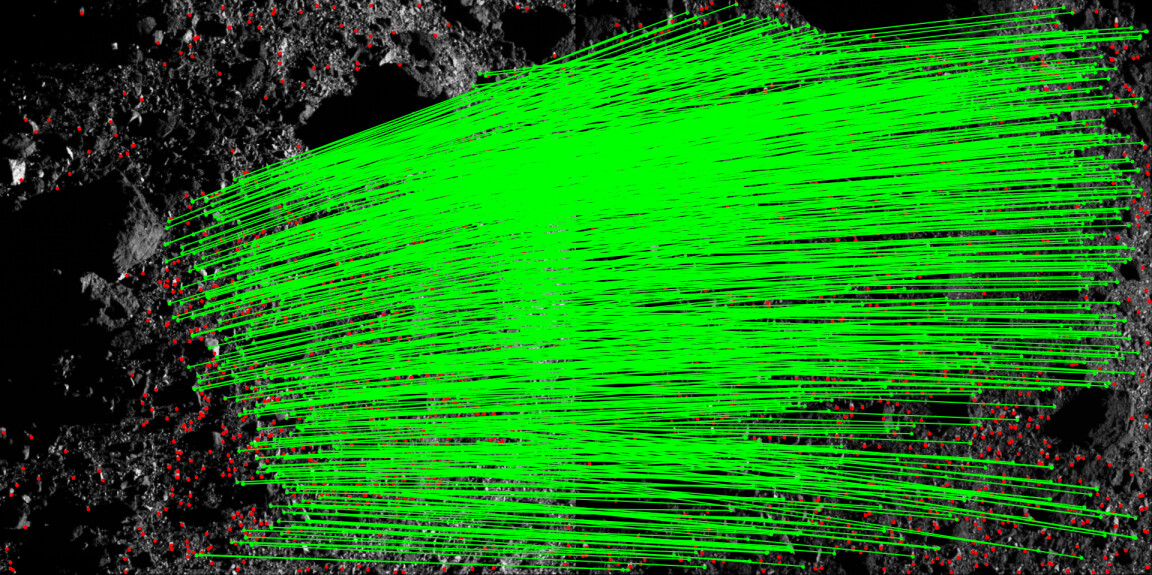}
  \end{subfigure}\\
  \vspace{4pt}
  \begin{subfigure}[t]{\linewidth}
    \includegraphics[width=.935\linewidth]{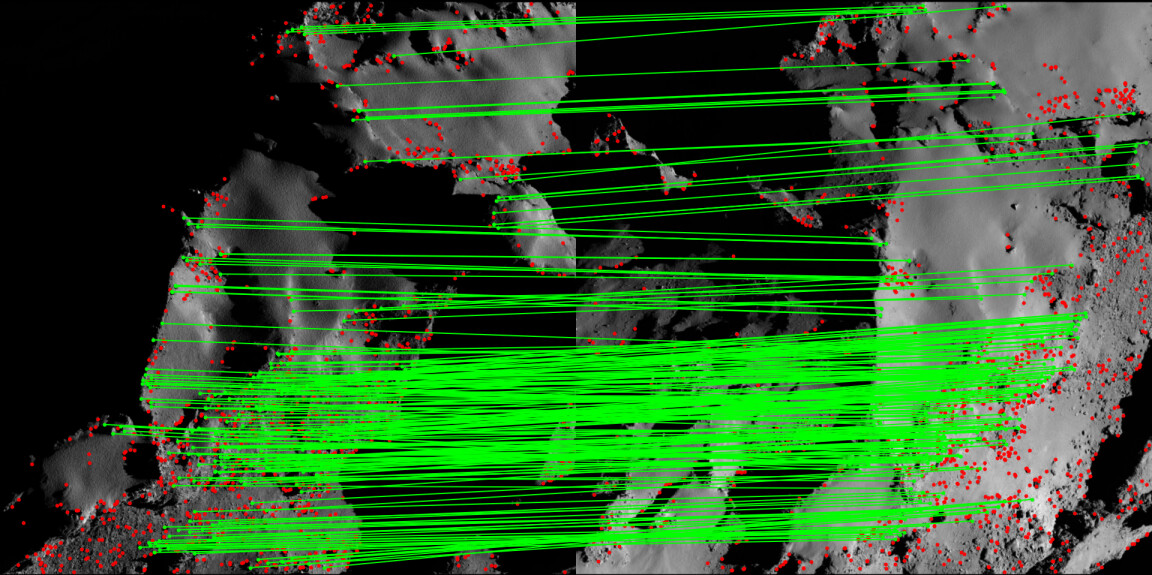}
  \end{subfigure}\\
  \vspace{1pt}
  \begin{subfigure}[t]{\linewidth}
    \includegraphics[width=.935\linewidth]{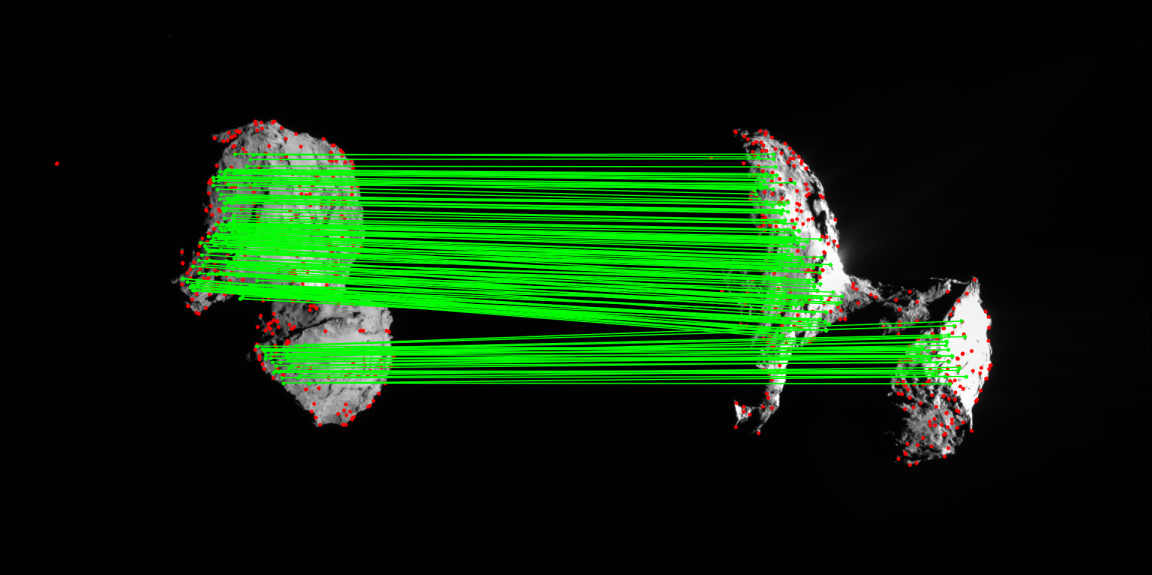}
  \end{subfigure}\\
    \vspace{4pt}
  \begin{subfigure}[t]{\linewidth}
    \includegraphics[width=.935\linewidth]{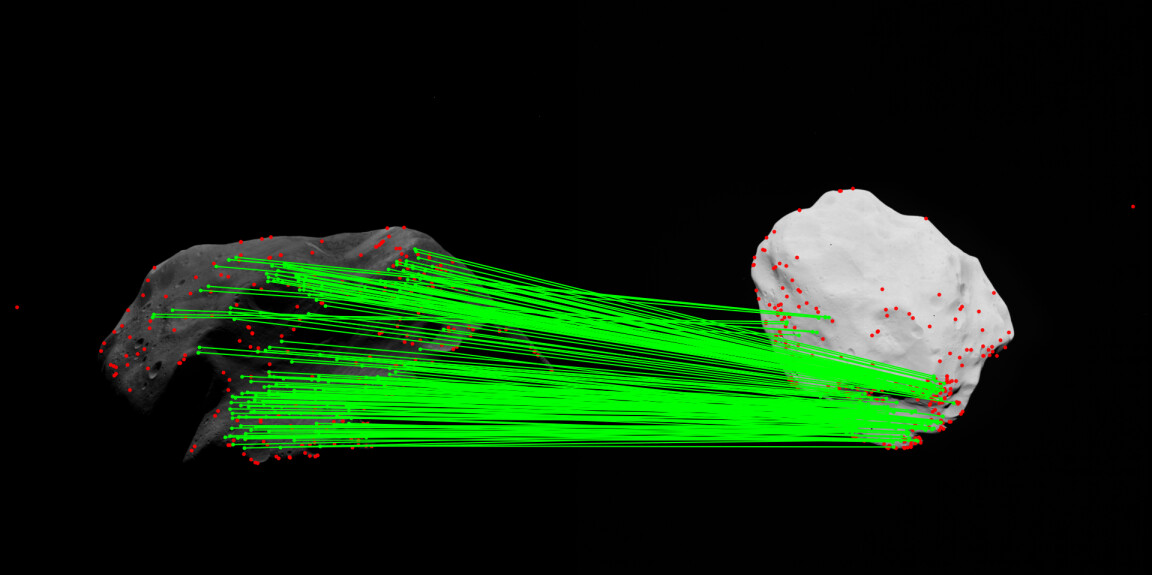}
  \end{subfigure}%
  \caption*{\large{ASLFeat}}
\end{subfigure}%
\caption{\textbf{Qualitative comparison of feature matching.} Correct matches are drawn in green, and the keypoints of incorrect matches are drawn in red.}
\label{fig:eval-qual-compare}
\end{figure*}

\subsection{Results \& Discussion}

We evaluated both handcrafted (i.e., ORB and SIFT) and data-driven (i.e., SuperPoint, R2D2, and ASLFeat) feature detection and description algorithms. 
These results are summarized in Table \ref{tab:matching-metrics-all}, and qualitative comparisons are provided in Figure \ref{fig:eval-qual-compare}. 
We also list the mean and median ground sample distance (GSD) for each dataset, i.e., the distance on the surface of the body covered by each pixel. 
SIFT demonstrates competitive performance on the Dawn and Cassini datasets, outperforming many of the data-driven methods, but suffers when applied to datasets with harsher illumination (i.e., Rosetta @ 67P, OSIRIS-REx @ 101955 Bennu). The efficacy of the orientation encoding of SIFT in certain scenarios can be seen in Figure \ref{fig:eval-epim}, although this behavior does not seem to be typical (see Figure \ref{fig:qerr-vs-precision}).
Superpoint achieves high recall but low precision and generally underperforms with respect to all other methods except ORB. 
Although R2D2 demonstrates high precision and accuracy, we found that the feature matches generally result in poor pose estimates. 
Finally, ASLFeat exhibits high precision, recall, and accuracy, which translates into generally superior relative pose estimates as indicated by the AUC score, and consistently ranks among the top performing methods with respect to all datasets. 
Therefore, we selected ASLFeat network for end-to-end training using the AstroVision data products. 
This is detailed in the next section.

We recognize the very low AUC values for all methods on the Cassini @ Mimas dataset. 
The relatively symmetric and homogeneous surface topology of Mimas generally resulted in low matching precision, and image pairs with high inlier ratios usually corresponded to pairs with relatively low baseline with respect to the radial imaging depth (e.g., Figure \ref{fig:eval-mimas}) resulting in spurious relative translation estimates given even small amounts of measurement noise. 
Indeed, the Cassini @ Mimas images have a mean GSD of 1,176.2 m/pixel, almost four times that of the next highest value. 
We also observed correspondence configurations that resulted in ambiguous essential matrix estimates. 
This suggests that the points may lie close to a so-called \textit{critical surface}~\cite{luong1996ijcv}, special surfaces which yield multiple essential matrix estimates that satisfy Equation \eqref{eq:ess_mat}. 
Detection (e.g., via the iterative method presented in \cite{torr1998cviu}) and rectification (e.g., by considering more views in a full SfM solution) of these degenerate configurations will be the subject of future work. 



\section{Learning Features from Small Body Imagery}\label{sec:results}
In this section, we leverage the AstroVision dataset to train a deep feature detection and description network. 


\begin{figure}[tb!]
    \centering
    \includegraphics[width=\linewidth]{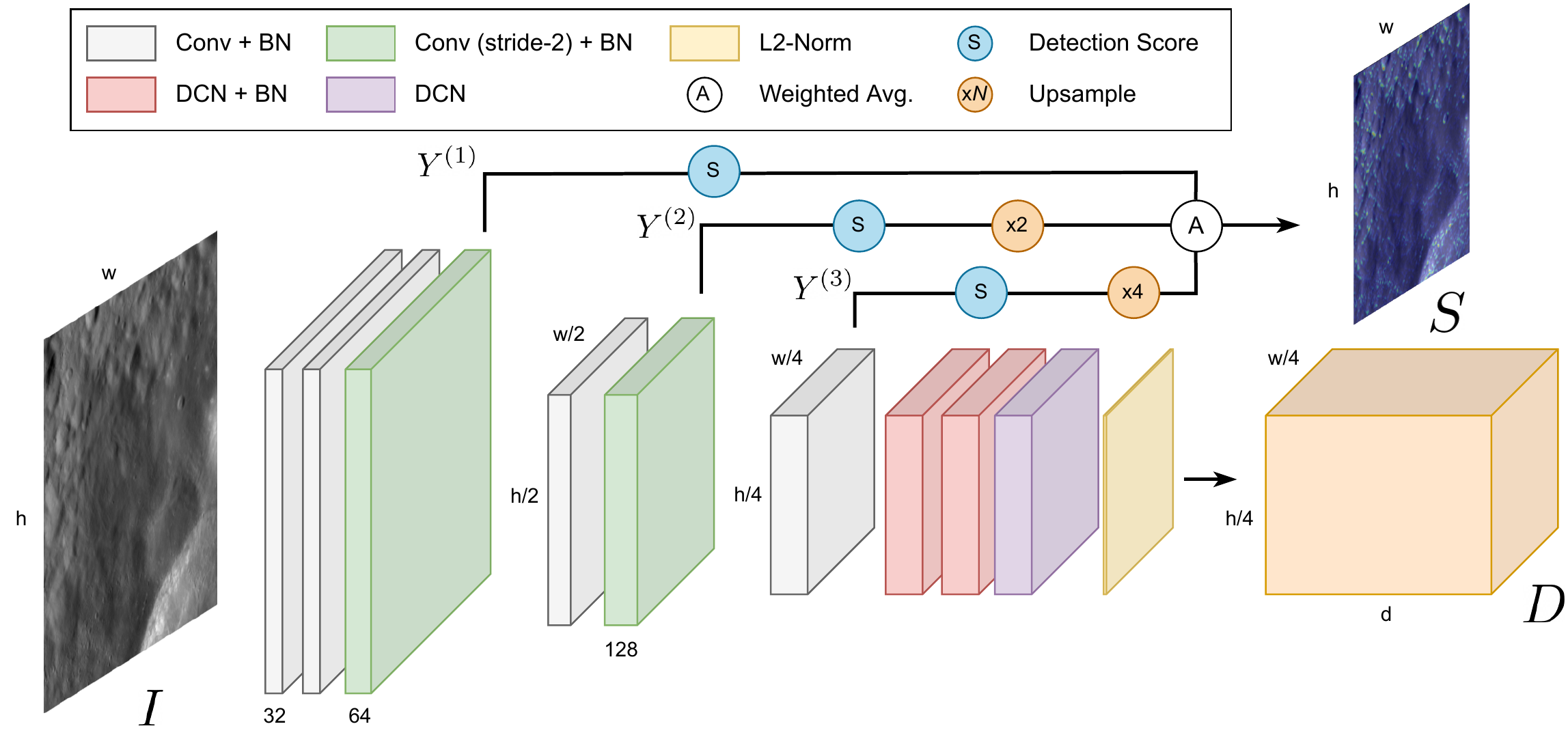}
    \caption{\textbf{ASLFeat architecture.}}
    \label{fig:aslfeat-arch}
\end{figure}

\subsection{Network Architecture} \label{sec:architecture}

Predicated on our evaluation benchmarks, we leverage the ASLFeat~\cite{luo2020cvpr} network architecture. 
Given an image $I\in \mathbb{R}^{h\times w\times c}$, ASLFeat uses a single deep CNN to generate both a detection score (saliency) map $S\in \mathbb{R}^{h \times w}$ and a dense descriptor volume $D\in \mathbb{R}^{h/4 \times w/4 \times d}$.

The score map $S$ is computed through aggregation of elements in intermediate feature maps $Y^{(\ell)}\in \mathbb{R}^{h_\ell\times w_\ell\times b_\ell}$, $\ell = 1,\ldots, 3$. 
Specifically, local peakiness over the channels $Y_c^{(\ell)}$, $c = 1,\ldots, b_\ell$, of the descriptor volume is used to compute channel-wise detection scores (dropping the $\ell$ subscript and superscript for conciseness):
\begin{equation}
    \beta^c_{ij} = \texttt{softplus}\left(y^c_{ij} - \frac{1}{b}\sum_t y^t_{ij}\right)
\end{equation}
where $y^c_{ij}$ is the element at pixel $(i,j)\in \{1,\ldots,h\}\times \{1,\ldots,w\}$ in $Y_c$ and $\texttt{softplus}(x) = \log(1 + \exp(x))$. 
Next, the local detection score is defined as
\begin{equation}
    \alpha^c_{ij} = \texttt{softplus}\left(y^c_{ij} - \frac{1}{|\mathcal{N}(i,j)|}\sum_{(i',j')\in \mathcal{N}(i,j)} y^c_{i'j'}\right)
\end{equation}
where $\mathcal{N}(i,j)$ is the set of 9 neighbors of the pixel $(i,j)$ (including itself).
The elements of the $\ell^{th}$ score map $S^{(\ell)}$ are computed as $s^{(\ell)}_{ij} = \max_c(\alpha^c_{ij}, \beta^c_{ij})$.
Finally, each score map is bilinearly upsampled to the spatial resolution of the input image, and the elements in the final score map $S$ are computed via a weighted average
\begin{equation}
  S = \frac{1}{\sum_{\ell} w_\ell}\sum_{\ell} w_\ell S^{(\ell)},
\end{equation}
where the weights $w_1, w_2, w_3$ have been empirically set to $1, 2, 3$, respectively.

Given correspondences $\mathcal{M} := \{(k, \tau(k))\ |\ \tau: K \leftrightarrow K'\}$ between keypoints $\{\mathbf{p}_k\}_{k\in K}$ and $\{\mathbf{p}_{k'}'\}_{k'\in K'}$ extracted from images $I$ and $I'$, respectively, the total loss is formulated as 
\begin{equation} \label{eq:loss}
    L(D, D', S, S';\mathcal{M}) = \frac{1}{|\mathcal{M}|}\sum_{(l, l') \in \mathcal{M}}\frac{s_l s_{l'}'}{\sum_{(k, k') \in \mathcal{M}} s_k s_{k'}'}m(\mathbf{d}_l, \mathbf{d}_{l'}').
\end{equation}
where $s_k$ ($s_{k'}'$) is the detection score and $\mathbf{d}_k$ ($\mathbf{d}_{k'}'$) is the descriptor at keypoint $\mathbf{p}_k$ ($\mathbf{p}_{k'}'$), and $m(\cdot,\cdot)$ is the descriptor reliability loss.
Note that descriptors and detection scores at \textit{subpixel} locations can be computed through (e.g., bilinear) interpolation of the score map $S$ ($S'$) and descriptor volume $D$ ($D'$). 
ASLFeat leverages a hardest-contrastive margin ranking loss~\cite{choy2019iccv} to enforce descriptor reliability:
\begin{multline}
    m(\mathbf{d}_l, \mathbf{d}_{l'}') = \max\left(\|\mathbf{d}_l - \mathbf{d}_{l'}'\| - M_p, 0\right)\ + \\ 
    \max\left(M_n - \min\left(\min_{k' \neq l'}\|\mathbf{d}_l - \mathbf{d}_{k'}'\|, \min_{k \neq l}\|\mathbf{d}_k - \mathbf{d}_{l'}'\|\right), 0\right),
\end{multline}
where $M_p$ and $M_n$ are the margins for positive and negative pairs, respectively.

The formulated loss $L$ in Equation \eqref{eq:loss} produces a weighted average of the margin terms $m$ over all matches based on their detection scores. 
Thus, in order for the loss to be minimized, the most distinctive correspondences (with a lower margin term) will get higher relative detection scores and vice versa.


\subsection{Implementation Details}

We train ASLFeat using a procedure similar to the original implementation~\cite{luo2020cvpr}. 
The train/test split is shown in Table~\ref{tab:train-test-split}, where we use an approximate 90/10 split.


\begin{table}[tb!]
\footnotesize
\centering
\ra{1.5}
\caption{\textbf{Train/test split.}}
\begin{tabular}{@{}lr@{}}
\toprule
 Dataset                            & \# Images \\
\midrule
\textbf{Train} & \\
 Cassini @ Dione (Saturn IV) (\textbf{D})        & 1381    \\ 
 Cassini @ Janus (Saturn X) (\textbf{J})         & 184       \\
 Cassini @ Phoebe (Saturn IX) (\textbf{P})       & 96        \\
 Cassini @ Rhea (Saturn V) (\textbf{R})          & 665       \\
 Cassini @ Tethys (Saturn III) (\textbf{T})      & 751            \\
 Dawn @ 1 Ceres (\textbf{C})                     & 34916          \\
 Dawn @ 4 Vesta (\textbf{V})                     & 15498          \\
 Hayabusa2 @ 162173 Ryugu (\textbf{U})           & 788          \\
 Mars Express @ Phobos (\textbf{M})              & 890            \\
 NEAR @ 433 Eros (\textbf{E})                    & 11156          \\
 OSIRIS-REx @ 101955 Bennu (\textbf{B})          & 14829          \\
 Rosetta @ 67P (\textbf{G})                      & 23275          \\
\midrule
 TOTAL                              & 104429            \\
\bottomrule
\textbf{Test} & \\
 Cassini @ Epimetheus (Saturn XI)             & 133               \\
 Cassini @ Mimas (Saturn I)                   & 307               \\
 Dawn @ 1 Ceres                               & 3624              \\
 Dawn @ 4 Vesta                               & 2006              \\
 Hayabusa @ 25143 Itokawa                     & 603               \\
 OSIRIS-REx @ 101955 Bennu                    & 1789              \\
 Rosetta @ 21 Lutetia                         & 40                \\
 Rosetta @ 67P                                & 3039              \\
\midrule
 TOTAL                                        & 11541           \\
\bottomrule
\end{tabular}
\label{tab:train-test-split}
\end{table}


\paragraph{Training}

The model is trained from scratch with ground truth cameras and depths from our AstroVision dataset. 
The relative perspective change between an image pair is limited during training, where the angle of rotation between the orientation quaternions of the respective images with respect to the body-fixed frame, as defined by 
Equation~(\ref{eq:err_q}), is used as a metric for the relative perspective change between two images. 
We ignore image pairs with a value greater than $\epsilon_q(\vvec{q}_{\fC_i\fB}, \vvec{q}_{\fC_j\fB}) = 60^{\circ}$. 
The training consumes $\sim$800k image pairs resized to $480\times 480$ using a batch size of 2. 
Ground truth matches for training are computed by first querying the landmark map for sparse correspondences. 
Dense matching is performed on image pairs with at least 128 shared landmarks by projecting a uniform grid of coordinates in the first image into the second image using the ground truth depth and calibration. 
Additionally, the visibility masks are used to remove matches that have keypoints in occluded regions of either image. 
Learning gradients are computed for image pairs that have at least 128 matches, while a maximum of 512 randomly selected matches are used for back propagating gradients. 
Each input image is standardized to have zero mean and unit standard deviation. 
The SGD optimizer is used with momentum of 0.9, and an exponentially decaying learning rate is used with an initial value of 0.1.
We use a two-stage training procedure as suggested by \cite{luo2020cvpr}. 
Specifically, all regular convolutions are trained for 400k iterations in the first stage of training. 
In the second stage, the DCNs are trained with the initial learning rate of 0.01 for another 400k iterations. 


\paragraph{Testing}

Non-maximum suppression is applied (sized 3) to remove detections that are spatially too close.
The position of the detected keypoints is improved using a local refinement and edge-elimination procedure over the detection score map following the approach used in SIFT~\cite{Lowe_2004IJCV}. 
The descriptors are then bilinearly interpolated at the refined (subpixel) positions.
We select the top-$k$ keypoints (nominally $k = 5000$) with respect to their detection scores, and empirically discard those whose scores are lower than 0.50.


\begin{figure}[tb]
    \centering
    \includegraphics[width=\linewidth]{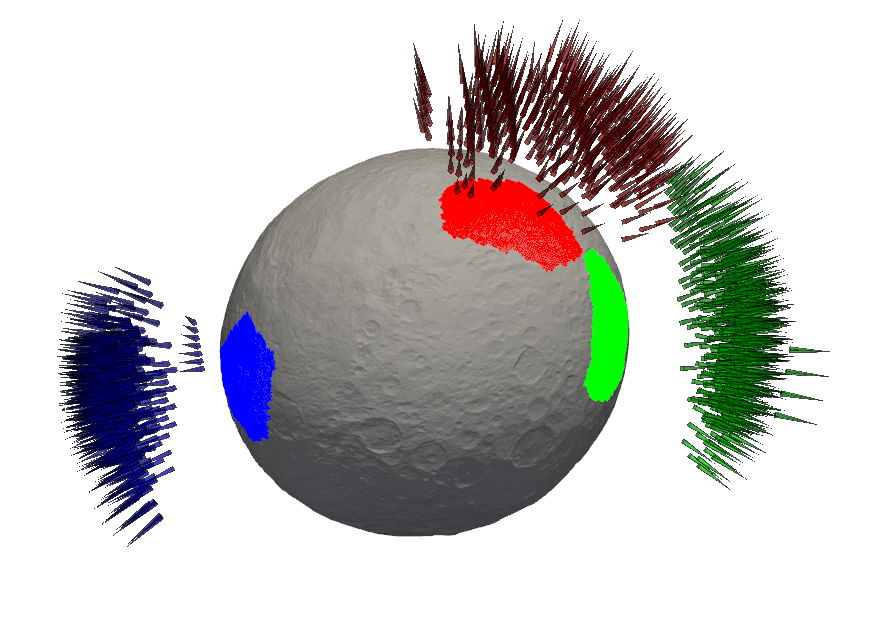}
    \caption{\textbf{Example image clusters.} Three representative clusters from the Dawn @ 1 Ceres dataset where the camera frustums and observed surface area of each cluster are color coded. Only cameras below an altitude of 1,000 km are drawn.}
    \label{fig:ceres-clusters-example}
\end{figure}

\subsection{Experiments}

We withheld data corresponding to 4 different small bodies with variable surface characteristics from training, i.e., Cassini @ Epimetheus, Cassini @ Mimas, Hayabusa @ 25143 Itokawa, and Rosetta @ 21 Lutetia.
In doing so, we test the networks ability to reliably compute features upon arrival at a previously unexplored small body. 
The network was also tested on held-out images of small bodies it saw during training.
This emulates a scenario in which images obtained during earlier stages of a mission could be used to train the network for feature extraction in later phases of the mission. 
In order to minimize overlap between the train and test sets, we cluster images within each dataset according to the backward-projected 3D coordinates of the principle point in each image using $k$-means~\cite{vassilvitskii2006acm} with a value of $k = 64$. 
Seven of these clusters are held-out for testing while the remaining are used during training. 
A visualization of a subset of the clusters for the Dawn @ 1 Ceres dataset in shown in Figure \ref{fig:ceres-clusters-example}.
Matching and verification is conducted using the procedure described in Section \ref{sec:eval_implementation}. 


\begin{table*}[tb!]
\footnotesize
\ra{1.5}
\caption{\textbf{AstroVision-trained model compared to pretrained}. Performance of the AstroVision-trained ASLFeat model compared to pretrained with respect to precision (P), recall (R), accuracy (A), and pose AUC in percentages. See Section \ref{sec:metrics} for metric definitions.}
\begin{adjustbox}{width=\textwidth}
\begin{tabular}{@{}lrlrrrr|rrr@{}}
\toprule
                                &             &               &            &        &       &        & \multicolumn{3}{c}{AUC}                    \\
\cmidrule(lr){8-10}
 Dataset                        & \# Images   & Feature       & \# Matches &    P &   R &   A & @$5^{\circ}$ & @$10^{\circ}$ & @$20^{\circ}$ \\
\midrule 
 Cassini @ Epimetheus (Saturn XI)$^\dagger$ & 133         & ASLFeat               &       386 &  27.4 &  \textbf{29.0} &   \textbf{74.7} &       \textbf{2.7} &        8.2 &       13.7 \\
                                            &             & ASLFeat-CVGBEDTRPJMU  &       \textbf{396} &  \textbf{28.9} &  27.5 &   74.1 &       \textbf{2.7} &        \textbf{8.6} &       \textbf{14.0} \\
\midrule 
 Cassini @ Mimas (Saturn I)$^\dagger$       & 307         & ASLFeat               &       \textbf{372} &  21.8 &  \textbf{15.7} &   65.3 &       \textbf{0.2} &        \textbf{0.2} &        \textbf{0.3} \\
                                            &             & ASLFeat-CVGBEDTRPJMU  &       328 &  \textbf{23.6} &  14.9 &   \textbf{67.1} &       0.0 &        0.1 &        0.2 \\
\midrule 
 Dawn @ 1 Ceres                             & 3624        & ASLFeat               &      \textbf{1535} &  48.4 &  67.8 &   80.2 &      12.9 &       27.1 &       42.4 \\
                                            &             & ASLFeat-CVGBEDTRPJMU  &      1514 &  \textbf{52.8} &  \textbf{71.5} &   \textbf{82.1} &      \textbf{15.9} &       \textbf{31.6} &       \textbf{46.9} \\
\midrule 
 Dawn @ 4 Vesta                             & 2006        & ASLFeat               &      \textbf{1524} &  59.0 &  66.1 &   84.3 &      \textbf{17.5} &       31.9 &       46.0 \\
                                            &             & ASLFeat-CVGBEDTRPJMU  &      1412 &  \textbf{70.3} &  \textbf{69.7} &   \textbf{87.4} &      \textbf{17.5} &       \textbf{33.0} &       \textbf{48.7} \\
\midrule 
 Hayabusa @ 25143 Itokawa$^\dagger$         & 603         & ASLFeat               &       338 &  13.5 &  \textbf{11.3} &   47.5 &       2.2 &        4.2 &        7.6 \\
                                            &             & ASLFeat-CVGBEDTRPJMU  &       \textbf{363} &  \textbf{15.2} &  11.0 &   \textbf{53.7} &       \textbf{2.9} &        \textbf{5.0} &        \textbf{8.8} \\
\midrule 
 OSIRIS-REx @ 101955 Bennu                  & 1789        & ASLFeat               &      \textbf{1378} &  33.1 &  \textbf{30.9} &   68.7 &       \textbf{8.0} &       \textbf{14.4} &       \textbf{20.9} \\
                                            &             & ASLFeat-CVGBEDTRPJMU  &       858 &  \textbf{34.2} &  28.4 &   \textbf{79.5} &       6.7 &       12.6 &       19.3 \\
\midrule 
 Rosetta @ 67P                              & 3039        & ASLFeat               &      \textbf{1147} &  25.0 &  \textbf{24.0} &   62.8 &       3.4 &        6.4 &       10.6 \\
                                            &             & ASLFeat-CVGBEDTRPJMU  &       837 &  \textbf{30.4} &  23.9 &   \textbf{69.8} &       \textbf{4.2} &        \textbf{7.9} &       \textbf{13.4} \\
\midrule 
 Rosetta @ 21 Lutetia$^\dagger$             & 40          & ASLFeat               &       \textbf{970} &  \textbf{42.9} &  \textbf{35.0} &   71.9 &       6.0 &        12.1 &       \textbf{23.8} \\
                                            &             & ASLFeat-CVGBEDTRPJMU  &       778 &  41.3 &  31.1 &   \textbf{76.3} &                 \textbf{8.4} &        \textbf{13.2} &       22.3 \\
\bottomrule
\end{tabular}
\end{adjustbox}\\
\vspace{10pt}
\footnotesize{$^\dagger$ No images of this body were included in the training set}\\
\label{tab:matching-metrics}
\end{table*}

\begin{figure*}[tbp!]
\centering
\input{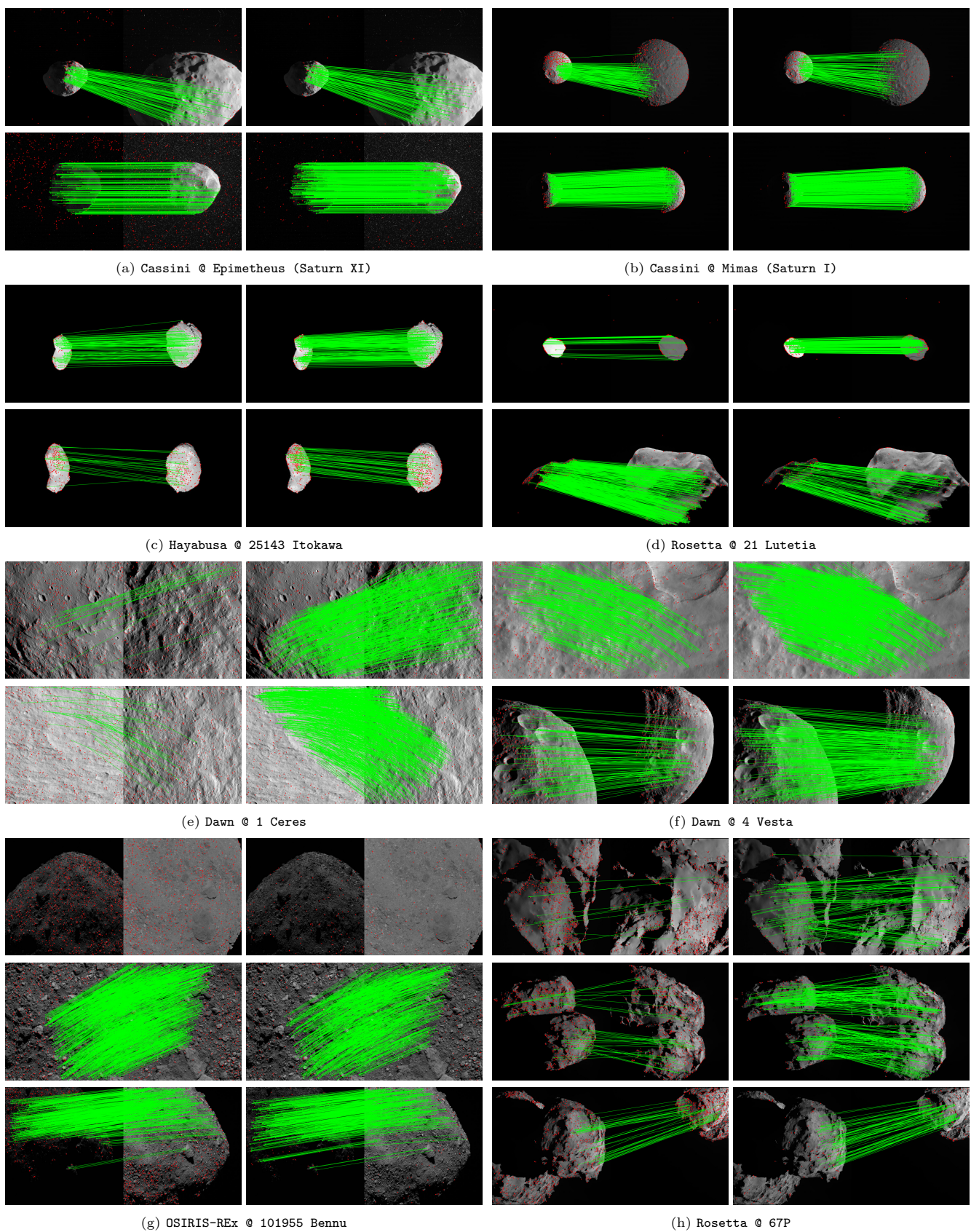}
\caption{\textbf{Qualitative comparison between pretrained (left) and AstroVision-trained (right) model feature matches.} Correct matches are drawn in green, and the keypoints of incorrect matches are drawn in red.}
\label{fig:qual-pre-vs-astro}
\end{figure*}

\begin{figure}[tb!]
\centering
\includegraphics[width=\linewidth]{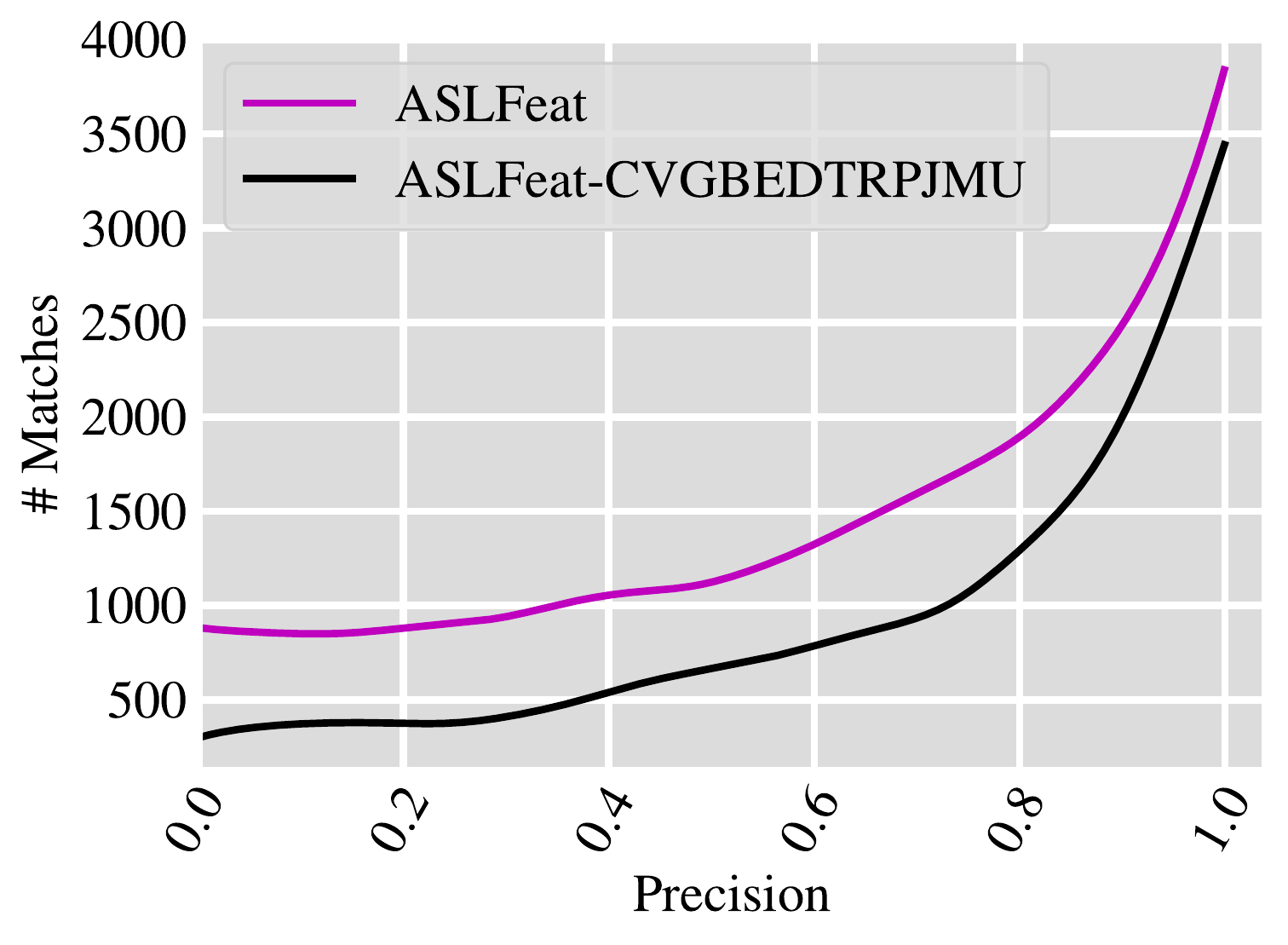}
\caption{\textbf{Precision versus number of matches on the OSIRIS-REx @ 101955 Bennu test set.}}
\label{fig:precision-vs-matches}
\end{figure}

\subsection{Results \& Discussion}

The ASLFeat model trained on AstroVision data, i.e., ASLFeat-CVGBEDTRPJMU, is compared against the pretrained model. 
These results are shown in Table \ref{tab:matching-metrics} and qualitative comparisons are shown in Figure \ref{fig:qual-pre-vs-astro}. 
The model trained on AstroVision consistently outperforms the pretrained model with respect to precision, recall, accuracy, and AUC. 
Importantly, the AstroVision-trained model achieves increased matching performance on many of the novel testing instances, i.e., Cassini @ Epimetheus, Cassini @ Mimas, and Hayabusa @ 25143 Itokawa. 
Indeed, very little is known about the surface characteristics of a small body prior to arrival. 
Our model obtains higher precision and accuracy on all novel test instances with the exception of Rosetta @ 21 Lutetia. 
Despite the lower precision and recall on Rosetta @ 21 Lutetia, we are able to achieve significantly better pose estimates as indicated by the pose AUC metric. 
This is most likely due to the more uniform distribution of matches on the surface of the body, whereas the pretrained network primarily computes matches on the boundary of the body (see Figure \ref{fig:qual-astro-lut}). 
Our model generally exhibits slightly lower recall, but achieves higher AUC on all novel test instances excluding Cassini @ Mimas. 

Moreover, ASLFeat-CVGBEDTRPJMU demonstrates impressive performance on the held-out images of the small bodies it saw during training. 
Our model demonstrates considerably higher performance with respect to all metrics on the Dawn @ 1 Ceres and Dawn @ 4 Vesta test sets. 
Intuitively, training on AstroVision data results in more conservative feature matching on the difficult OSIRIS-REx @ 101955 Bennu and Rosetta @ 67P test sets, as indicated by the higher precision and accuracy and lower recall and number of matches, which exhibit hard and rapidly changing illumination, significant perspective changes, and repetitive surface characteristics. 
We achieve slightly lower pose AUC as compared to the pretrained model for the OSIRIS-REx@ 101955 Bennu test set despite having higher precision and significantly higher accuracy. 
This is most likely due to the reduced number of matches, although this is primarily restricted to low precision image pairs as shown in Figure \ref{fig:precision-vs-matches}.
Indeed, for difficult image pairs with precision close to zero, ASLFeat-CVGBEDTRPJMU features typically result in an order of magnitude fewer incorrect matches compared to the pretrained model. 
An example of this is provided in Figure \ref{fig:qual-astro-bennu}. 


\begin{table}[htp!]
\footnotesize
\ra{1.5}
\caption{\textbf{ASLFeat-B Benchmark performance.} Performance of ASLFeat-B, i.e., ASLFeat trained on OSRIS-REx @ 101955 Bennu data only, with respect to precision (P), recall (R), accuracy (A), and pose AUC in percentages. See Section \ref{sec:metrics} for metric definitions.}
\begin{adjustbox}{width=\linewidth}
\begin{tabular}{@{}lrrrr|rrr@{}}
\toprule
                                 &            &        &       &        & \multicolumn{3}{c}{AUC}                      \\
\cmidrule(lr){6-8}
 Dataset                          & \# Matches &      P &     R &      A & @$5^{\circ}$ & @$10^{\circ}$ & @$20^{\circ}$ \\
\midrule 
 Cassini @ Epimetheus (Saturn XI) &       475 &  28.6 &  26.7 &   60.6 &       2.4 &        7.2 &       12.1 \\
\midrule 
 Cassini @ Mimas (Saturn I)       &       306 &  12.1 &   8.7 &   58.1 &       0.0 &        0.0 &        0.1 \\
\midrule 
 Dawn @ 1 Ceres                   &      1631 &  48.8 &  61.7 &   76.3 &      12.2 &       26.4 &       41.6 \\
\midrule 
 Dawn @ 4 Vesta                   &      1430 &  48.2 &  54.9 &   78.2 &      11.7 &       23.0 &       34.9 \\
\midrule 
 Hayabusa @ 25143 Itokawa         &       337 &   9.6 &   6.9 &   43.3 &       1.6 &        2.8 &        4.9 \\
\midrule 
 OSIRIS-REx @ 101955 Bennu        &      1400 &  35.4 &  34.7 &   70.7 &       7.9 &       14.9 &       21.9 \\
\midrule 
 Rosetta @ 67P                    &      1075 &  22.2 &  20.9 &   58.6 &       2.4 &        4.8 &        8.2 \\
\midrule 
 Rosetta @ 21 Lutetia             &       561 &  25.8 &  16.3 &   67.3 &       1.0 &        3.3 &        9.2 \\
\bottomrule
\end{tabular}
\end{adjustbox}\\
\label{tab:bennu-ablation}
\end{table}

We experimented with training the network on OSIRIS-REx @ 101955 Bennu data only, referred to as ASLFeat-B, as we suspected the network may be prioritizing discrimination of other feature classes more relevant to the other training instances due to the unique and challenging surface features of Bennu and the lower number of training images relative to some of the other missions (e.g., Dawn @ 1 Ceres, Rosetta @ 67P). 
Benchmarking results for this experiment are presented in Table \ref{tab:bennu-ablation}. 
ASLFeat-B achieves increased performance with respect to all metrics compared to the pretrained model on the OSIRIS-REx @ 101955 Bennu dataset.
We postulate that adding more small body instances with similar surface characteristics will increase performance. 



\begin{figure*}[htb!]
\centering
\input{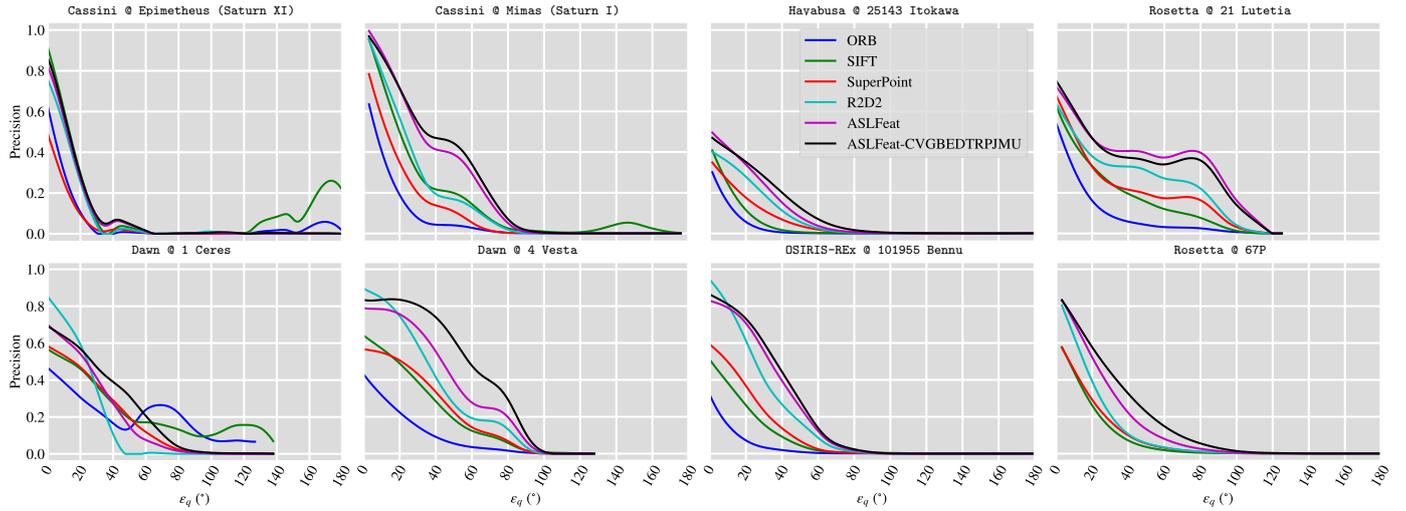}
\caption{\textbf{Perspective change versus precision.} Perspective change is measured by $\epsilon_q(\vvec{q}_{\fC_i\fB}, \vvec{q}_{\fC_j\fB})$ as defined in Equation \eqref{eq:err_q}, i.e., the minimum rotation angle between the respective cameras.}
\label{fig:qerr-vs-precision}
\end{figure*}


\begin{figure}[htp!]
\begin{subfigure}[t]{1.0\linewidth}
  \centering
  \caption*{\normalsize{\texttt{OSIRIS-REx @ 101955 Bennu}}}
  \vspace{-7pt}
  \includegraphics[width=\linewidth]{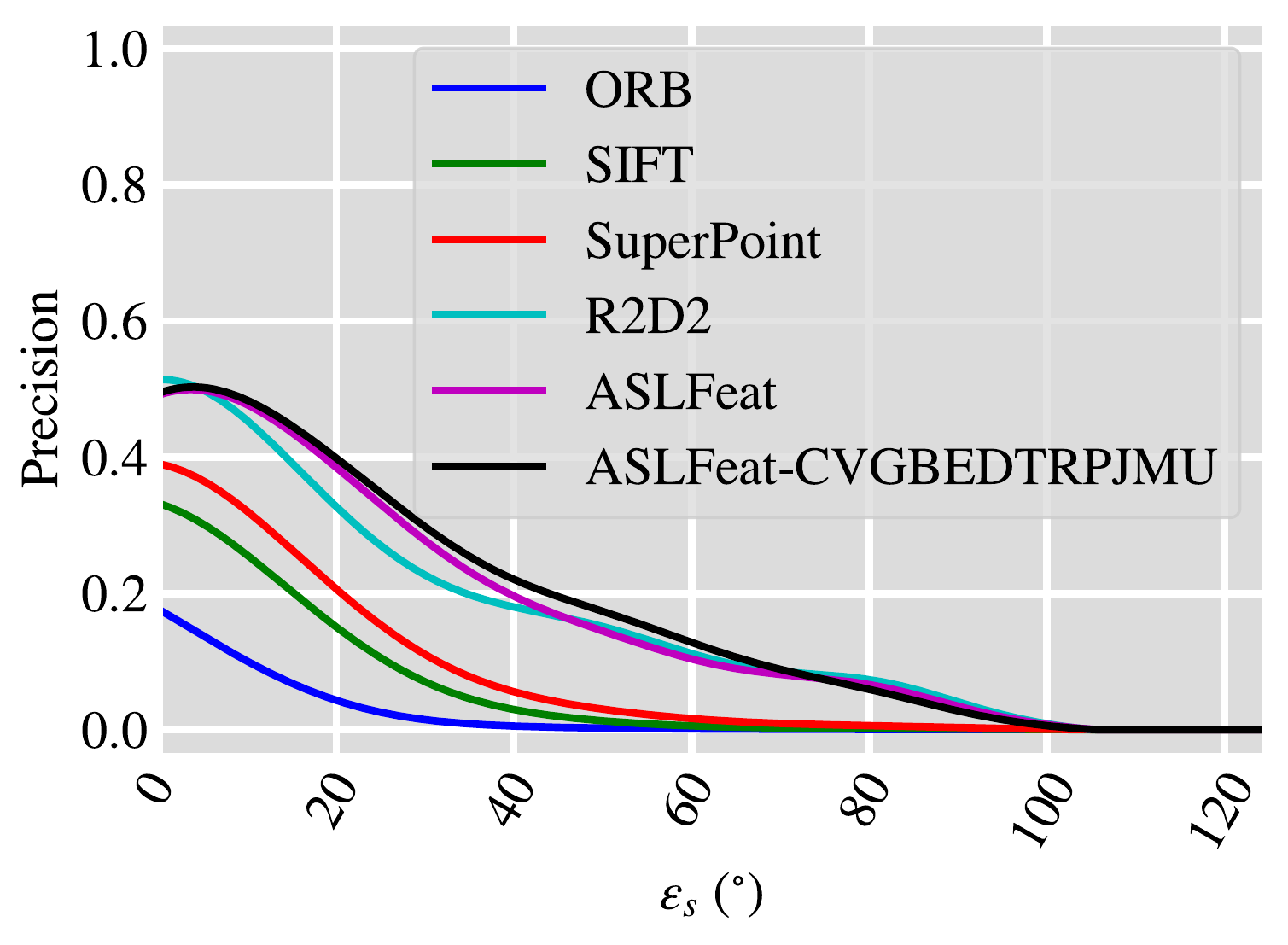}
\end{subfigure}\\
\begin{subfigure}[t]{1.0\linewidth}
  \centering
  \caption*{\normalsize{\texttt{Rosetta @ 67P}}}
  \vspace{-7pt}
  \includegraphics[width=\linewidth]{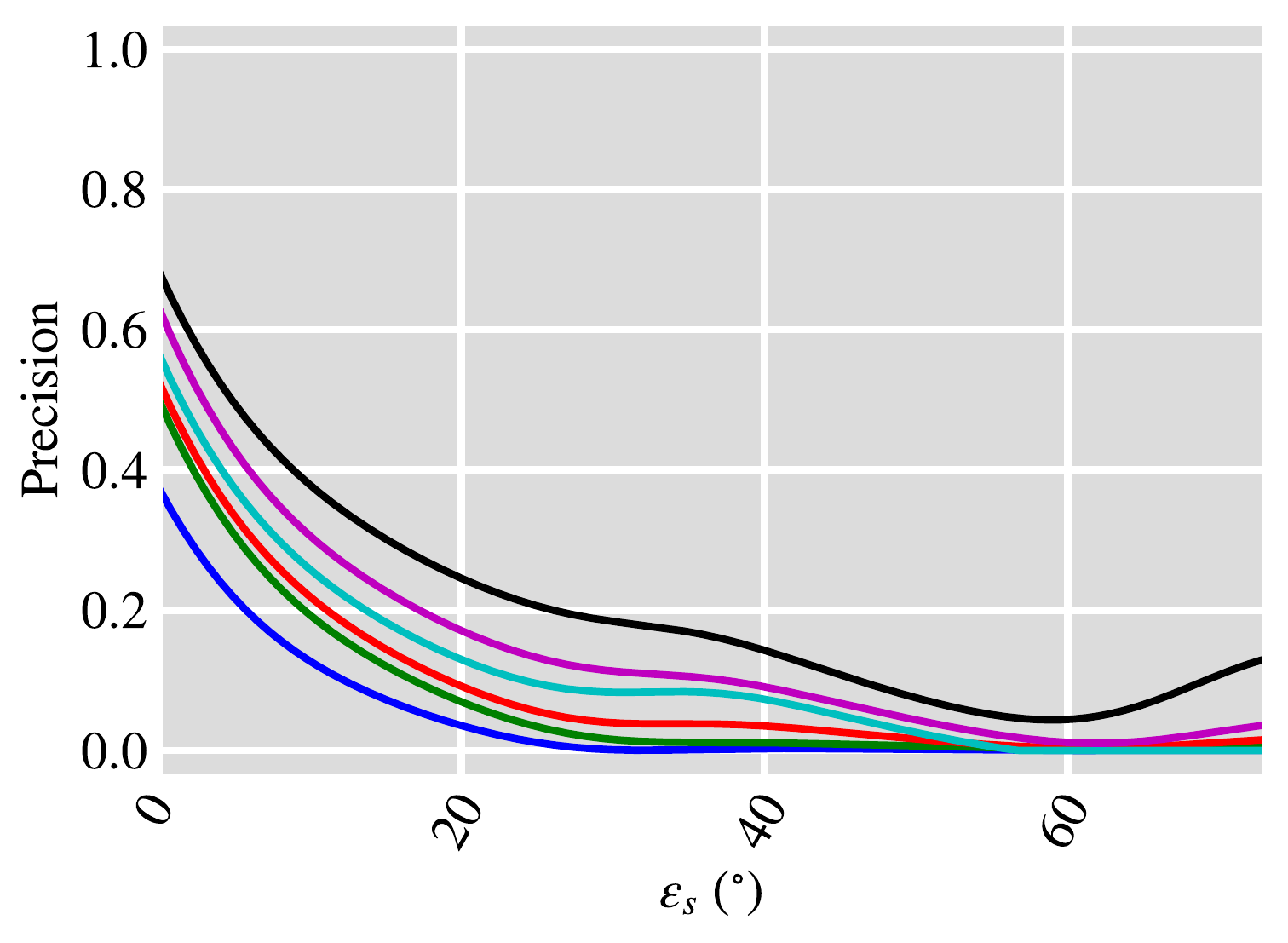}
\end{subfigure}%
\caption{\textbf{Illumination change versus precision.} Illumination change is measured by $\epsilon_s$ as defined in Equation \eqref{eq:err_s}, i.e., the angle between the Sun vectors in the respective cameras.}
\label{fig:illumination-vs-precision}
\end{figure}

We also compared matching precision against perspective and illumination changes in Figures \ref{fig:qerr-vs-precision} and \ref{fig:illumination-vs-precision}. 
We leverage Equation \eqref{eq:err_q} as a measure for perspective change, and
\begin{equation} \label{eq:err_s}
    \epsilon_s := \cos^{-1}(\Hat{\vvec{s}}^{\fC_i}\cdot\Hat{\vvec{s}}^{\fC_j})
\end{equation}
as a measure of illumination change, where $\Hat{\vvec{s}}^{\fC_i}$ and $\Hat{\vvec{s}}^{\fC_j}$ denote the (unit) Sun vector in $\fC_i$ and $\fC_j$, respectively. 
Our model exhibits superior invariance to both perspective and illumination changes for all test sets. 


\begin{figure*}[tb!]
\centering
\input{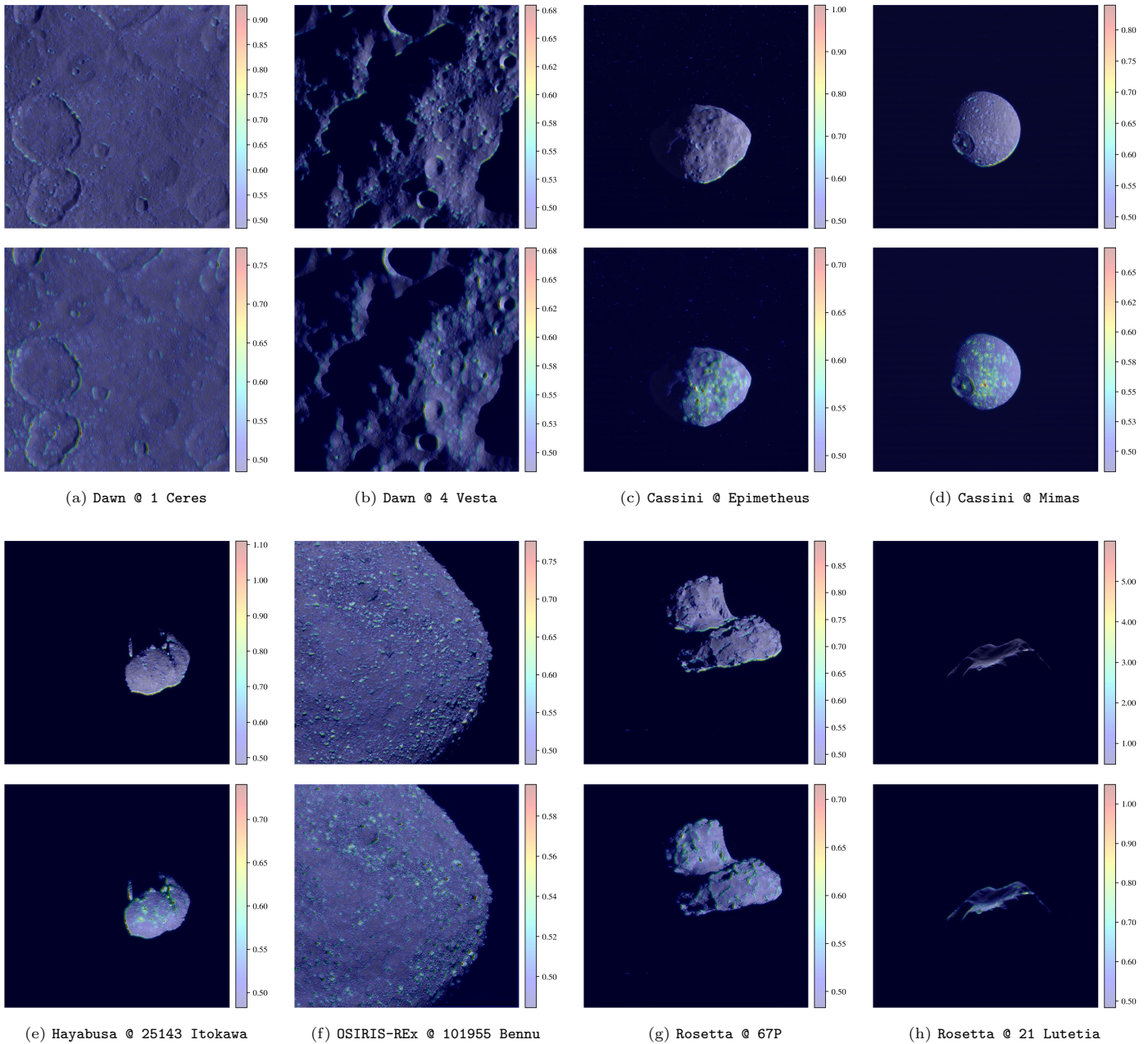}
\caption{\textbf{Detection score maps.} Qualitative comparison of detection score maps for the pretrained (top) and ASLFeat-CVGBEDTRPJMU (bottom) models. The color bar indicates the models confidence in the feature corresponding to that pixel.}
\label{fig:score-maps}
\end{figure*}

The detection score maps $S$ for the respective models, as described in Section \ref{sec:architecture}, are visualized in Figure \ref{fig:score-maps}. 
It can been seen that the pretrained model repeatably places a high confidence to edges formed from hard shadowing and to features on the boundary between the body and deep space. 
Features in these regions are known to be relatively unreliable and not repeatable, as the appearance of these features can change dramatically due to deformation of the shadows, or become completely occluded, as the body rotates about its axis~\cite{morrell2020}. 
However, the model trained on AstroVision learns to assign a low confidence to these regions and gives a higher confidence to features corresponding to salient topographic structures such as rocky outcroppings and crater rims.


\subsection{Ablation Study for Masking} \label{sec:masking}

We found that utilizing the visibility masks during training led to faster convergence and greatly increased overall performance. 
Specifically, a grid of image coordinates from the first image was projected into the second image using the ground truth calibrations of each camera and the depth map to generate a collection of ground truth matches during training. 
We mask keypoints according to the visibility masks of each image and ignore matches with keypoints in occluded regions during training. 
The results in Table \ref{tab:matching-metrics-masking} demonstrate how utilizing the visibility masks during training greatly improves the overall performance. 
It can be seen that there is a slight degradation in precision for the Dawn @ 1 Ceres dataset, which has significantly fewer shadowing occlusions as compared to the other datasets. 
This could indicate that exposing the network to training instances in occluded regions could benefit matching performance. 
Investigating training strategies that allow the network to effectively learn from occluded matches will be the subject of future work.

\begin{table}[tb!]
\footnotesize
\ra{1.5}
\caption{\textbf{Masking ablation study results.} Performance of ASLFeat-CVGBEDTRPJMU trained with and without masking with respect to precision (P), recall (R), and accuracy (A). See Section \ref{sec:metrics} for metric definitions.}
\begin{adjustbox}{width=\linewidth}
\begin{tabular}{@{}llrrrr|rrr@{}}
\toprule
                                &               &            &        &       &        & \multicolumn{3}{c}{AUC}                    \\
\cmidrule(lr){7-9}
 Dataset                        & Feature       & \# Matches &    P &   R &   A & @$5^{\circ}$ & @$10^{\circ}$ & @$20^{\circ}$ \\
\midrule 
 Cassini @ Epimetheus (Saturn XI)$^\dagger$ & masking      &       \textbf{396} &  \textbf{28.9} &  \textbf{27.5} &   \textbf{74.1} &       2.7 &        8.6 &       14.0 \\
                                            & w/o masking  &       391 &  28.1 &  25.8 &   63.1 &       \textbf{2.8} &        \textbf{8.8} &       \textbf{14.7} \\
\midrule 
 Cassini @ Mimas (Saturn I)$^\dagger$       & masking      &       \textbf{328} &  \textbf{23.6} &  \textbf{14.9} &   \textbf{67.1} &       0.0 &        \textbf{0.1} &        \textbf{0.2} \\
                                            & w/o masking  &       330 &  15.0 &  11.3 &   57.6 &       0.0 &        0.0 &        0.0 \\
\midrule 
 Dawn @ 1 Ceres                             & masking      &      1514 &  52.8 &  \textbf{71.5} &   \textbf{82.1} &      \textbf{15.9} &       \textbf{31.6} &       \textbf{46.9} \\
                                            & w/o masking  &      \textbf{1683} &  \textbf{56.9} &  71.3 &   80.2 &      13.5 &       29.0 &       45.3 \\
\midrule 
 Dawn @ 4 Vesta                             & masking      &      1412 &  \textbf{70.3} &  69.7 &   \textbf{87.4} &      \textbf{17.5} &       \textbf{33.0} &       \textbf{48.7} \\
                                            & w/o masking  &      \textbf{1494} &  63.7 &  \textbf{70.9} &   84.9 &      14.2 &       28.0 &       43.2 \\
\midrule 
 Hayabusa @ 25143 Itokawa$^\dagger$         & masking      &       363 &  \textbf{15.2} &  11.0 &   \textbf{53.7} &       \textbf{2.9} &        \textbf{5.0} &        \textbf{8.8} \\
                                            & w/o masking  &       \textbf{552} &  13.8 &  \textbf{11.6} &   39.8 &       2.4 &        4.5 &        7.8 \\
\midrule 
 OSIRIS-REx @ 101955 Bennu                  & masking      &       858 &  \textbf{34.2} &  28.4 &   \textbf{79.5} &       \textbf{6.7} &       \textbf{12.6} &       \textbf{19.3} \\
                                            & w/o masking  &      \textbf{1076} &  32.9 &  \textbf{30.1} &   75.9 &       6.0 &       11.9 &       18.5 \\
\midrule 
 Rosetta @ 67P                              & masking      &       837 &  \textbf{30.4} &  \textbf{23.9} &   \textbf{69.8} &       \textbf{4.2} &        \textbf{7.9} &       \textbf{13.4} \\
                                            & w/o masking  &       \textbf{952} &  26.4 &  23.8 &   61.1 &       3.4 &        6.3 &       10.6 \\
\midrule 
 Rosetta @ 21 Lutetia$^\dagger$             & masking      &       778 &  \textbf{41.3} &  \textbf{31.2} &   \textbf{76.3} &       \textbf{8.4} &       \textbf{13.2} &       \textbf{22.3} \\
                                            & w/o masking  &       \textbf{943} &  33.7 &  27.9 &   70.7 &       2.5 &        5.7 &       13.8 \\
\bottomrule
\end{tabular}
\end{adjustbox}\\
\vspace{10pt}
\tiny{$^\dagger$ No images of this body were included in the training set}\\
\label{tab:matching-metrics-masking}
\end{table}

\section{Conclusion}

In this paper we presented a first-of-a-kind dataset composed of densely annotated images of small celestial bodies acquired during past and ongoing missions. 
The AstroVision dataset was leveraged to develop a novel benchmark suite for evaluation of feature detection and description methods on \textit{real} remote imagery of small bodies. 
Moreover, we showed that leveraging the Astrovision data for training a deep feature detection and description network increases matching and pose estimation performance on small bodies with a wide variety of surface characteristics, including on bodies completely unseen during training. 
We believe that feature extraction based on deep learning is a promising alternative to current human-in-the-loop practices used in state-of-the-practice small body 3D shape reconstruction methods, e.g., SPC~\cite{gaskell2008}. 
Furthermore, pending ongoing advancements in space-grade multi-core processors~\cite{lentaris2017,george2018,kosmidis2020,kothari2020final}, deep learning approaches to feature extraction could feasibly be implemented for autonomous relative navigation onboard future spacecraft. 
Finally, we postulate that the use of AstroVision will extend beyond feature detection and description and enable the deployment of a variety of new deep learning methods for deep space applications, ultimately leading to a significant increase in small body science mission capabilities. 
The code, data, and trained models will be made available to the public at \href{https://github.com/astrovision}{\texttt{https://github.com/astrovision}}. 


\section*{Acknowledgements}

This work was supported by a NASA Space Technology Graduate Research Opportunity.
The authors would like to thank Kenneth Getzandanner and Andrew Liounis from NASA Goddard Space Flight Center for several helpful discussions, and Robert Gaskell from the Planetary Science Institute for providing detailed SPC shape models.


\appendix


\section{Photometric Calibration Details} \label{sec:photo_calib}

Table \ref{tab:calibration} defines the different types of photometric calibration applied to each of the datasets. 
\textbf{Bias + Dark + Smear} indicates that sensor bias subtraction, dark current (warm pixel) removal, and readout smear correction have been applied to the images. 
\textbf{Radiometric} indicates radiometric calibration was conducted to convert the raw sensor measurements to units of radiance or reflectance. 
\textbf{Deblurred} refers to applying a deblurring filter to the radiometrically calibrated images.
More details can be found in the technical reports for the respective instrumentation: Cassini Imaging Science Subsystem (ISS)~\cite{cassini_iss}, Dawn Framing Camera~\cite{dawn_fc}, NEAR Multispectral Imager (MSI)~\cite{murchie2002}, OSIRIS-REx Camera Suite (OCAMS)~\cite{orex_ocams}, Rosetta NavCam~\cite{ros_navcam}, and Mars Express High Resolution Stereo Camera (HRSC)~\cite{mexp_hrsc}.

\begin{table*}[htb!]
\footnotesize
\centering
\ra{1.5}
\caption{\textbf{Photometric calibration specifications.}}
\begin{adjustbox}{width=\textwidth}
\begin{tabular}{@{}lccccccc@{}}
\toprule
Calibration type    & Cassini    & Dawn       & Hayabusa   & Mars Express & NEAR       & OSIRIS-REx & Rosetta   \\
\midrule
\textbf{Bias + Dark + Smear} & \checkmark & \checkmark &            & \checkmark   & \checkmark & \checkmark & \checkmark \\
\textbf{Radiometric}         & \checkmark & \checkmark &            & \checkmark   & \checkmark & \checkmark & \checkmark \\
\textbf{Deblurred}           &            &            &            &              & \checkmark &            &            \\
\bottomrule
\end{tabular}
\end{adjustbox}
\label{tab:calibration}
\end{table*}


\bibliography{mybibfile}

\end{document}